\documentclass[11pt,titlepage,a4paper]{article}
\usepackage{bozhomac}   
\usepackage{bozhlogo}   
\usepackage{cite}   

\title{\bfseries    \vspace*{-1.678902345in}
{\huge Connection theory
\\[1.22ex]  on differentiable fibre bundles:         }
\\ \vspace{1.1ex}  {\LARGE  A pedagogical introduction    }
}

\vspace{1.7ex}

\author{
Bozhidar Z.\ Iliev
\thanks{Laboratory of Mathematical Modeling in Physics,
Institute for Nuclear Research and \mbox{Nuclear} Energy,
Bulgarian Academy of Sciences,
Boul.\ Tzarigradsko chauss\'ee~72, 1784 Sofia, Bulgaria}
\thanks{E-mail address: bozho@inrne.bas.bg}
\thanks{URL: http://theo.inrne.bas.bg/$\sim$bozho/}
}

\date{
 \vspace{2.27ex}\ShortTitle{Connection theory: A pedagogical introduction}  \\[0.27ex]
 \vspace{3.27ex}
\small
    \begin{tabular}{r@{$\colon\to~$}l}
 \vspace{0.27ex} Produced   & \fbox{\today} \\[0.27ex]
    \end{tabular} \\[3.27ex]
\normalsize
\small
    \begin{tabular}{r@{$\colon~$}l}
\normalsize\sffamily\bfseries
 \vspace{0.27ex} http://www.arXiv.org e-Print archive No. &
\normalsize\sffamily\bfseries
math-ph/0510004				\\[1.27ex]
    \end{tabular} \\[-0.27ex]
\normalsize
 \vspace{4.27ex}{\Huge \BOZHO}  \\[4.27ex]
    \begin{tabular}{r@{\hspace{0.512em}}|@{\hspace{0.512em}}l}
 \vspace{0.27ex}   
&
 \vspace{0.27ex} 
    \end{tabular} \\[1.27ex]
 \vspace{0.27ex}    \\[0.27ex]
}

\begin{document}        

\renewcommand{\thepage}{\roman{page}}

\renewcommand{\thefootnote}{\fnsymbol{footnote}} 
\maketitle              
\renewcommand{\thefootnote}{\arabic{footnote}}   

\tableofcontents        


\begin{abstract}

The paper contains a partial review on the general connection theory on
differentiable fibre bundles. Particular attention is paid on (linear)
connections on vector bundles. The (local) representations of connections in
frames adapted to holonomic and arbitrary frames is considered.

\end{abstract}

\renewcommand{\thepage}{\arabic{page}}


\section {Introduction}
\label{Introduction}

	This is a partial review of the connection theory on differentiable
fibre bundles. From different view\ndash points, this theory can be found in
many works, like%
~\cite{Kobayashi-1957,K&N-1,Sachs&Wu,Nash&Sen,Warner,Bishop&Crittenden,
Yano&Kon,Steenrod,Sulanke&Wintgen,Bruhat,Husemoller,Mishchenko,R_Hermann-1,
Greub&et_al.-1,Atiyah,Dandoloff&Zakrzewski,Tamura,Hicks,Sternberg,Rahula,
Mangiarotti&Sardanashvily}.
The presentation of the material in sections~\ref{Sect2}--\ref{Sect5},
containing the grounds of the connection theory, follows some of the main ideas
of~\cite[chapters~1 and~2]{Rahula}, but their realization here is quite
different and follows the modern trends in differential geometry. Since in the
physical literature one can find misunderstanding or not quite rigorous
applications of known mathematical definitions and results, the text is written
in a way suitable for direct application in some regions of theoretical
physics.

	The work is organized as follows.

	In Sect.~\ref{Sect2} is collected some introductory material, like
the notion of Lie derivatives and distributions on manifolds, needed for our
exposition. Here some of our notation is fixed too.

	Section~\ref{Sect3} is devoted to the general connection theory on
bundles whose base and bundles spaces are differentiable manifolds. In
Subsect.~\ref{Subsect3.1} are reviewed some coordinates and frames/bases on
the bundle space which are compatible with the fibre structure of a bundle.
Subsect.~\ref{Subsect3.2} deals with the general connection theory. A connection
on a bundle is defined as a distribution on its bundle space which is
complimentary to the vertical distribution on it. The notion of parallel
transport generated by connection and of specialized frame are introduced.
The fibre coefficients and fibre components of the curvature of a connection
are defined via part of the components of the anholonomy object of a
specialized frame. Frames adapted to local bundle coordinates are introduced
and the local (2\ndash index) coefficients in them of a connection are
defined; their transformation law is derived and it is proved that a
geometrical object with such transformation law uniquely defines a
connection. The parallel transport equation in their terms is derived and it
is demonstrated how from it the equation of geodesics on a manifold can be
obtained.

	In Sect.~\ref{Sect4}, the general connection theory from
Sect.~\ref{Sect3} is specified on vector bundles. The most important
structures in/on them are the ones that are consistent/compatible with the
vector space structure of their fibres. The vertical lifts of sections of
a vector bundle and the horizontal lifts of  vector fields on its base are
investigated in more details in Subsect.~\ref{Subsect4.1}. The general
results are specified on the (co)tangent bundle over a manifold in
Subsect.~\ref{Subsect4.2}. Subsect.~\ref{Subsect4.3} is devoted to linear
connections on vector bundles, \ie connections such that the assigned to
them parallel transport is a linear mapping. It is proved that the 2\ndash
index coefficients of a linear connection are linear in the fibre
coordinates, which leads to the introduction of the (3\ndash index)
coefficients of the connection; the latter coefficients being defined on the
base space. The transformations of different objects under a change of vector
bundle coordinates are explored. The covariant derivatives are introduced and
investigated in Subsect.~\ref{Subsect4.4}. They are defined via the Lie
derivatives and a mapping realizing an isomorphism between the
vertical vector fields on the bundle space and the sections of the bundle.
The equivalence of that definition with the widespread one, defining them as
mappings on the module of sections of the bundle with suitable properties, is
proved. Some properties of the covariant derivatives are explored. In
Subsect.~\ref{Subsect4.5}, the affine connections on vector bundles are
considered briefly.

	Section~\ref{Sect5} deals briefly with morphisms between bundles with
connections defined on them.

	In section~\ref{Sect6}, some of the results of the previous sections
are generalized when frames more general than the ones generated by local
coordinates on the bundle space are employed. The most general such frames,
compatible with the fibre structure, and the frames adapted to them are
investigated. The main differential\ndash geometric objects, introduced in
the previous sections, are considered in such general frames. Particular
attention is paid to the case of a vector bundle. In vector bundles, a
bijective correspondence between the mentioned general frames and pairs of
bases, in the vector fields over the base and in the sections of the bundle,
is proved. The (3\ndash index) coefficients of a connection in such pairs of
frames and their transformation laws are considered. The covariant
derivatives are also mentioned on this context.

	Section~\ref{Conclusion} closes the paper with some concluding
remarks.


\section {Preliminaries}
\label{Sect2}

        This section contains an introductory material, notation etc.\ that
will be needed for our exposition. The reader is referred for details to
standard books on differential geometry, like~\cite{K&N,Warner,Poor}.

        A differentiable finite-dimensional manifold over a field $\field$
will be denoted typically by $M$. Here $\field$ stands for the field
$\field[R]$ of real or the field $\field[C]$ of complex numbers,
$\field=\field[R],\field[C]$. The manifolds we consider are supposed to be
smooth of class $C^2$.~%
\footnote{~%
Some of our definitions or/and results are valid also for $C^1$ or even $C^0$
manifolds, but we do not want to overload the material with continuous
counting of the required degree of differentiability of the manifolds
involved. Some parts of the text admit generalizations on more general
spaces, like the topological ones, but this is out of the subject of the
present work.%
}
The sets of vector fields, realized as first order differential operators,
and of differential k\ndash forms, $k\in\field[N]$, over $M$ will be denoted by
$\mathcal{X}(M)$ and $\Lambda^k(M)$, respectively. The space tangent (resp.\
cotangent) to $M$ at $p\in M$ is $T_p(M)$ (resp.\ $T_p^*(M)$) and
$(T(M),\pi_T,M)$ (resp. $(T^*(M),\pi_{T^*},M)$) will stand for the tangent
(resp.\ cotangent) bundle over $M$. The value of $X\in\mathcal{X}(M)$ at
$p\in M$ is $X_p\in T_p(M)$ and the action of $X$ on a $C^1$ function
$\varphi\colon M\to\field$ is a function $X(\varphi)\colon M\to \field$ with
$X(\varphi)|_p:=X_p(\varphi)\in\field$.

        If $M$ and $\bar{M}$ are manifolds and $f\colon \bar{M}\to M$ is  a
$C^1$ mapping, then $f_*:=\od f:=T(f)\colon T(\bar{M})\to T(M)$ denotes the
induced tangent mapping (or differential) of $f$ such that, for $p\in M$,
 $f_*|_p :=\od f|_p := T_p(f)\colon T_p(\bar{M})\to T_{f(p)}(M)$
and, for a $C^1$ function $g$ on $M$,
 $(f_*(X))(g):=X(g\circ f)\colon p\mapsto f_*|_p(g)=X_p(g\circ f)$, with
$\circ$ being the composition of mappings sign. Respectively, the induced
cotangent mapping  is $f^*:=T^*(f)\colon T^*(M)\to T^*(\bar{M})$.
If $h\colon N\to \bar{M}$, $N$ being a manifold, we have the chain rule
$\od (f\circ h)=\od f\circ \od h$, which is an abbreviation for
$\od (f\circ h)_q=(\od f)_{f(q)}\circ (\od h)_q$ for $q\in N$.

	By $J\subseteq\field[R]$ will be denoted an arbitrary real interval
that can be open or closed at one or both its ends.
	The notation $\gamma\colon J\to M$ represents an arbitrary path in $M$.
	For a $C^1$ path $\gamma\colon J\to M$, the vector tangent to $\gamma$
at $s\in J $ will be denoted by
\(
\dot\gamma(s)
:=\frac{\od} {\od t}\Big|_{t=s}(\gamma(t))
=\gamma_*\bigl( \frac{\od}{\od r} \big|_{s} \bigr) \in T_{\gamma(s)}(M) ,
\)
where $r$ in $\frac{\od }{\od r}\big|_s$ is the standard coordinate function
on $\field[R]$, \ie $r \colon \field[R]\to \field[R]$ with $r(s):=s$ for all
$s\in\field[R]$ and hence $r=\id_{\field[R]}$ is the identity mapping of
$\field[R]$.
If $s_0\in J$ is an end point of $J$ and $J$ is closed at $s_0$, the
derivative in the definition of $\dot{\gamma}(s_0)$ is regarded as a one\ndash
sided derivative at $s_0$.

        The Lie derivative relative to $X\in \mathcal{X}(M)$ will be denoted
by $\Lied_X$. It is defined on arbitrary geometrical objects on
$M$~\cite{Yano/LieDerivatives}, but bellow we shall be interested in its
action on tensor fields~\cite[ch.~I, \S~2]{K&N-1} (see also~\cite{KMS-1993}).
If $f$, $Y$, and $\theta$ are $C^1$ respectively function, vector field and
1\ndash form on $M$, then
        \begin{subequations}        \label{2.1}
        \begin{align}   \label{2.1a}
\Lied_X(f) &= X(f)
\\                      \label{2.1b}
\Lied_X(Y) &= [X,Y]_{\_}
\\                      \label{2.1c}
(\Lied_X(\theta))(Y)
&= X(\theta(Y)) - \theta([X,Y]_{\_})
= (\od\theta)(X,Y) + Y(\theta(X)) ,
        \end{align}
        \end{subequations}
where $[A,B]_{\_}=A\circ B- B\circ A$ is the commutator of operators $A$ and
$B$ (with common domain) and $\od$ denotes the exterior derivative operator.

        Since $\Lied_X$ is a derivation of the tensor algebra over the vector
fields on $M$, for a tensor field
 \(
S \colon\Lambda^1(M)\times \dots \times \Lambda^1(M)
 \times \mathcal{X}(M)\times\dots\times \mathcal{X}(M) ,
\)
we have
        \begin{equation}        \label{2.2}
(\Lied_XS) (\theta,\ldots;Y,\ldots)
\!=\! X(S(\theta,\ldots;Y,\ldots))
  - S(\Lied_X\theta,\ldots;Y,\ldots) - \dots
  - S(\theta,\ldots;\Lied_XY,\ldots)) - \dotsb    ,
        \end{equation}
 which defines $\Lied_XS$ explicitly, due to~\eref{2.1}.

        Let the Greek indices $\lambda,\mu,\nu,\dots$ run over the range
$1,\dots,\dim M$ and $\{E_\mu\}$ be a $C^1$ frame in $T(M)$, \ie
$E_\mu\in\mathcal{X}(M)$ be of class $C^1$ and, for each $p\in M$, the set
$\{E_\mu|_p\}$ be a basis of the vector space $T_p(M)$.~%
\footnote{~%
There are manifolds, like the even-dimensional spheres $\mathbb{S}^{2k}$,
$k\in\field[N]$, which do not admit global, continuous (and moreover $C^k$
for $k\ge1$), and nowhere vanishing vector fields~\cite{Spivak-1}. If this
is the case, the considerations must be localized over an open subset of $M$
on which such fields exist. We shall not overload our exposition with such
details.%
}
Let $\{E^\mu\}$ be the coframe dual to $\{E_\mu\}$, \ie
$E^\mu\in\Lambda^1(M)$,  $\{E^\mu|_p\}$ be a basis in $T_p^*(M)$, and
$E^\mu(E_\nu)=\delta_\nu^\mu$ with $\delta_\mu^\nu$ being the Kronecker deltas
($\delta_\mu^\nu=1$ for $\mu=\nu$ and $\delta_\mu^\nu=0$ for $\mu\not=\nu$).
Assuming the Einstein's summation convention (summation on indices repeated
on different levels over the whole range of their values), we define the
\emph{components} $(\Gamma_X)\Sprindex{\nu}{\mu}$ of $\Lied_X$ in (relative
to) $\{E_\mu\}$ via the expansion
        \begin{gather}        \label{2.3}
\Lied_XE_\mu =: (\Gamma_X)\Sprindex{\mu}{\nu} E_\nu
\\\intertext{which is equivalent to}
                                \label{2.3'}
                               \tag{\protect\ref{2.3}$^\prime$}
\Lied_XE^\mu = - (\Gamma_X)\Sprindex{\nu}{\mu} E^\nu
        \end{gather}
by virtue of $E^\mu(E_\nu)=\delta_\nu^\mu$ and the commutativity of the Lie
derivatives and contraction operators.~%
\footnote{~%
The sign before $(\Gamma_X)\Sprindex{\nu}{\mu}$ in~\eref{2.3} or~\eref{2.3'}
is conventional and we have chosen it in a way similar to the accepted
convention for the components of a covariant derivative (or, equivalently,
the coefficients of a linear connection --- see Sect.~\ref{Sect4}).%
}
Sometimes, it is convenient~\eref{2.3} and~\eref{2.3'} to be written in a
matrix form,
        \begin{equation}        \label{2.4}
\Lied_X E = E\cdot \Gamma_X
\qquad
\Lied_X E^* = -\Gamma_X\cdot E^*,
        \end{equation}
where $\Gamma_X:=[(\Gamma_X)\Sprindex{\nu}{\mu}]_{\mu,\nu=1}^{\dim M}$,
 $E:=(E_1,\dots,E_{\dim M})$, and $E^*:=(E^!,\dots,E^{\dim M})^\top$, with
$\top$ being the matrix transposition sign, and the matrix multiplication is
explicitly denoted by centered dot $\cdot$ as otherwise  $E\cdot\Gamma_X$ may
be confused with
\(
E\Gamma_X
= E(\Gamma_X)
= (E_1(\Gamma_X),\ldots)
= ([E_1((\Gamma_X)\Sprindex{\mu}{\nu})],\ldots).
\)
From~\eref{2.3} and~\eref{2.1b}, we get
        \begin{equation}        \label{2.5}
(\Gamma_X)\Sprindex{\mu}{\nu}
= -E_\mu(X^\nu) - C_{\mu\lambda}^{\nu} X^\lambda ,
        \end{equation}
in $\{E_\mu\}$, where $X=X^\mu E_\mu$  and the functions
$C_{\mu\lambda}^{\nu}$, known as the \emph{components of the anholonomy
object of} $\{E_\mu\}$, are defined by
        \begin{gather}        \label{2.6}
[E_\mu,E_\nu]_{\_} =: C_{\mu\nu}^{\lambda} E_\lambda
\\\intertext{or, equivalently, by its dual (see~\eref{2.1c})}
                                \label{2.6'}
                                \tag{\protect\ref{2.6}$^\prime$}
dE^\lambda = -\frac{1}{2} C_{\mu\nu}^{\lambda} E^\mu\wedge E^\nu ,
        \end{gather}
with $\wedge$ being the exterior (wedge) product sign.~%
\footnote{~%
If $M$ is a Lie group and $\{E_\mu\}$ is a basis of its Lie algebra
($:=\{\text{left invariant vector fields in } \mathcal{X}(M)\}$),
then $C_{\mu\nu}^{\lambda}$ are constants, called structure constants of $M$,
and~\eref{2.6} and~\eref{2.6'} are known as the structure equations of $M$.%
}
For a tensor field $S$ of type $(r,s)$, $r,s\in\field[N]\cup\{0\}$, with
components
 $S_{\nu_1,\dots,\nu_s}^{\mu_1,\dots,\mu_r}$ relative to the tensor frame
induced by $\{E_\mu\}$ and $\{E^\mu\}$, we get, from~\eref{2.2}, the
components of $\Lied_XS$ as
        \begin{multline}        \label{2.7}
\bigl( \Lied_X S \bigr)_{\nu_1,\dots,\nu_s}^{\mu_1,\dots,\mu_r}
= X\bigl(S_{\nu_1,\dots,\nu_s}^{\mu_1,\dots,\mu_r}\bigr)
+ \sum_{a=1}^{r}(\Gamma_X)\Sprindex{\lambda}{\mu_a}
  S_{\nu_1,\dots,\nu_s}^{\mu_1,\dots,\mu_{a-1},\lambda,\mu_{a+1},\dots,\mu_r}
\\
- \sum_{b=1}^{s}(\Gamma_X)\Sprindex{\nu_b}{\lambda}
S^{\mu_1,\dots,\mu_r}_{\nu_1,\dots,\nu_{b-1},\lambda,\nu_{b+1},\dots,\nu_s} .
        \end{multline}

        A frame $\{E_\mu\}$ or its dual coframe $\{E^\mu\}$ is called
\emph{holonomic} (\emph{anholonomic}) if $C_{\mu\nu}^{\lambda}=0$
($C_{\mu\nu}^{\lambda}\not=0$) for all (some) values of the indices $\mu$,
$\nu$, and $\lambda$. For a holonomic frame always exist local coordinates
$\{x^\mu\}$ on $M$ such that \emph{locally} $E_\mu=\frac{\pd}{\pd x^\mu}$ and
$E^\mu=\od x^\mu$. Conversely, if $\{x^\mu\}$ are local coordinates on $M$,
then the local frame $\bigl\{\frac{\pd}{\pd x^\mu}\bigr\}$ and local coframe
$\{\od x^\mu\}$ are defined and holonomic on the domain of $\{x^\mu\}$.

        A straightforward calculation by means of~\eref{2.6} reveals that a
change
        \begin{gather}        \label{2.7-0}
\{E_\mu\} \to \{\bar E_\mu = B_\mu^\nu E_\nu\}
\\\intertext{of the frame $\{E_\mu\}$, where $B=[B_\mu^\nu]$ is a
non-degenerate matrix-valued function, entails the transformation}
                                \label{2.7-1}
C_{\mu\nu}^{\lambda} \mapsto \bar C_{\mu\nu}^{\lambda}
= (B^{-1})_\varrho^\lambda
\bigl( B_\mu^\sigma E_\sigma(B_\nu^\varrho)
     - B_\nu^\sigma E_\sigma(E_\mu^\varrho)
+B_\mu^\sigma B_\nu^\tau C_{\sigma\tau}^{\varrho}
\bigr) .
\\\intertext{Besides, from~\eref{2.5} and~\eref{2.7-1}, we see that the
quantities $(\Gamma_X)\Sprindex{\mu}{\nu}$ undergo the change}
                                \label{2.7-2}
(\Gamma_X)\Sprindex{\mu}{\nu} \mapsto (\bar \Gamma_X)\Sprindex{\mu}{\nu}
= (B^{-1})_\varrho^\mu
\bigl( (\Gamma_X)\Sprindex{\sigma}{\varrho} B_\nu^\sigma + X(B_\nu^\sigma)
\bigr)
\\\intertext{when~\eref{2.7-0} takes place. Setting
$\Gamma_X:=[(\Gamma_X)\Sprindex{\mu}{\nu}]$  and
$\bar\Gamma_X:=[(\bar\Gamma_X)\Sprindex{\mu}{\nu}]$, we can
rewrite~\eref{2.7-2} in a more compact matrix form as}
                                \label{2.7-3}
\Gamma_X\mapsto \bar\Gamma_X
= B^{-1} \cdot (\Gamma_X \cdot B + X(B)) .
        \end{gather}

        If $n\in \field[N]$ and $n\le\dim M$, an  $n$-dimensional
\emph{distribution} $\Delta$ on $M$ is defined as a mapping
$\Delta\colon p\mapsto \Delta_p$ assigning to each $p\in M$ an $n$\ndash
dimensional subspace $\Delta_p$ of the tangent space $T_p(M)$ of $M$ at $p$,
$\Delta_p\subseteq T_p(M)$. A \emph{solution} (resp.\ \emph{first integral})
of a distribution $\Delta$ on $M$ is an immersion $\varphi\colon N\to M$
(resp.\ submersion $\psi\colon M\to N$), $N$ being a manifold, such that
$\Image\varphi_*\subseteq\Delta$ (resp.\ $\Ker\psi_*\supset\Delta$), i.e., for
each $q\in N$, (resp.\ $p\in M$),
$\varphi_*(T_q(N))\subseteq\Delta_{\varphi(q)}$
(resp.\ $\psi_*(\Delta_p)=0_{\psi(q)}\in T_{\psi(q)}(N)$). A distribution is
\emph{integrable} if there is a submersion $\psi\colon M\to N$ such that
$\Ker \psi_*=\Delta$; a necessary and locally sufficient condition for the
 integrability of $\Delta$ is the commutator of every two vector fields in
$\Delta$ to be in $\Delta$. We say that a vector field $X\in\mathcal{X}(M)$
is in $\Delta$ and write $X\in\Delta$, if $X_p\in\Delta_p$ for all $p\in M$.
A \emph{basis on $U\subseteq M$ for} $\Delta$ is a set  $\{X_1,\dots,X_n\}$ of
$n$ linearly independent (relative to functions $U\to\field$) vector fields in
$\Delta|_U$, \ie $\{X_1|_p,\dots,X_n|_p\}$ is a basis for $\Delta_p$ for all
$p\in U$.

        A distribution is convenient to be described in terms of (global)
frames or/and coframes over $M$. In fact, if $p\in M$ and
$\varrho=1,\dots,n$, in each $\Delta_p\subseteq T_p(M)$, we can choose a basis
$\{X_\varrho|_p\}$ and hence a frame $\{X_\varrho\}$,
$X_\varrho\colon p\mapsto X_\varrho|_p$, in
$\{\Delta_p : p\in M\} \subseteq T(M) $; we say that $\{X_\rho\}$ is a basis
for/in $\Delta$. Conversely, any collection of $n$ linearly independent
(relative to functions $M\to\field$) vector fields $X_\varrho$ on $M$ defines
a distribution
\(
 p\mapsto\bigl\{ \sum_{\varrho=1}^{n} f^\varrho X_\varrho|_p :
f^\varrho\in\field \bigr\}.
\)
Consequently, a frame in $T(M)$ can be formed by
adding to a basis for $\Delta$ a set of $(\dim M-n)$ new linearly independent
vector fields (forming a frame in $T(M)\setminus\{\Delta_p : p\in M\}$)  and
v.v., by selecting  $n$ linearly independent vector fields on $M$, we can
define a distribution $\Delta$ on $M$. Equivalently, one can use $\dim M-n$
linearly independent 1\ndash forms $\omega^a$, $a=n+1,\dots,\dim M$, which are
annihilators for it, $\omega^a|_{\Delta_p}=0$ for all $p\in M$. For instance,
if $\{X_\mu:\mu=1,\dots,\dim M\}$ is a frame in $T(M)$ and
$\{X_\varrho:\varrho=1,\dots,n\}$ is a basis for $\Delta$, then one can
define $\omega^a$ to be elements in the coframe $\{\omega^\mu\}$ dual to
$\{X_\mu\}$. We call $\{\omega^a\}$ a \emph{cobasis} for $\Delta$.


\section {Connections on bundles}
\label{Sect3}

	Before presenting the general connection theory in
Subsect.~\ref{Subsect3.2}, we at first fix some notation and concepts
concerning fibre bundles in Subsect.~\ref{Subsect3.1}.


\subsection{Frames and coframes on the bundle space}
\label{Subsect3.1}

         Let $(E,\pi,M$) be a bundle with bundle space $E$, projection
$\pi\colon E\to M$, and base space $M$.
	Suppose that the spaces $M$ and $E$ are
manifolds of finite dimensions $n\in\field[N]$ and $n+r$, for some
$r\in\field[N]$, respectively; so the dimension of the fibre $\pi^{-1}(x)$,
with $x\in M$, \ie the fibre dimension of $(E,\pi,M)$, is $r$.
Besides, let these manifolds be $C^1$ differentiable, if the opposite is not
stated explicitly.~%
\footnote{~%
Most of our considerations are valid also if $C^1$ differentiability is
assumed and even some of them hold on $C^0$ manifolds. By assuming $C^2$
differentiability, we skip the problem of counting the required
differentiability class of the whole material that follows. Sometimes, the
$C^2$  differentiability is required explicitly, which is a hint that a
statement or definition is not valid otherwise. If we want to emphasize that
some text is valid under a $C^1$ differentiability assumption, we indicate
that fact explicitly.%
}

         Let the Greek indices $\lambda,\mu,\nu,\ldots$ run from 1 to
$n=\dim M$, the Latin indices $a,b,c,\ldots$  take  the  values  from  $n+1$
to  $n+r=\dim E$, and the uppercase Latin indices
${I},{J},{K},\ldots$ take values in the whole set $\{1,\ldots,n+r\}$. One may
call these types of indices respectively base, fibre, and bundle indices.

        Suppose $\{u^{I}\}=\{u^\mu,u^a\}=\{u^1,\dots,u^{n+r}\}$ are local
\emph{bundle coordinates} on an open set $U\subseteq E$, \ie on the set
$\pi(U)\subseteq M$ there are local coordinates $\{x^\mu\}$ such that
$u^\mu=x^\mu\circ \pi$;~%
\footnote{~%
On a bundle or fibred manifold, these coordinates are known also as adapted
coordinates~\cite[definition~1.1.5]{Saunders}.%
}
the coordinates $\{u^\mu\}$ (resp.\ $\{u^a\}$) are
called \emph{basic} (resp.\ \emph{fibre}) \emph{coordinates}~\cite{Poor}.~%
\footnote{~%
If $(U,v)$ is a bundle chart, with $v\colon U\to\field^n\times\field^r$ and
$e^a\colon\field^r\to\field$ are such that $e^a(c_1,\dots,c_r)=c_a\in\field$,
then one can put $u^a=e^a\circ \pr_2\circ v$, where
$\pr_2\colon\field^n\times\field^r\to\field^r$ is the projection on  the second
multiplier $\field^r$.%
}

        Further only coordinate changes
        \begin{subequations}    \label{3.1}
        \begin{equation}        \label{3.1a}
\{u^\mu,u^a\} \mapsto \{\tilde{u}^\mu,\tilde{u}^a\}
        \end{equation}
on $E$ which respect the fibre structure, \viz the division into basic and
fibre coordinates, will be considered. This means that
        \begin{equation}        \label{3.1b}
        \begin{split}
\tilde{u}^\mu(p) &= f^\mu(u^1(p),\dots,u^n(p))\\
\tilde{u}^a(p)   &= f^a(u^1(p),\dots,u^n(p),u^{n+1}(p),\dots,u^{n+r}(p))
        \end{split}
        \end{equation}
        \end{subequations}
for $p\in E$ and some functions $f^{I}$. The bundle coordinates
$\{u^\mu,u^a\}$ induce the (local) frame
$\bigl\{ \pd_ \mu:=\frac{\pd}{\pd u^\mu}, \pd_a:=\frac{\pd}{\pd u^a} \bigr\}$
and coframe $\{\od u^\mu,\od u^a\}$ over $U$ in respectively the tangent
$T(E)$ and cotangent $T^*(E)$ bundle spaces of the tangent and cotangent
bundles over the bundle space $E$. Since a change~\eref{3.1} of the
coordinates on $E$ implies
\(
\pd_{I}\mapsto\tilde{\pd}_{I}
:=\frac{\pd}{\pd\tilde{u}^{I}}
=\frac{\pd u^{J}}{\pd\tilde{u}^{I}} \pd_{J}
\)
and
\(
\od u^{I} \mapsto \pd\tilde{u}^{I}
=\frac{\pd\tilde{u}^{I}}{\pd u^{J}} \od u^{J} ,
\)
the transformation~\eref{3.1} leads to
        \begin{subequations}    \label{3.2}
        \begin{align}   \label{3.2a}
(\pd_\mu,\pd_a) &\mapsto
(\tilde{\pd}_\mu,\tilde{\pd}_a) = (\pd_\nu,\pd_b) \cdot A
\\                      \label{3.2b}
(\od u^\mu,\od u^a)^\top &\mapsto
(\od\tilde{u}^\mu,\od\tilde{u}^a)^\top
= A^{-1}\cdot(\od u^\nu,\od u^b)^\top .
        \end{align}
        \end{subequations}
Here expressions like $(\pd_\mu,\pd_a)$ are shortcuts for ordered
$(n+r)$\ndash tuples like
$(\pd_1,\dots,\pd_{n+r}) = \bigl([\pd_\mu]_{\mu=1}^n,[\pd_a]_{a=n+1}^{n+r}$),
$\top$ is the matrix transpositions sign, the centered dot $\cdot$ stands for
the matrix multiplication, and the transformation matrix $A$ is
        \begin{equation}        \label{3.3}
A
:= \Bigl[\frac{\pd u^{I}}{\pd\tilde{u}^{J}}\Bigr]_{{I},{J}=1}^{n+r}
=
        \begin{pmatrix}
\bigl[\frac{\pd u^\nu}{\pd\tilde{u}^\mu}\bigr] & 0_{n\times r}
\\
\bigl[\frac{\pd u^b}{\pd\tilde{u}^\mu}\bigr] &
\bigl[\frac{\pd u^b}{\pd\tilde{u}^a}\bigr]
        \end{pmatrix}
=:
        \begin{bmatrix}
\frac{\pd u^\nu}{\pd\tilde{u}^\mu} & 0
\\
\frac{\pd u^b}{\pd\tilde{u}^\mu} &
\frac{\pd u^b}{\pd\tilde{u}^a}
        \end{bmatrix}
\ ,
        \end{equation}
where $0_{n\times r}$ is the $n\times r$ zero matrix.
	Besides, in expressions of the form $\pd_I a^I$, like the one in the
r.h.s.\ of~\eref{3.2a}, the summation excludes differentiation, \ie
$\pd_Ia^I:=a^I\pd_I=\sum_Ia^I\pd_I$; if a differentiation really takes place,
we write $\pd_I(a^I):=\sum_I\pd_I(a^I)$. This rule allows a lot of formulae to
be writtem in a compact matrix form, like~\eref{3.2a}.
	The explicit form of the matrix inverse to~\eref{3.3} is
 $A^{-1}=\bigl[\frac{\pd\tilde{u}^{I}}{\pd u^{J}}\bigr]=\ldots$ and it
is obtained from~\eref{3.3} via the change $u\leftrightarrow\tilde{u}$.

        The formulae~\eref{3.2} can be generalized for arbitrary frame
$\{e_{I}\}=\{e_\mu,e_a\}$ in $T(E)$ and its dual coframe
$\{e^{I}\}=\{e^\mu,e^a\}$ in $T^*(E)$ which respect the fibre structure in
a sense that their \emph{admissible changes} are given by
        \begin{subequations}        \label{3.4}
        \begin{align}   \label{3.4a}
(e_{I})=(e_\mu,e_a)
&\mapsto
(\tilde{e}_{I})=(\tilde{e}_\mu,\tilde{e}_a)
= (e_\nu,e_b)\cdot A
\\                      \label{3.4b}
        \begin{pmatrix}  e^\mu \\ e^a  \end{pmatrix}
&\mapsto
        \begin{pmatrix}  \tilde{e}^\mu \\ \tilde{e}^a  \end{pmatrix}
=
A^{-1}\cdot
        \begin{pmatrix}  e^\nu \\ e^b  \end{pmatrix} .
        \end{align}
        \end{subequations}
Here $A=[A_{J}^{I}]$ is a nondegenerate matrix-valued function with a block
structure similar to~\eref{3.3}, \viz
        \begin{subequations}        \label{3.5}
        \begin{align}   \label{3.5a}
A =
        \begin{pmatrix}
[A_\mu^\nu]_{\mu,\nu=1}^{n} & 0_{n\times r}
\\
\bigl[A_\mu^b\bigr]_{
           \begin{subarray}{l}
           \mu=1,\dots,n \\
           b=n+1,\dots,n+r
           \end{subarray}
          }
&
[A_a^b]_{a,b=n+1}^{n+r}
        \end{pmatrix}
=:
        \begin{bmatrix}
A_\mu^\nu & 0
\\
A_\mu^b   & A_a^b
        \end{bmatrix}
\\\intertext{with inverse matrix}
                        \label{3.5b}
A^{-1}
=
        \begin{pmatrix}
[A_\mu^\nu]^{-1} & 0
\\
- [A_b^a]^{-1} \cdot [A_\mu^a]\cdot [A_\mu^\nu]^{-1}
&
[A_b^a]^{-1}
        \end{pmatrix}\ .
        \end{align}
        \end{subequations}
Here $A_\mu^a\colon U\to\field$ and $[A_\mu^\nu]$ and $[A_b^a]$
are non-degenerate matrix\ndash valued functions on $U$ such that
$[A_\mu^\nu]$ is constant on the fibres of $E$, i.e., for $p\in E$,
$A_\mu^\nu(p)$ depends only on $\pi(p)\in M$, which is equivalent to any one
of the equations
        \begin{equation}        \label{3.6}
A_\mu^\nu = B_\mu^\nu \circ \pi
\qquad
\frac{\pd A_\mu^\nu }{\pd u^a} = 0 ,
        \end{equation}
with $[B_\mu^\nu]$ being a nondegenerate matrix-valued function on
$\pi(U)\subseteq M$. Obviously,~\eref{3.2} corresponds
to~\eref{3.4} with
$e_I=\frac{\pd}{\pd{u}^I}$,
$\tilde{e}_I=\frac{\pd}{\pd\tilde{u}^I}$, and
$A_{I}^{J}=\frac{\pd u^{J}}{\pd\tilde{u}^{I}}$.

	All frames on $E$ connected via~\eref{3.4}--\eref{3.5}, which are
(locally) obtainable from holonomic ones, induced by bundle coordinates, via
admissible changes, will be referred as \emph{bundle frames}. Only such frames
will be employed in the present work.

        If we deal with a \emph{vector} bundle $(E,\pi,M)$ endowed with
vector bundle coordinates $\{u^{I}\}$~\cite{Poor}, then the new fibre
coordinates $\{\tilde{u}^a\}$ in~\eref{3.1} must be \emph{linear and
homogeneous} in the old ones  $\{u^a\}$, \ie
        \begin{equation}        \label{3.7}
\tilde{u}^a = (B_b^a\circ\pi)\cdot u^b
\text{ and }
u^a = ( (B^{-1})_b^a \circ\pi ) \cdot \tilde{u}^b ,
        \end{equation}
with $B=[B_b^a]$ being a non-degenerate matrix-valued function on
$\pi(U)\subseteq M$. In that case, the matrix~\eref{3.3} and its inverse take
the form
        \begin{equation}        \label{3.8}
A=
        \begin{bmatrix}
\frac{\pd u^\mu}{\pd \tilde{u}^\nu} & 0
\\
\bigl(\frac{\pd (B^{-1})_b^a} {\pd \tilde{x}^\nu} \circ \pi \bigr)
                                                \cdot \tilde{u}^b
&
(B^{-1})_a^b\circ\pi
        \end{bmatrix}
\qquad
A^{-1}=
        \begin{bmatrix}
\frac{\pd \tilde{u}^\nu}{\pd u^\mu} & 0
\\
\bigl(\frac{\pd B_a^b}{\pd x^\mu} \circ \pi \bigr) \cdot u^a
&
B_a^b\circ\pi
      \end{bmatrix}
\ .
        \end{equation}
More generally, in the vector bundle case, admissible are
transformations~\eref{3.4} with matrices like
        \begin{equation}        \label{3.9}
A=
        \begin{pmatrix}
[A_\nu^\mu] & 0
\\
[A_{c\mu}^{b} \tilde{u}^c]  & [A_b^a]
        \end{pmatrix}
\qquad
A^{-1}=
        \begin{pmatrix}
[A_\nu^\mu]^{-1} & 0
\\
- [A_b^a]^{-1} \cdot[A_{c\mu}^{b} \tilde{u}^c]\cdot [A_\nu^\mu]^{-1}
								 & [A_b^a]^{-1}
        \end{pmatrix}
        \end{equation}
with $A_{b\mu}^{a}\colon U\to\field$ being functions on $U$ which are
constant on the fibres of $E$,
        \begin{equation}        \label{3.9-1}
A_{b\mu}^a = B_{b\mu}^a \circ \pi
\qquad
\frac{\pd A_{b\mu}^a }{\pd u^c} = 0
        \end{equation}
for some functions $B_{b\mu}^{a}\colon \pi(U)\to\field$. Obviously,~\eref{3.9}
corresponds to~\eref{3.5} with $A_{\mu}^{b}=A_{c\mu}^{b}\tilde{u}^c$ and the
setting $A_{I}^{J}=\frac{\pd u^{J}}{\pd\tilde{u}^{I}}$. reduces~\eref{3.9}
to~\eref{3.8} due to~\eref{3.7}.



\subsection{Connection theory}
\label{Subsect3.2}

	From a number of equivalent definitions of a connection on
differentiable manifold~\cite[sections~2.1
and~2.2]{Mangiarotti&Sardanashvily}, we shall use the following one.

    \begin{Defn}    \label{Defn3.1}
A \emph{connection on a bundle} $(E,\pi,M)$ is an $n=\dim M$
dimensional distribution $\Delta^h$ on $E$ such that, for each $p\in E$
and the \emph{vertical distribution} $\Delta^v$ defined by
    \begin{equation}    \label{3.9-2}
\Delta^v\colon p \mapsto \Delta^v_p
:= T_{\imath(p)}\bigl( \pi^{-1}(\pi(p)) \bigr)
\cong T_{p}\bigl( \pi^{-1}(\pi(p)) \bigr),
    \end{equation}
with $\imath\colon\pi^{-1}(\pi(p))\to E$ being the inclusion mapping, is
fulfilled
    \begin{equation}    \label{3.9-3}
\Delta^v_p\oplus \Delta^h_p = T_p(E) ,
    \end{equation}
where
\(
\Delta^h\colon p \mapsto \Delta^h_p \subseteq T_{p}(E)
\)
and $\oplus$ is the direct sum sign. The distribution $\Delta^h$ is
called \emph{horizontal} and symbolically we write
$\Delta^v\oplus\Delta^h=T(E)$.
    \end{Defn}

    A \emph{vector} at a point $p\in E$ (resp. a \emph{vector field} on $E$)
is said to be \emph{vertical} or \emph{horizontal} if it (resp.\ its value at
$p$) belongs to $\Delta^h_p$ or $\Delta^v_p$, respectively, for the given
(resp.\ any) point $p$.
	A vector $Y_p\in T_p(E)$ (resp.\ vector field $Y\in\mathcal{X}(E)$) is
called a \emph{horizontal lift of a vector} $X_{\pi(p)}\in T_{\pi(p)}(M)$
(resp.\ \emph{vector field} $X\in\mathcal{X}(M)$ on $M=\pi(E)$) if
$\pi_*(Y_p)=X_{\pi(p)}$ for the given (resp.\ any) point $p\in E$. Since
$\pi_*|_{\Delta_p^h}\colon \Delta_p^h\to T_{\pi(p)}(M)$ is a vector space
isomorphism for all $p\in E$~\cite[sec.~1.24]{Poor}, any vector in
$T_{\pi(p)}(M)$ (resp.\ vector field in $\mathcal{X}(M)$) has a unique
horizontal lift in $T_p(E)$ (resp.\ $\mathcal{X}(E)$).

	As a result of~\eref{3.9-3}, any vector $Y_p\in T_p(E)$ (resp.\ vector
field $Y\in\mathcal{X}(E)$) admits a unique representation
 $Y_p=Y_p^v\oplus Y_p^h$ (resp.\ $Y=Y^v\oplus Y^h$) with
 $Y_p^v\in\Delta_p^v$ and $Y_p^h\in\Delta_p^h$
(resp.\ $Y^v\in\Delta^v$ and $Y^h\in\Delta^h$). If the distribution
$p\mapsto\Delta_p^h$ is differentiable of class $C^m$,
$m\in\field[N]\cup\{0,\infty,\omega\}$, it is said that the \emph{connection
$\Delta^h$ is (differentiable) of class} $C^m$. A connection $\Delta^h$ is of
class $C^m$ if and only if, for every $C^m$ vector field $Y$ on $E$, the
vertical $Y^v$ and horizontal $Y^h$ vector fields are of class $C^m$.

	A $C^1$ \emph{path} $\beta\colon J\to E$ is called \emph{horizontal}
(\emph{vertical}) if its tangent vector $\dot{\beta}$ is horizontal
(vertical) vector along $\beta$, \ie $\dot{\beta}(s)\in\Delta_{\beta(s)}^h$
($\dot{\beta}(s)\in\Delta_{\beta(s)}^v$) for all $s\in J$.
    A \emph{lift} $\bar\gamma\colon J\to E$ of a path $\gamma\colon J\to M$,
\ie $\pi\circ\bar\gamma=\gamma$, is called \emph{horizontal} if $\bar\gamma$
is a horizontal path, \ie when the vector field $\dot{\bar{\gamma}}$ tangent
to $\bar\gamma$ is horizontal or, equivalently, if $\dot{\bar{\gamma}}$ is
a horizontal lift of $\dot\gamma$.
    Since $\pi^{-1}(\gamma(J))$ is an $(r+1)$ dimensional submanifold of
$E$, the distribution
 $p\mapsto \Delta_p^h\cap T_p(\pi^{-1}(\gamma(J)))$ is one\ndash dimensional
and, consequently, is integrable. The integral paths of that distribution are
horizontal lifts of $\gamma$ and,
\emph{for each $p\in\pi^{-1}(\gamma(J))$, there
is (locally) a unique horizontal lift $\bar{\gamma}_p$ of $\gamma$ passing
through} $p$.~%
\footnote{~%
In this sense, a connection $\Delta^h$ is an Ehresmann
connection~\cite[p.~314]{Greub&et_al.-1} and \emph{vice
versa}~\cite[pp.~85--89]{Saunders}.%
}

    \begin{Defn}    \label{Defn3.2}
Let $\gamma\colon[\sigma,\tau]\to M$, with $\sigma,\tau\in\field[R]$ and
$\sigma\le \tau$, and
$\bar\gamma_p$ be the unique horizontal lift of $\gamma$ in $E$ passing
through $p\in\pi^{-1}(\gamma([\sigma,\tau]))$.
The \emph{parallel transport (translation, displacement)} generated by
(assigned to, defined by) a connection $\Delta^h$ is a mapping
$\Psf\colon\gamma\mapsto\Psf^\gamma$, assigning to the path $\gamma$ a mapping
    \begin{equation}    \label{3.9-4}
\Psf^\gamma\colon \pi^{-1}(\gamma(\sigma)) \to \pi^{-1}(\gamma(\tau))
\qquad \gamma\colon[\sigma,\tau]\to M
    \end{equation}
such that, for each $p\in\pi^{-1}(\gamma(\sigma))$,
    \begin{equation}    \label{3.9-5}
\Psf^\gamma(p) := \bar\gamma_p(\tau).
    \end{equation}
    \end{Defn}

    In vector bundles are important the \emph{linear} connections for
which is required the parallel transport assigned to them to be linear in a
sense that the mapping~\eref{3.9-4} is linear for every path $\gamma$ (see
Subsect.~\ref{Subsect4.3} below).

    Let us now look on the connections $\Delta^h$ on a bundle $(E,\pi,M)$
from a view point of (local) frames and their dual coframes on $E$. Let
$\{e_\mu\}$ be a basis for $\Delta^h$, \ie $e_\mu\in\Delta^h$ and
$\{e_\mu|_p\}$ is a basis for $\Delta_p^h$ for all $p\in E$, and $\{e^a\}$
be the coframe for $\Delta^h$, \ie a collection of 1\ndash forms $e^a$,
$a=n+1,\dots,n+r$, which are linearly independent (relative to functions
$E\to \field$) and such that $e^a(X)=0$ if $X\in\Delta^h$.

    \begin{Defn}    \label{Defn3.3}
A frame $\{e_I\}$ in $T(E)$ over $E$ is called \emph{specialized} for a
connection $\Delta^h$ if the first $n=\dim M$ of its vector fields
$\{e_\mu\}$ form a basis for the horizontal distribution $\Delta^h$ and its
last $r=\dim\pi^{-1}(x)$, $x\in M$, vector fields $\{e_a\}$ form a basis for
the vertical distribution $\Delta^v$. Respectively, a coframe $\{e^I\}$ on
$E$ is called \emph{specialized} if $\{e^a\}$ is a cobasis for $\Delta^h$ and
$\{e^\mu\}$ is a cobasis for $\Delta^v$.
    \end{Defn}

    The horizontal lifts of vector fields and 1\ndash forms can easily be
described in specialized (co)frames. Indeed, let $\{e_{I}\}$ and
$\{e^{I}\}$ be respectively a specialized frame and its dual coframe.
Define a frame $\{E_\mu\}$ and its dual coframe $\{E^\mu\}$ on $M$ which  are
$\pi$\ndash related to $\{e_{I}\}$ and $\{e^{I}\}$, \ie
$E_\mu:=\pi_*(e_\mu)$ and $e^\mu:=\pi^*(E^\mu)=E^\mu\circ\pi_*$.~%
\footnote{~%
Recall, $\pi_*|_{\Delta_p^h}\colon \Delta_p^h\to T_{\pi(p)}(M)$ is a vector
space isomorphism.%
}
If $Y=Y^\mu E_\mu\in\mathcal{X}(M)$ and
$\phi=\phi_\mu e^\mu\in\Lambda^1(M)$, then their horizontal lifts (from
$M$ to $E$) respectively are
    \begin{equation}    \label{3.9-6}
\bar{Y}=(Y^\mu\circ\pi) e_\mu
\qquad
\bar{\phi}=(\phi_\mu\circ\pi) e^\mu .
    \end{equation}

    The specialized (co)frames transform into each other according to the
general rules~\eref{3.4} in which the transformation matrix and its inverse
have the following block structure:
    \begin{equation}    \label{3.10}
A=
    \begin{pmatrix}
[A_\mu^\nu]     & 0_{n\times r} \\
0_{r\times n}   & [A_a^b]
    \end{pmatrix}
\qquad
A^{-1}=
    \begin{pmatrix}
[A_\mu^\nu]^{-1}    & 0_{n\times r} \\
0_{r\times n}       & [A_a^b]^{-1}
    \end{pmatrix}
,
    \end{equation}
where $A_\mu^\nu,A_a^b\colon E\to\field$ and the functions $A_\mu^\nu$ are
constant on the fibres of the bundle $(E,\pi,M)$, that is, we have
     \begin{alignat}{2}        \label{3.10-1}
&A_\mu^\nu = B_\mu^\nu \circ \pi
&\quad\text{or}\quad
&\frac{\pd A_\mu^\nu }{\pd u^a} =0
\\\intertext{for some nondegenerate matrix-valued function $[B_\mu^\nu]$ on
$M$. Besides, in a case of vector bundle, the functions $A_b^a$ are also
constant on the fibres of the bundle $(E,\pi,M)$, \ie}
				\label{3.10-2}
&A_a^b = B_a^b \circ \pi
&\quad\text{or}\quad
&\frac{\pd A_a^b }{\pd u^a} = 0
        \end{alignat}
for some nondegenerate matrix-valued function $B=[B_a^b]$ on $M$. Changes
like~\eref{3.4}, with $A$ given by~\eref{3.10}, respect the fibre as well as
the connection structure of the bundle.

	Let $E$ be a $C^2$ manifold and $\Delta^h$ a $C^1$ connection on
$(E,\pi,M)$.
    The components $C_{{I}{J}}^{{K}}$ of the anholonomy object of
a \emph{specialized} frame $\{e_{I}\}$ are (local) functions on $E$
defined by (see~\eref{2.6})
    \begin{equation}    \label{3.11}
[e_{I},e_{J}]_{\_} =: C_{{I}{J}}^{{K}} e_{K}
    \end{equation}
and are naturally divided into the following six groups
(cf.~\cite[p.~21]{Rahula}:
    \begin{equation}    \label{3.12}
\{ C_{\mu\nu}^{\lambda} \}, \quad
\{ C_{\mu\nu}^{a} \}, \quad
\{ C_{\mu b}^{\lambda} = 0 \}, \quad
\{ C_{ab}^{\lambda} = 0 \}, \quad
\{ C_{\mu b}^{c} \}, \quad
\{ C_{ab}^{c} \} .
    \end{equation}
The functions $C_{\mu\nu}^{\lambda}$ are constant on the fibres of
$(E,\pi,M)$, precisely $C_{\mu\nu}^{\lambda}=f_{\mu\nu}^{\lambda}\circ\pi$
where $f_{\mu\nu}^{\lambda}$ are the components of the anholonomy object for
the $\pi$\ndash related frame $\{\pi_*(e_\mu)\}$ on $M$, as the commutators of
$\pi$\ndash related vector fields are $\pi$\ndash
related~\cite[sec.~1.55]{Warner}. Besides, since the vertical distribution
$\Delta^v$ is integrable (the space $\Delta_p^v$ is the space tangent to the
fibre through $p\in E$ at $p$), we have
    \begin{equation}    \label{3.12-1}
[e_a,e_b]_{\_} = C_{ab}^{c} e_c
    \end{equation}
(so that $C_{ab}^{\lambda}=0$), due to which $C_{ab}^{c}$ are called
components of the vertical anholonomy object.
	To prove that $C_{\mu b}^{\lambda}=0$, one should expand $\{e_I\}$
along $\bigl\{\pd_I=\frac{\pd}{\pd u^I}\bigr\}$, with $\{u^I\}$ being some
bundle coordinates, \viz
 $e_\mu=e_\mu^\nu\pd_\nu+e_\mu^b\pd_b$ and $e_a=e_a^b\pd_b$, with some
functions $e_\mu^\nu$, $e_\mu^b$ and $e_a^b$, and to notice that $e_\mu^\nu$
are constant on the fibres, \ie $\pd_a(e_\mu^\nu)=0$.

    The non-trivial mixed ``vertical-horizontal'' components
between~\eref{3.12}, \viz $C_{\mu\nu}^{a}$ and $C_{\mu b}^{a}$, are important
 characteristics of the connection $\Delta^h$. The functions
    \begin{subequations}    \label{3.13}
    \begin{align}   \label{3.13a}
\lindex[\Gamma]{}{\circ}_{b\mu}^{a} &:= + C_{b\mu}^{a} = - C_{\mu b}^{a}
\\          \label{3.13b}
R_{\mu\nu}^{a} 			    &:= + C_{\mu\nu}^{a} = - C_{\nu\mu}^{a} ,
    \end{align}
    \end{subequations}
which enter into the commutators
    \begin{subequations}    \label{3.14}
    \begin{align}   \label{3.14a}
\Lied_{e_\mu} e_b
= [e_\mu,e_b]_{\_}
& = \lindex[\Gamma]{}{\circ}_{b\mu}^{a} e_a
\\          \label{3.14b}
 [e_\mu,e_\nu]_{\_}
& =  R_{\mu\nu}^{a} e_a  + C_{\mu\nu}^{\lambda} e_\lambda ,
    \end{align}
    \end{subequations}
are called respectively  the
\emph{fibre coefficients of} $\Delta^h$ (or
\emph{components of the connection object of $\Delta^h$}) and
\emph{fibre components of the curvature of} $\Delta^h$ (or
\emph{components of the curvature (object) of} $\Delta^h$) in $\{e_{I}\}$.
Under a change~\eref{3.4}, with a matrix~\eref{3.10}, of the specialized
frame, the functions~\eref{3.13} transform into respectively
    \begin{subequations}    \label{3.15}
    \begin{align}   \label{3.15a}
\lindex[\tilde{ \Gamma}]{}{\circ}_{b\mu}^{a}
& =
A_\mu^\nu \bigl([A_e^f]^{-1}\bigr)_d^a
\bigl( \lindex[\Gamma]{}{\circ}_{c\nu}^{d} A_b^c + e_\nu(A_b^d) \bigr)
\\          \label{3.15b}
\tilde{R}_{\mu\nu}^{a}
& =
\bigl([A_e^f]^{-1}\bigr)_b^a  A_\mu^\lambda A_\nu^\varrho R_{\lambda\varrho}^{b} ,
    \end{align}
    \end{subequations}
which formulae are direct consequences of~\eref{3.14}. If we put
$\bar{A} := [A_a^b]$,
 $\lindex[\Gamma]{}{\circ}_\nu := [ \lindex[\Gamma]{}{\circ}_{c\nu}^{d} ]$ ,
and
\(
\lindex[\tilde{\Gamma}]{}{\circ}_\nu
:= [ \lindex[\tilde{\Gamma}]{}{\circ}_{c\nu}^{d} ]
\), then~\eref{3.15a} is tantamount to
    \begin{equation}    \label{3.16}
    \begin{split}
\lindex[\tilde{\Gamma}]{}{\circ}_\mu
&= A_\mu^\nu \bar{A}^{-1}\cdot
  ( \lindex[\Gamma]{}{\circ}_\nu \cdot \bar{A} + e_\nu(\bar{A}) )
\\
&= A_\mu^\nu
( \bar{A}^{-1} \cdot
\lindex[\Gamma]{}{\circ}_\nu - e_\nu(\bar{A}^{-1}) ) \cdot \bar{A} .
    \end{split}
    \end{equation}
Up to a meaning of the matrices $[A_\mu^\nu]$ and $\bar{A}$ and the size of
the matrices $\lindex[\Gamma]{}{\circ}_\nu$ and $\bar{A}$, the last equation
is identical with the one expressing the transformed matrices of the
coefficients of a linear connection (covariant derivative operator) in a
vector bundle~\cite[eq.~(3.5)]{bp-NF-D+EP} on which we shall return later in
this work (see Sect.~\ref{Sect4}, in particular equation~\eref{4.25'} in it).
Equation~\eref{3.15b} indicates that $R_{\mu\nu}^{a}$ are components of a
tensor, \viz
    \begin{equation}    \label{3.17}
\Omega := \frac{1}{2} R_{\mu\nu}^{a} e_a\otimes e^\mu \wedge e^\nu ,
    \end{equation}
called \emph{curvature tensor} of the connection $\Delta^h$. By~\eref{3.14a},
the horizontal distribution $\Delta^h$ is (locally) integrable iff its
curvature tensor vanishes, $\Omega=0$.

    \begin{Defn}    \label{Defn3.4}
    A connection with vanishing curvature tensor is called \emph{flat},
or \emph{integrable}, or \emph{curvature free}.
    \end{Defn}

    \begin{Prop}    \label{Prop3.4}
    The flat connections are the only ones that may admit holonomic
specialized frames.
    \end{Prop}

    \begin{Proof}
    See definition~\ref{Defn3.4} and~\eref{3.14b}.
    \end{Proof}

    The above considerations of specialized (co)frames for a connection
$\Delta^h$ on a bundle $(E,\pi,M)$ were \emph{global} as we supposed that
these (co)frames are defined  on the whole bundle space $E$, which is always
possible if no smoothness conditions on $\Delta^h$ are imposed. Below we
shall show how \emph{local} specialized (co)frames can be defined via local
bundle coordinates on $E$.

    Let $\{u^{I}\}$ be local bundle coordinates on an open set
$U\subseteq E$. They define on $T(U)\subseteq T(E)$ the local basis
$\bigl\{ \pd_{I} := \frac{\pd}{\pd u^{I}} \bigr\}$, so that any vector
can be expended along its vectors. In particular, we can do so with any basic
vector field $e_{I}^U$ of a \emph{specialized} frame $\{e_{I}\}$ restricted to
$U$, $e_{I}^U:=e_{I}|_{U}$. Since $\{\pd_a|_p\}$, with $p\in U$, is a
basis for $\Delta_p^v$, we can write
    \begin{equation}    \label{3.18}
(e_\mu^U,e_a^U )
= (A_\mu^\nu\pd_\mu + A_\mu^a\pd_a , A_a^b\pd_b )
=
(\pd_\nu,\pd_b) \cdot
    \begin{pmatrix}
[A_\mu^\nu] & 0 \\
[A_\mu^b] & [A_a^b]
    \end{pmatrix} \ ,
    \end{equation}
where $[A_\mu^\nu]$ and  $[A_a^b]$ are non-degenerate matrix-valued functions
on $U$.~%
\footnote{~%
The non-degeneracy of $[A_\mu^\nu]$ follows from the fact that the vector
fields
\(
\pi_*|_{\Delta^h} (e_\mu^U)
= A_\mu^\nu\pi_*\bigl( \frac{\pd}{\pd u^\mu}  \bigr)
\)
form a basis for $\mathcal{X}(\pi(U))\subseteq\mathcal{X}(M)$.%
}

    \begin{Defn}    \label{Defn3.5}
A frame $\{X_{I}\}$ over $U$ in $T(U)$ is called
\emph{adapted (to the coordinates $\{u^{I}\}$ in $U$)}
for a connection $\Delta^h$ if it is the specialized frame obtained
from~\eref{3.18} via admissible transformation~\eref{3.4} with the matrix
\(
 A=
    \Bigl(\begin{smallmatrix}
[A_\mu^\nu]^{-1} & 0 \\
0	 & [A_a^b]^{-1}
    \end{smallmatrix}\Bigr)  .
\)
    \end{Defn}

    \begin{Exrc}    \label{Exrc3.1}
An arbitrary \emph{specialized} frame $\{e_I^U\}$ in $T(E)$ over $U$ enters in
the definition of a frame $\{X_I\}$ adapted to bundle coordinates $\{u^I\}$ on
$U$. Prove that $\{X_I\}$ is independent of the particular choice of the frame
$\{e_I^U\}$. (Hint: apply definition~\ref{Defn3.5} and~\eref{3.4a} with $A$
given by~\eref{3.10}.) The below\ndash derived equality~\eref{3.20-3} is an
indirect proof of that fact too.
    \end{Exrc}

    According to~\eref{3.4} and definition~\ref{Defn3.5}, the adapted frame
$\{X_{I}\}$ and the corresponding to it \emph{adapted coframe} $\{\omega^{I}\}$
are given by
    \begin{subequations}    \label{3.19}
    \begin{alignat}{2}  \label{3.19a}
X_\mu &= \pd_\mu + \Gamma_\mu^a \pd_a  &\qquad X_a &= \pd_a
\\			  \label{3.19b}
\omega^\mu &= \od u^\mu & \omega^a & = \od u^a - \Gamma_\mu^a \od u^\mu .
    \end{alignat}
    \end{subequations}
Here the functions $\Gamma_\mu^a\colon U\to \field$ are defined via
    \begin{equation}    \label{3.19-1}
[\Gamma_\mu^a] = +[A_\nu^a] \cdot [A_\mu^\nu]^{-1}
    \end{equation}
and are called \emph{(2-index) coefficients} of $\Delta^h$. In a
matrix form, the equations~\eref{3.19} can be written as
    \begin{equation}    \label{3.20}
(X_\mu,X_a)
= (\pd_\nu,\pd_b) \cdot
    \begin{bmatrix}
\delta_\mu^\nu & 0 \\
+ \Gamma_\mu^b & \delta_a^b
    \end{bmatrix}
\quad
\begin{pmatrix} \omega^\mu \\ \omega^a  \end{pmatrix}
=
    \begin{bmatrix}
\delta^\mu_\nu & 0 \\
- \Gamma_\nu^a & \delta^a_b
    \end{bmatrix}
\cdot
\begin{pmatrix} \od u^\nu \\ \od u^b    \end{pmatrix}
\ .
    \end{equation}
The operators $X_\mu=\pd_\mu + \Gamma_\mu^a\pd_a$ are known as \emph{covariant
derivatives on} $T(U)$ and the plus sing in~\eref{3.19a} before
$\Gamma_\mu^a$ (hence in the r.h.s.\ of~\eref{3.19-1}) is conventional.

    If $\{u^{I}\}$ and $\{\tilde{u}^{I}\}$ are local coordinates on
open sets $U\subseteq E$ and  $\tilde{U}\subseteq E$, respectively, and
$U\cap\tilde{U}\not=\varnothing$, then, on the overlapping set
$U\cap\tilde{U}$, a problem arises: how are connected the adapted frames
corresponding to these coordinates? Let us mark with a tilde all quantities
that refer to the coordinates $\{\tilde{u}^{I}\}$. Since the adapted
frames are, by definitions, specialized ones, we can write (see~\eref{3.4})
    \begin{subequations}    \label{3.21}
    \begin{equation}    \label{3.21a}
(\tilde{X}_\mu,\tilde{X}_a) = (X_\nu,X_b) \cdot A \qquad
\begin{pmatrix} \tilde{\omega}^\mu \\ \tilde{\omega}^a  \end{pmatrix}
= A^{-1} \cdot
\begin{pmatrix} \omega^\nu \\ \omega^b   \end{pmatrix} ,
    \end{equation}
where the transformation matrix $A$ and its inverse have the
form~\eref{3.10}.  Recalling~\eref{3.2} and~\eref{3.3}, from these
equalities, we get
    \begin{equation}    \label{3.21b}
A = \diag
\Bigl( \Bigl[ \frac{\pd u^\nu}{\pd \tilde{u}^\mu} \Bigr] ,
      \Bigl[ \frac{\pd u^b}{\pd \tilde{u}^a} \Bigr]
\Bigr)
=
    \begin{pmatrix}
\bigl[ \frac{\pd u^\nu}{\pd \tilde{u}^\mu} \bigr]   & 0 \\
0	&	\bigl[ \frac{\pd u^b}{\pd \tilde{u}^a} \bigr]
    \end{pmatrix} \ .
    \end{equation}
    \end{subequations}
Combining~\eref{3.19-1} and~\eref{3.21}, one can easily prove

    \begin{Prop}    \label{Prop3.3}
A change  $\{u^{I}\}\mapsto \{\tilde{u}^{I}\}$ of the local bundle
coordinates implies the following transformation of the 2-index coefficients
of the connection:
    \begin{equation}    \label{3.22}
\Gamma_\mu^a \mapsto \tilde{\Gamma}_\mu^a
=
\Bigl( \frac{\pd \tilde{u}^a}{\pd u^b} \Gamma_\nu^b +
       \frac{\pd \tilde{u}^a}{\pd u^\nu}
\Bigr)  \frac{\pd u^\nu}{\pd \tilde{u}^\mu} .
    \end{equation}
    \end{Prop}

	It is obvious, a connection $\Delta^h$ is of class $C^m$,
$m\in\field[N]\cup\{0\}$, if and only if its coefficients $\Gamma_\mu^a$ are
$C^m$ functions on $U$, provided $\pd_I$ are $C^m$ vector fields on $U$
(which is the case when $E$ is a $C^{m+1}$ manifold). By virtue
of~\eref{3.22}, the $C^{m+1}$ changes of the local bundle coordinates
preserve the $C^m$ differentiability of $\Gamma_\mu^a$. Thus the $C^{m+1}$
differentiability of the base $M$ and bundle $E$ spaces is a necessary
condition for existence of $C^m$ connections on $(E,\pi,M)$; as we assumed
$m=1$ in this work, the connections considered here can be at most of
differentiability class $C^1$.

    The next proposition states that a connection on a bundle is
locally equivalent to a geometric object whose components transform
like~\eref{3.22}.

    \begin{Prop}    \label{Prop3.1}
To any connection $\Delta^h$ in a bundle $(E,\pi,M)$ can be assigned,
according to the procedure described above, a geometrical object on $E$ whose
components $\Gamma_\mu^a$ in bundle coordinates $\{u^{I}\}$ on $E$
transform according to~\eref{3.22} under a change
$\{u^{I}\}\mapsto \{\tilde{u}^{I}\}$ of the bundle coordinates on the
intersection of the domains of $\{u^{I}\}$ and $\{\tilde{u}^{I}\}$.
Conversely, given a geometrical object on $E$ with local transformation
law~\eref{3.22}, there is a unique connection $\Delta^h$ in $(E,\pi,M)$ which
generates the components of that object as described above.
    \end{Prop}

    \begin{Proof}
The first part of the statement was proved above, when we constructed the
adapted frame~\eref{3.19a} and derived~\eref{3.22}. To prove the second part,
choose a point $p\in E$ and some local coordinates $\{u^{I}\}$ on an open
set $U$ in $E$ containing $p$ in which the geometrical object mentioned has
local components $\Gamma_\mu^a$. Define a local frame
$\{X_{I}\}=\{X_\mu,X_a\}$ on $U$ by~\eref{3.19a}. The required connection
is then \( \Delta^h\colon q \mapsto \Delta_q^h := \{ r^\mu X_\mu|_q :
r^\mu\in\field\} \) for any $q\in U$, which means that $\Delta_q^h$ is the
linear cover of $\{X_\mu|_q\}$. The transformation law~\eref{3.22} insures the
independence of $\Delta^h$ from the local coordinates employed in its
definition.
    \end{Proof}

    From the construction of an adapted frame $\{X_{I}\}$, as well as
from the proof of proposition~\ref{Prop3.1}, follows that the set of vectors
$\{X_\mu\}$ is a basis for the horizontal distribution $\Delta^h$ and the set
$\{X_a\}$ is a basis for the vertical distribution $\Delta^v$. The matrix of
the restricted tangent projection $\pi_*|_{\Delta^h}$  in bundle coordinates
$\{u^\mu=x^\mu\circ\pi,u^a\}$ on $E$, where $\{x^\mu\}$ are local coordinates
on $M$, is the identity matrix as
\(
( \pi_*|_{\Delta_p^h})_\mu^\nu
=\frac{\pd(x^\mu\circ \pi)} {\pd u^\mu} \Big|_p
=\delta_\mu^\nu
\)
for any point $p$ in the domain of $\{u^{I}\}$. Hereof we get
    \begin{equation}    \label{3.20-1}
\pi_*|_{\Delta^h} (X_\mu) = \frac{\pd}{\pd x^\mu}
\qquad
\Bigl( \iff
\pi_*|_{\Delta_p^h} (X_\mu|_p) = \frac{\pd}{\pd x^\mu} \Big|_{\pi(p)}
\Bigr) .
    \end{equation}
In particular, from here follows that
$\pi_*|_{\Delta_p^h}\colon \Delta_p^h\to T_{\pi(p)}(M) $ is a vector space
isomorphism. The inverse to equation~\eref{3.20-1}, \viz
	\begin{equation}	\label{3.20-3}
X_\mu = (\pi_*|_{\Delta^h})^{-1}\Bigl(\frac{\pd}{\pd x^\mu}\Bigr)
   = (\pi_*|_{\Delta^h})^{-1} \circ \pi_* \Bigl(\frac{\pd}{\pd u^\mu}\Bigr)  ,
	\end{equation}
can be used in an equivalent definition of a frame $\{X_I\}$ adapted to local
coordinates $\{u^I\}$, namely, this is the frame
\(
\bigl(
(\pi_*|_{\Delta^h})^{-1} \circ \pi_* \bigl(\frac{\pd}{\pd u^\mu}\bigr) ,
\frac{\pd}{\pd u^a}
\bigr) .
\)
If one accepts such a definition of an adapted frame for $\Delta^h$, the
(2\ndash index) coefficients of $\Delta^h$ have to be defined via the
expansion~\eref{3.19a}; the only changes this may entail are in the proofs of
some results, like~\eref{3.21} and~\eref{3.22}.

	It is useful to be recorded also the simple fact that, by
construction, we have
    \begin{equation}    \label{3.20-2}
\pi_*(X_a) = 0 .
    \end{equation}

    Let $E$ be a $C^2$ manifold and $\Delta^h$ be a $C^1$ connection.
 The \emph{adapted frames are generally anholonomic} as the commutators
between the basic vector fields of the adapted frame~\eref{3.19a} are
(cf.~\eref{3.12} and~\eref{3.13})
    \begin{equation}    \label{3.23}
[X_\mu,X_\nu]_{\_} =  R_{\mu\nu}^{a} X_a
\quad
[X_\mu,X_b]_{\_} =  \lindex{}{\circ}\Gamma_{b\mu}^{a} X_a
\quad
[X_a,X_b]_{\_} = 0 ,
    \end{equation}
with
    \begin{subequations}    \label{3.24}
    \begin{align}   \label{3.24a}
R_{\mu\nu}^{a}
& = \pd_\mu(\Gamma_\nu^a) - \pd_\nu(\Gamma_\mu^a)
 + \Gamma_\mu^b\pd_b(\Gamma_\nu^a)
 - \Gamma_\nu^b\pd_b(\Gamma_\mu^a)
 = X_\mu(\Gamma_\nu^a) -  X_\nu(\Gamma_\mu^a)
\\          \label{3.24b}
\lindex{}{\circ}\Gamma_{b\mu}^{a}
& = -\pd_b(\Gamma_\mu^a) = -X_b(\Gamma_\mu^a)
    \end{align}
    \end{subequations}
being the fibre components of the curvature and fibre coefficients of the
connection. An obvious corollary from~\eref{3.24} is

    \begin{Prop}    \label{Prop3.2}
An adapted frame is holonomic if and only if
    \begin{equation}    \label{3.25}
R_{\mu\nu}^{a} = 0 \quad(\iff\ \Omega=0)\quad
\lindex{}{\circ}\Gamma_{b\mu}^{a} = 0 .
    \end{equation}
    \end{Prop}

    Therefore only the flat (integrable) $C^1$ connections, for which
$\Omega=0$, may admit holonomic adapted frames. Besides, as a consequence
of~\eref{3.24b} and~\eref{3.25}, such connections admit holonomic adapted
frames on $U\subseteq E$ if and only if there are local coordinates on $U$ in
which the coefficients $\Gamma_\mu^a$ are constant on the fibres passing
through $U$, \ie iff $\Gamma_\mu^a=G_\mu^a\circ\pi$ for some functions
$G_\mu^a\colon \pi(U)\to\field$, which is equivalent to
$\pd_b(\Gamma_\mu^a)=0$.

    \begin{Exmp}[\normalfont horizontal lifting of a path\bfseries]
		\label{Exmp3.1}
    Recall, the horizontal lift of a $C^1$ path $\gamma\colon J\to M$
passing through a point $p\in\pi^{-1}(\gamma(t_0))$ for some $t_0\in J$ is
the unique path $\bar{\gamma}_p\colon J\to E$ such that
$\pi\circ\bar{\gamma}_p=\gamma$, $\bar{\gamma}_p(t_0)=p$, and
$\dot{\bar{\gamma}}_p(t) \in\Delta_{\bar{\gamma}_p(t)}^h$ for all $t\in J$.
As in a specialized frame $\{e_{I}\}$ the relation $X_p\in\Delta^h_p$ is
equivalent to $e^a(X)=0$ for any $X\in\mathcal{X}(M)$, in an adapted
coframe, given by~\eref{3.19b}, the horizontal lift $\bar{\gamma}_p$ of
$\gamma$ is the unique solution of the initial value problem
    \begin{subequations}    \label{3.26}
    \begin{align}   \label{3.26a}
\omega^a(\dot{\bar{\gamma}}_p) &=  0
\\          \label{2.26b}
\bar{\gamma}_p(t_0) & =p
    \end{align}
    \end{subequations}
which is tantamount to any one of the initial-value problems ($t\in J$)
    \begin{subequations}    \label{3.26'}
    \begin{align}
       \tag{\protect\ref{3.26}$^{\prime}$a} \label{3.26'a}
\dot{\bar{\gamma}}_p^a(t)
   - \Gamma_\mu^a(\bar{\gamma}_p(t)) \dot{\bar{\gamma}}_p^\mu(t) &= 0
\\     \tag{\protect\ref{3.26}$^{\prime}$b} \label{3.26'b}
 \bar{\gamma}_p^{I}(t_0) &= p^{I} :=u^{I}(p)
        \end{align}
    \end{subequations}
\vspace{-2.4ex}
    \begin{subequations}    \label{3.26''}
    \begin{align}
       \tag{\protect\ref{3.26}$^{\prime\prime}$a}   \label{3.26''a}
\frac{\od (u^a\circ\bar{\gamma}_p(t) ) }{\od t}
   - \Gamma_\mu^a(\bar{\gamma}_p(t))
\frac{\od(x^\mu\circ\gamma(t))}{\od t} &= 0
\\     \tag{\ref{3.26}$^{\prime\prime}$b}   \label{3.26''b}
u^{I}( \bar{\gamma}_p(t_0) ) &= u^{I}(p) ,
    \end{align}
    \end{subequations}
\addtocounter{equation}{-2}
where $\{x^\mu\}$ are the local coordinates in the base that
induce the basic coordinates $\{u^\mu\}$ on the bundle space,
$u^\mu=x^\mu\circ\pi$. (Note that the quantities
$\frac{\od(x^\mu\circ\gamma(t))}{\od t}$, entering
into~\eref{3.26''a}, are the components of the vector
$\dot{\gamma}$ tangent to $\gamma$ at parameter value $t.)$ One
may call~\eref{3.26a}, or any one of its versions~\eref{3.26'a}
or~\eref{3.26''a}, the \emph{parallel transport equation} in an
adapted frame as it uniquely determines the parallel transport
along the restriction of $\gamma$ to any closed subinterval in $J$
(see definition~\ref{Defn3.2}).
    \end{Exmp}

    \begin{Exmp}[\normalfont the equation of geodesic paths\bfseries]
		\label{Exmp3.2}
    A connection $\Delta^h$ on the tangent bundle $(T(M),\pi_T,M)$ of a
manifold $M$ is called a \emph{connection on} $M$. In this case,
equation~\eref{3.26} defines also the geodesics (relative to $\Delta^h$) in
$M$. A $C^2$ path $\gamma\colon J\to M$ in a $C^2$ manifold $M$ is called a
\emph{geodesic path} if its tangent vector field $\dot\gamma$ undergoes
parallel transport along the same path $\gamma$, \ie
$\Psf^{\gamma|[\sigma,\tau]}(\dot\gamma(\sigma))=\dot\gamma(\tau)$ for all
$\sigma,\tau\in J$, which means that the lifted path
$\dot\gamma\colon J\to T(M)$ is a \emph{horizontal} lift of $\gamma$
(relative to $\Delta^h$). So, if $\{x^\mu\}$ are local coordinates on
$\pi(U)\in M$ and the bundle coordinates on $U\subseteq E$ are such
that~\cite[sect.~1.25]{Warner} $u^\mu=x^\mu\circ\pi$ and $u^{n+\mu}=\od
x^\mu$ ($\mu,\nu,\dots=1,\dots,n=\dim M$), then~\eref{3.26''a} transforms
into the \emph{geodesic equation} (on $M$)
    \begin{equation}    \label{3.27}
\frac{\od^2(x^\mu\circ\gamma(t))}{\od t^2}
-
\Gamma_\nu^{n+\mu}(\dot\gamma(t)) \frac{\od(x^\nu\circ\gamma(t))}{\od t} = 0
\qquad t\in J ,
    \end{equation}
which (locally) defines all geodesics in $M$. (With obvious
renumbering of the indices, one usually writes $\Gamma_\nu^{\mu}$
for $\Gamma_\nu^{n+\mu}$.) Of course, a particular geodesic is
specified by fixing some initial values for $\gamma(t_0)$ and
$\dot\gamma(t_0)$ for some $t_o\in J$. Notice,
equation~\eref{3.27} is an equation for a path $\gamma$ in $M$,
while~\eref{3.26''a} is an equation for the lifted path
$\bar\gamma$ in $T(M)$ provided the path $\gamma$ in $M$ is known;
for a geodesic path, evidently, we have $\bar\gamma=\dot\gamma$.
    \end{Exmp}


\section {Connections on vector bundles}
\label{Sect4}

    In this section, by $(E,\pi,M)$ we shall denote an arbitrary
\emph{vector} bundle~\cite{Poor}. A peculiarity of such bundles is that their
fibres are isomorphic vector spaces, which leads to a natural description of
the vertical distribution $\Delta^v$ on their fibre spaces, as well as to existence of a
kind of differentiation of their sections (known as covariant
differentiation).

    In the vector bundles are used, as we shall do in this section, the
so-called vector bundle coordinates which are linear on their fibres
and are constructed as follows (cf.~\cite[p.~30]{Saunders}).

    Let $\{e_a\}$ be a frame in $E$ over a subset
$U_M\subseteq M$, \ie $\{e_a(x)\}$ to be a basis in $\pi^{-1}(x)$ for all
$x\in U_M$. Then, for each $p\in\pi^{-1}(U_M)$, we have a unique expansion
$p=p^a e_a(\pi(p))$ for some numbers $p^a\in\field$. The \emph{vector fibre
coordinates} $\{u^a\}$ on $\pi^{-1}(U_M)$ induced (generated) by the frame
$\{e_a\}$ are defined via $u^a(p):=p^a$ and hence can be identified with the
elements of the coframe $\{e^a\}$ dual to $\{e_a\}$, \ie $u^a=e^a$. Conversely,
if $\{u^I\}$ are coordinates on $\pi^{-1}(U_M)$ for some $U_M\subseteq M$
which are linear on the fibres over $U_M$, then there is a unique frame
$\{e_a\}$ in $\pi^{-1}(U_M)$ which generates $\{u^a\}$ as just described;
indeed, considering $u^{n+1},\dots,u^{n+r}$ as 1\ndash forms on
$\pi^{-1}(U_M)$, one should define the frame $\{e_a\}$ required as a one whose
dual is $\{u^a\}$, \ie via the conditions $u^a(e_b)=\delta_b^a$.

    A collection $\{u^{I}\}$ of basic coordinates $\{u^\mu$\} and
vector fibre coordinates $\{u^{a}\}$ on $\pi^{-1}(U_M)$ is called \emph{vector
bundle coordinates} on $\pi^{-1}(U_M)$. Only such coordinates on $E$ will be
employed in this section.


\subsection{Vertical lifts}
    \label{Subsect4.1}

The idea of describing the vertical distribution $\Delta^v$ on a vector
bundle is that, if $L$ is a vector space, then to any $Y\in L$ there
corresponds a `vertical' vector field
$Y^v\in\mathcal{X}(L)=\Sec(T(L),\pi_T,L)$ whose value at $X\in L$ is the
vector tangent to the path $t\mapsto X+tY\in L$, with $t\in\field[R]$, at
$t=0$, \ie $Y^v|_X:=\frac{\od}{\od t}\big|_{t=0}(X+tY)$. Here and below,
with $\Sec(E,\pi,M)$ (resp.\ $\Sec^m(E,\pi,M)$ with $m\in\field[N]\cup\{0\}$)
we denote the module of sections (resp.\ $C^m$ sections) of a bundle
$(E,\pi,M)$ (resp.\ of a $C^{m+1}$ bundle $(E,\pi,M)$).

    Let $(E,\pi,M)$ be a vector bundle and $\Delta^v$ the vertical
distribution on it, viz., for each $p\in E$,
$\Delta^v\colon p\mapsto\Delta_p^v:=T_p(\pi^{-1}(\pi(p)))$. To every
$Y\in\Sec(E,\pi,M)$, we assign a \emph{vertical} vector field
$Y^v\in\Delta^v$ on $E$ such that, for $p\in E$,
    \begin{equation}    \label{4.1}
Y_p^v:=Y^v|_p := \frac{\od}{\od t}\Big|_{t=0}(p+tY|_{\pi(p)}).
    \end{equation}
(The mapping $(p,Y_{\pi(p)})\mapsto Y_p^v$ defines an isomorphism from the
pullback bundle $\pi^*E$ into the vertical bundle $\mathcal{V}(E)$ ---
see~\cite[sections~1.27 and~1.28]{Poor}
and also~\cite[p.~41, exercises~2.2.1 and~2.2.2]{Saunders}.)

    \begin{Lem} \label{Lem4.1}
Let $\{u^a\}$ be vector fibre coordinates generated by a frame $\{e_a\}$ on
$M$. If $Y\in\Sec(E,\pi,M)$ and $Y=Y^ae_a$, then
    \begin{equation}    \label{4.2}
Y^v = ( Y^a\circ\pi ) \frac{\pd}{\pd u^a} .
    \end{equation}
    \end{Lem}

    \begin{Proof}
Using the definition~\eref{4.1}, we get for $p\in E$:
    \begin{multline*}
Y^v|_p
= \frac{\od}{\od t}\Big|_{t=0}(p+tY|_{\pi(p)})
= \frac{ \od(u^a(p+tY|_{\pi(p)})) }{\od t} \Big|_{t=0}
                        \frac{\pd}{\pd u^a}\Big|_p
\\
= \frac{ \od(p^a+tY^a(\pi(p))) }{\od t} \Big|_{t=0}
                        \frac{\pd}{\pd u^a}\Big|_p
= Y^a(\pi(p)) \frac{\pd}{\pd u^a}\Big|_p
= \Bigl((Y^a\circ\pi) \cdot \frac{\pd}{\pd u^a}\Bigr)\Big|_p .
    \end{multline*}
    \end{Proof}

    If $Y\in\Sec(E,\pi,M)$, the vector field $Y^v:=v(Y)\in\Delta^v$,
defined via~\eref{4.1}, is called the \emph{vertical lift of the section} $Y$
and, in vector bundle coordinates, is (locally) given by~\eref{4.2}.

    \begin{Prop}    \label{Prop4.1}
The mapping
    \begin{equation}    \label{4.1-1}
    \begin{split}
v & \colon\Sec(E,\pi,M)  \to \{ \text{\textup{vector fields in }}\Delta^v \}
\\
v & \colon Y  \mapsto Y^v \colon p\mapsto
            Y^v|_p  := \frac{\od}{\od t}\Big|_{t=0}(p+tY_{\pi(p)})
    \end{split}
    \end{equation}
is a bijection and it and its inverse are linear mapping.
    \end{Prop}

    \begin{Proof}
The linearity and injectivity of $v$ follow directly from~\eref{4.1}. Now we
shall prove that, for each $Z\in\Delta^v$, there is a $Y\in\Sec(E,\pi,M)$
such that $Y^v=Z$, \ie $v$ is also surjective. Let $Z=Z^a\frac{\pd}{\pd u^a}$,
with $\{u^{I}\}$ being (local) vector bundle coordinates on $E$ and the
functions $Z^a$ being constant on the fibres of $E$, that is
$Z^{I}=z^{I}\circ\pi$ for some functions $z^I$ on $M$. Define $Y=z^ae_a$ with
$\{e_a\}$ being the frame in $E$ over $M$ generating $\{u^I\}$. By
lemma~\ref{Lem4.1}, we have
$Y^v=(z^a\circ\pi) \frac{\pd}{\pd u^a} = Z^a \frac{\pd}{\pd u^a}=Z$. The
linearity of $v^{-1}$ follows from here too.
    \end{Proof}

    Consider a section $\omega$ of the bundle dual to
$(E,\pi,M)$~\cite{Poor}. Its \emph{vertical lift} $\omega_v$ is a 1\ndash
form on $\Delta^v$ such that, for $Z\in\Delta^v$ and $p\in E$,
$\omega_v(Z)|_p=\omega(Y)|_{\pi(p)}$ for the
unique section $Y\in\Sec(E,\pi,M)$ with $Y^v=Z$ (see
proposition~\ref{Prop4.1}), \ie we have
$\omega_v(\cdot)|_{p}=(\omega\circ v^{-1}(\cdot))|_{\pi(p)}$ which means that
    \begin{equation}    \label{4.4}
\omega_v(Z) = (\omega\circ v^{-1}(Z))\circ\pi
\quad\text{or}\quad
\omega_v(Y^v)|_p = \omega(Y)|_{\pi(p)}\quad (=\omega_{\pi(p)}(Y_{\pi(p)})) .
    \end{equation}

    If $\{e^a=u^a\}$ is the coframe dual to $\{e_a\}$, and
$\omega=\omega_ae^a$, then in the coframe $\{\od u^a\}$ dual to
$\bigl\{\frac{\pd}{\pd u^a}\bigr\}$, we can write (cf.~\eref{4.2})
    \begin{equation}    \label{4.3}
\omega_v = (\omega_a\circ \pi) \od  u^a .
    \end{equation}

    It should be mentioned, `vertical' lifts of vector fields or 1\ndash
forms over the base space are generally not defined unless $E=T(M)$ or
$E=T^*(M)$, respectively.~%
\footnote{~%
Since $\pi_*(\Delta^v_p)=0_{\pi(p)}\in T_{\pi(p)}(M)$, $p\in E$, we can say
that only the zero vector field over $M$ has vertical lifts relative to $\pi$
and any vector field in $\Delta^v$ is its vertical lift. This conclusion is
independent of the existence of a connection on $(E,\pi,M)$ and depends only
on the fibre structure of $E$ induced by $\pi$.%
}

    Let $\Delta^h$ be a connection on $(E,\pi,M)$ and
$\varphi\colon E\to\field$ be a  $C^1$ mapping. Since any
$X\in\mathcal{X}(E)$ can uniquely be written as a direct sum
$X=v(X)\oplus h(X)$, with $v(X)\in\Delta^v$ and $h(X)\in\Delta^h$, we have
 $\varphi_*(X)=\varphi_*(v(X))+\varphi_*(h(X))\in\mathcal{X}(M)$.
If $\{Z_{I}\}$ is a specialized frame in $T(E)$ and $\{Z^{I}\}$ is its
dual coframe   of one\ndash forms on $\mathcal{X}(E)$, we immediately get
    \begin{equation}    \label{4.5}
\varphi_*
= (\varphi_*(Z_a))Z^a + (\varphi_*(Z_\mu))Z^\mu
= (Z_a(\varphi))Z^a + (Z_\mu(\varphi))Z^\mu
    \end{equation}
as $X=X^{I} Z_{I}$ entails $v(X)=X^aZ_a$ and $h(X)=X^\mu Z_\mu$; in
particular,~\eref{4.5} holds in any adapted (co)frame~\eref{3.19} and/or a
section $\varphi$ of the bundle $(E,\pi,M)$. If $\{u^{I}\}$ are vector
bundle coordinates, in the (co)frame~\eref{3.19} adapted to them, we have
$Z_\mu=X_\mu$, $Z_a=\pd_a$, $Z^\mu=\omega^\mu=\od u^\mu$, $Z^a=\omega^a$, and
we can write the expansion $\varphi=\varphi_au^a$ with $\varphi_a\colon
E\to\field$. Thus~\eref{4.5} takes the form
\[
\varphi_*
= \varphi_a\omega^a + (X_\mu(\varphi_a u^a)) \omega^\mu
= \varphi_v + (X_\mu(\varphi_a u^a)) \omega^\mu,
\]
where~\eref{4.3} was applied.

    A \emph{section $Y$ of} $(E,\pi,M)$ and \emph{section $\omega$   of
the bundle dual to} $(E,\pi,M)$ can be lifted \emph{vertically} via the
mappings
    \begin{subequations}    \label{4.6-1}
    \begin{align}   \label{4.6-1a}
v\colon Y& \mapsto Y^v\in\Delta^v
\\          \label{4.6-1b}
\omega&\mapsto \omega_v
    \end{align}
    \end{subequations}
respectively given by~\eref{4.1-1} and~\eref{4.4} (see also~\eref{4.2}
and~\eref{4.3}). These mappings do not require a connection and arise only
from the fibre structure of the bundle space induced from the projection
$\pi\colon E\to M$.

    If a connection $\Delta^h$ on $(E,\pi,M)$ is given, it generates
\emph{horizontal lifts of the vector fields on the base space} $M$
and of the \emph{one\ndash forms on the same base space} $M$ into respectively
vector fields in $\Delta^h$ and linear mappings on the vector fields in
$\Delta^h$. Precisely, if $F\in\mathcal{X}(M)$ and $\phi\in\Lambda^1(M)$,
their \emph{horizontal lifts} are defined by the mappings~%
\footnote{~%
Alternatively, one may define $\phi_h'=\phi\circ\pi_*=\pi^*(\phi)$, which
expands the domain of $\phi_h$, defined by~\eref{4.6-2b}, on the whole space
$\mathcal{X}(E)$. Obviously, $\phi_h'(Z)=\phi_h(Z)$ for
$Z\in\Delta^h\subseteq\mathcal{X}(E)$ and $\phi_h'(Z)=0$ for
$Z\in\mathcal{X}(E)\setminus\{X\in\Delta^h\}$.%
}
    \begin{subequations}    \label{4.6-2}
    \begin{align}   \label{4.6-2a}
F &\mapsto F^h\in\Delta^h
\quad\text{with}\quad
F^h\colon p\mapsto F_p^h := (\pi_*|_{\Delta^h_p})^{-1} (F_{\pi(p)})
\qquad p\in E
\\          \label{4.6-2b}
\phi &\mapsto \phi_h
\quad\text{with}\quad
\phi_h := \phi\circ \pi_*|_{\Delta^h} \colon p\mapsto \phi_h|_p
    = \phi|_{\pi(p)} \circ (\pi_*|_{\Delta^h_p}) .
    \end{align}
    \end{subequations}
The horizontal lift $\phi_h$ of $\phi$ can also be defined alternatively via
    \begin{equation}    \label{4.6-3}
\phi_h(F^h)|_p = \phi(F)|_{\pi(p)}
    \end{equation}
which equation is tantamount to~\eref{4.6-2b}.

    Let $\{u^\mu=x^\mu\circ\pi,u^a\}$ be vector bundle coordinates and
$\{X_{I}$\} (resp.\ $\{\omega_{I}\}$) be the adapted to them frame
(resp.\ coframe) constructed from them according to~\eref{3.19}. If
$Y=Y^ae_a$, $\omega=\omega_ae^a$,
$F=F^\mu\frac{\pd}{\pd x^\mu}\in\mathcal{X}(M)$, and
$\phi=\phi_\mu\od x^\mu\in\Lambda^1(M)$, the equations~\eref{4.2}
and~\eref{4.3} imply
    \begin{equation}    \label{4.6-4}
Y^v = (Y^a\circ\pi) X_a \quad \omega_v = (\omega_a\circ\pi) \omega^a,
    \end{equation}
while from~\eref{4.6-2} and~\eref{3.20-1}, one gets
    \begin{equation}    \label{4.6-5}
F^h = (F^\mu\circ\pi) X_\mu \quad \phi_h = (\phi_\mu\circ\pi) \omega^\mu ,
    \end{equation}
which agree with~\eref{3.9-6}.



\subsection{The tangent and cotangent bundle cases}
    \label{Subsect4.2}

    As an example, in the present subsection is considered a connection
$\Delta^h$ on the tangent bundle $(T(M),\pi_T,M)$ over a manifold $M$.

	A vector field $Y\in\mathcal{X}(M)=\Sec(T(M),\pi_T,M)$ has unique
vertical lift $Y^v\in\Delta^v$ (which is independent of $\Delta^h$) and unique
\emph{horizontal} lift given by (see~\eref{4.1-1})
    \begin{equation}    \label{4.7}
Y^v:=v(Y) \in\Delta^v
\quad
Y^h:= ((\pi_T)_*|_{\Delta^h})^{-1}(Y) \in\Delta^h,
    \end{equation}
the last equality meaning that $Y_p^h:= ((\pi_T)_*|_{\Delta_p^h})^{-1}(Y_p)$, which
is correct as $(\pi_T)_*|_{\Delta_p^h}\colon\Delta_p^h\to T_{\pi(p)}(M)$ is an
isomorphism. Respectively, if $\omega$ is 1\ndash form on $M$, it has
vertical lift $\omega_v$ (which is independent of $\Delta^h$) and
\emph{horizontal} lift $\omega_h$, which is one\ndash form on $\Delta^h$,
defined by (see~\eref{4.4})
    \begin{equation}    \label{4.8}
\omega_v(Z) = (\omega\circ v^{-1}(Z))\circ\pi_T
\quad
\omega_h:=\omega\circ(\pi_T)_* =\pi_T^*(\omega) .
    \end{equation}
The horizontal lift of $\omega$ has the properties
    \begin{subequations}    \label{4.9}
    \begin{align}   \label{4.9a}
\omega_h(Y^v) &= 0
\quad\text{for}\quad Y\in\mathcal{X}(M)
\\          \label{4.9b}
\omega_h(Y^h) &= (\omega(Y))\circ \pi_T
\quad\text{for}\quad Y\in\mathcal{X}(M) ,
    \end{align}
    \end{subequations}
the first of which is equivalent to
    \begin{equation}
    \tag{\protect\ref{4.9a}$^\prime$}   \label{4.9a'}
 \omega_h(Z) = 0
\quad\text{for}\quad Z\in\Delta^v ,
    \end{equation}
due to proposition~\ref{Prop4.1}.

    Thus there arises a lift $\mathcal{X}(M)\to\mathcal{X}(T(M))$ such
that the \emph{lift} of $Y\in\mathcal{X}(M)$ is $\bar{Y}\in\mathcal{X}(T(M))$
with
    \begin{equation}    \label{4.10}
\bar{Y}:=Y^v\oplus Y^h  .
    \end{equation}
Obviously, this decomposition respects definition~\ref{Defn4.1} and
    \begin{equation}    \label{4.11}
(\pi_T)_*(\bar{Y}) = (\pi_T)_*(Y^h) = Y .
    \end{equation}
The dual \emph{lift}  $\omega\mapsto\bar{\omega}\in\Lambda^1(T(M))$ of a
1\ndash form $\omega\in\Lambda^1(M)$ is given by
    \begin{equation}    \label{4.12}
\bar{\omega} = \omega_v\oplus\omega_h.
    \end{equation}
As a result of~\eref{4.4} and~\eref{4.9}, we have
    \begin{equation}    \label{4.13}
\bar{\omega}(\bar{Y})
= \omega_v(Y^v) + \omega_h(Y^h)
= 2(\omega(Y)) \circ\pi_T .
    \end{equation}

    At last, let us look on the vertical and/or horizontal lifts from the
view point of local bases/frames.

    In a case of the tangent bundle $(T(M),\pi_T,M)$ (resp.\ cotangent
bundle $(T^*(M),\pi_T^*,M)$) over a manifold $M$, any coordinate system
$\{x^\mu\}$ on an open set $U_M\subseteq M$ induces natural vector bundle
coordinates in the bundle space~\cite[sec.~1.25]{Warner} (see
also~\cite[pp.~8,~43]{Saunders}). For the purpose, we put
$e_\mu=\frac{\pd}{\pd x^\mu}$, so that $e^\mu=\od x^\mu$ and we get
($\lambda,\mu,\ldots=1,\dots,\dim M$ and $a,b=\dim M+1,\dots,2\dim M$)
    \begin{subequations}    \label{4.6}
    \begin{gather}  \label{4.6a}
\{u^{I}\} = \{x^\mu\circ\pi_T,\od x^\nu\}
\quad\text{i.e.}\quad
u^\mu = x^\mu\circ\pi_T \quad u^a=\od x^{a-\dim M}
\\\intertext{on $\pi_T^{-1}(U_M)$ in the tangent bundle case, and}
                \label{4.6b}
\{u^{I}\}
= \Bigl \{x^\mu\circ\pi_{T^*},(\cdot)\Bigl(\frac{\pd}{\pd x^\nu}\Bigr) \Bigr\}
\quad\text{i.e.}\quad
u^\mu = x^\mu\circ\pi_{T^*}
\quad
u^{\dim M+\nu} \colon \xi\mapsto \xi \Bigl(\frac{\pd}{\pd x^\nu}\Bigr)
    \end{gather}
    \end{subequations}
on $\pi_{T^*}^{-1}(U_M)\ni\xi$, in the cotangent bundle case. In connection
with the higher order (co)tangent bundles, it is convenient the vector fibre
coordinates to be denoted also as $u_1^\mu:=\dot{x}^\mu:=\od x^\mu$ in $T(M)$
and by $u_\mu^1(\cdot)=(\cdot)\Bigl(\frac{\pd}{\pd x^\mu}\Bigr)$ in $T^*(M)$.

    Consider the vector bundle coordinates
$\{u^\mu=x^\mu\circ\pi_T,u_1^\nu=\od x^\nu\}$ on $\pi_T^{-1}(U_M)$. They
induce the frame
\(
\bigl\{\pd_\mu=\frac{\pd}{\pd u^\mu} ,
    \pd_\nu^1=\frac{\pd}{\pd u_1^\nu} \Bigr\}
\)
and the coframe $\{\od u^\mu,\od u_1^\nu\}$ on $\pi_{T}^{-1}(U_M)$ and
$\pi_{T^*}^{-1}(U_M)$, respectively. According to~\eref{3.20}, they induce
the following adapted frame and its dual coframe:
    \begin{subequations}    \label{4.14}
    \begin{gather}  \label{4.14a}
(X_\mu,X_\mu^1)
= (\pd_\nu,\pd_\nu^1) \cdot
    \begin{bmatrix}
\delta_\mu^\nu & 0 \\
+ \Gamma_\mu^\nu & \delta_\mu^\nu
    \end{bmatrix}
= (\pd_\mu + \Gamma_\mu^\nu \pd_\nu^1 , \pd_\mu^1)
\\          \label{4.14b}
\begin{pmatrix} \omega^\mu \\ \omega_1^\mu  \end{pmatrix}
=
    \begin{bmatrix}
\delta_\nu^\mu & 0 \\
- \Gamma_\nu^\mu & \delta_\nu^\mu
    \end{bmatrix}
\cdot
\begin{pmatrix} \od u^\nu \\ \od u_1^\mu    \end{pmatrix}
=
    \begin{pmatrix}
\od u^\mu \\
\od u_1^\mu - \Gamma_\nu^\mu \od u^\nu
    \end{pmatrix} \ ,
    \end{gather}
    \end{subequations}
where, as accepted in the (co)tangent bundle case, a fibre index, like $a$,
is replace with a base index, like $\mu$, according to
$a\mapsto\mu=a-\dim M$, which leads to identification like
$\Gamma_\mu^\nu:=\Gamma_\mu^{\dim M+\nu}$.

    Consider a vector field
$Y=Y^\mu\frac{\pd}{\pd x^\mu}\in\mathcal{X}(M)$ and  1\ndash form
 $\eta=\eta_\mu\od x^\mu \in \Lambda^1(M)$. According to~\eref{4.2}
and~\eref{4.3}, their vertical lifts are
    \begin{subequations}    \label{4.15}
    \begin{gather}  \label{4.15a}
Y^v =(Y^\mu\circ\pi_T) X_\mu^1\in\Delta^v
\quad
\eta^v = (\eta_\mu\circ\pi_{T^*} ) \omega_1^\mu
\\\intertext{ and similarly, due to~\eref{4.6-5}, the horizontal lifts of $Y$
and $\eta$ are}
            \label{4.15b}
Y^h =(Y^\mu\circ\pi_T) X_\mu\in\Delta^h
\quad
\eta^h = (\eta_\mu\circ\pi_{T^*} ) \omega^\mu .
    \end{gather}
    \end{subequations}



\subsection{Linear connections on vector bundles}
    \label{Subsect4.3}

    The most valued structures in/on vector bundles are the ones which
are compatible/consistent with the linear structure of the fibres of these
bundles. Since a distribution $\Delta\colon p\mapsto\Delta_p\subseteq
T_p(E)$, $p\in E$, on the bundle space $E$ of a (vector) bundle $(E,\pi,M)$
cannot be considered as a linear mapping without additional hypotheses, the
concept of a linear connection arises from the one of the parallel transport
assigned to a connection (see definition~\ref{Defn3.2}). (For an alternative
approach, see~\cite[p.~42]{Mangiarotti&Sardanashvily}.)

    \begin{Defn}    \label{Defn4.1}
A connection on a vector bundle is called \emph{linear} if the assigned to it
parallel transport is a linear mapping along every path in the base space, \ie
if the mapping~\eref{3.9-4} is linear for all paths
$\gamma\colon[\sigma,\tau]\to M$ in the base.
    \end{Defn}

    The restriction on a connection to be linear is quite severe and
is described locally by

    \begin{Thm} [\normalfont
			cf.~\protect{\cite[sec.~5.2]{Rahula}} \bfseries]
		\label{Thm4.1}
Let $(E,\pi,M)$ be a vector bundle, $\{u^{I}\}$ be vector bundle
coordinates on an open set $U\subseteq E$, and $\Delta^h$ be a connection on
it described in the frame $\{X_{I}\}$, adapted to $\{u^{I}\}$, by its
2\ndash index coefficients $\Gamma_\mu^a$ (see~\eref{3.18}--\eref{3.19-1}).
The connection $\Delta^h$ is linear if and only if, for each $p\in U$,
    \begin{equation}    \label{4.16}
\Gamma_\mu^a(p)
= - \Gamma_{b\mu}^{a}(\pi(p)) u^b(p)
= - \bigl( (\Gamma_{b\mu}^{a}\circ \pi )\cdot u^b\bigr) (p) ,
    \end{equation}
where $\Gamma_{b\mu}^{a}\colon\pi(U)\to\field$ are some functions on the set
$\pi(U)\subseteq M$ and the minus sign before $\Gamma_{b\mu}^{a}$
in~\eref{4.16} is conventional.
    \end{Thm}

    \begin{Proof}
Take a $C^1$ path $\gamma\colon[\sigma,\tau]\to\pi(U)$ and consider the
parallel transport equation~\eref{3.26''a}, \viz
    \begin{equation}    \label{4.17}
\frac{\od\bar{\gamma}_p^a(t)}{\od t}
=
\Gamma_\mu^a(\bar{\gamma}_p(t)) \dot\gamma^\mu(t),
    \end{equation}
where $\bar{\gamma}_p\colon[\sigma,\tau]\to U$ is the horizontal
lift of $\gamma$ through $p\in\pi^{-1}(\gamma(\sigma))$,
$\bar{\gamma}^a:=u^a\circ\bar{\gamma}$, and \( \dot\gamma^\mu(t)
=\frac{\od(x^\mu\circ\gamma(t))}{\od t}
=\frac{\od(u^\mu\circ\bar{\gamma}(t))}{\od t} \) as
$u^\mu=x^\mu\circ\pi$ for some coordinates $\{x^\mu\}$ on
$\pi(U)$.

    SUFFICIENCY.
    If~\eref{4.16} holds,~\eref{4.17} is transformed into
    \begin{equation}    \label{4.18}
\frac{\od\bar{\gamma}_p^a(t)}{\od t}
=
- \Gamma_{b\mu}^a(\gamma(t)) \bar{\gamma}_p^b(t) \dot\gamma^\mu(t),
    \end{equation}
which is a system of $r$ linear-first order ordinary differential equations
for the $r$ functions $\bar{\gamma}_p^{n+1},\dots,\bar{\gamma}_p^{n+r}$.
According to the general theorems of existence and uniqueness of the
solutions of such systems~\cite{Hartman}, it has a unique solution
    \begin{equation}    \label{4.19}
\bar{\gamma}_p^{a}(t) = Y_b^a(t) p^b
    \end{equation}
satisfying the initial condition $\bar{\gamma}_p^{a}(\sigma)=u^a(p)=:p^a$,
where $Y=[Y_b^a]$ is the fundamental solution of~\eref{4.18}, \ie
    \begin{equation}    \label{4.20}
\frac{\od Y(t)}{\od t}
=
- [\Gamma_{b\mu}^{a}(\gamma(t))\dot\gamma^\mu(t)]_{a,b=n+1}^{n+r} \cdot Y(t)
\quad
Y(\sigma)=\openone_{r\times r}=[\delta_b^a] .
    \end{equation}
The linearity of~\eref{3.9-4} in $p$ follows from~\eref{4.19} for $t=\tau$.

    NECESSITY.
    Suppose~\eref{3.9-4} is linear in $p$ for all paths
$\gamma\colon[\sigma,\tau]\to\pi(U)$. Then
$\bar{\gamma}_p(t):=\Psf^{\gamma|[\sigma,t]}(p)$ is the horizontal lift of
$\gamma|[\sigma,t]$ through $p$ and (cf.~\eref{4.19})
 $\bar{\gamma}_p^{a}(t)=A_b^a(\gamma(t))p^b$ for some $C^1$ functions
$A_b^a\colon\pi(U)\to\field$. The substitution of this equation
in~\eref{4.17} results into
\[
\frac{\pd A_b^a(x)}{\pd x^\mu} \Big|_{x=\gamma(t) =\pi(\bar{\gamma}_p(t))}
\cdot \dot{\gamma}^\mu  p^b
=
\Gamma_{\mu}^{a}(\bar{\gamma}_p(t)) \dot\gamma^\mu(t) .
\]
Since $\gamma\colon[\sigma,\tau]\to M$, we get equation~\eref{4.16} from here,
for $t=\sigma$, with
 $\Gamma_{b\mu}^{a}(x) = - \frac{\pd A_b^a(x)}{\pd x^\mu}$ for $x\in\pi(U)$.
    \end{Proof}

    The functions $\Gamma_{b\mu}^{a}\colon\pi(U)\to\field$ will be
referred as the \emph{(local) 3\ndash index coefficients} of the linear
connection $\Delta^h$ in the adapted frame $\{X_{I}\}$. If there is no
risk to confuse them with the 2\ndash index coefficients
$\Gamma_{\mu}^{a}\colon U\to\field$, they will be called simply coefficients
of $\Delta^h$. Note, the 2\ndash index coefficients of a linear connections
are defined on (a subset of) the bundle space $E$, while the 3\ndash index
ones are define on (a subset of) the base space $M$. The equation~\eref{4.18}
is simply the \emph{parallel transport equation} for the linear connection
considered.

    \begin{Exmp}    \label{Exmp4.1}
    Since $u^a$ is replaced by $u_1^\mu=\od x^\mu$ in the tangent bundle case
(see Subsect.~\ref{Subsect4.2}), the linear connections in $(T(M),\pi_T,M)$
have 2\ndash index coefficients of the form
    \begin{equation}    \label{4.29}
\Gamma_\mu^\nu
= - (\Gamma_{\lambda\mu}^{\nu}\circ\pi_T) \cdot u_1^\lambda
= - (\Gamma_{\lambda\mu}^{\nu}\circ\pi_T) \cdot \od x^\lambda
    \end{equation}
and, consequently, they can be regarded as 1-forms on $M$.
    \end{Exmp}

    Consider a linear connection $\Delta^h$ on a vector bundle
$(E,\pi,M)$. Let $\Gamma_\mu^a$ and $\Gamma_{b\mu}^{a}$ be its 2- and 3\ndash
index coefficients, respectively, in a frame $\{X_{I}\}$ adapted to
vector bundle coordinates $\{u^{I}\}$.

    \begin{Cor}    \label{Cor4.1}
	The 3-index coefficients $\Gamma_{b\mu}^{a}$ of a linear connection
$\Delta^h$ uniquely define the fibre coefficients of $\Delta^h$ in $\{X_{I}\}$
by
    \begin{equation}    \label{4.22}
\lindex[\Gamma]{}{\circ}{}_{b\mu}^{a}
= \Gamma_{b\mu}^{a}\circ\pi = \pi^*(\Gamma_{b\mu}^{a}),
    \end{equation}
that is the fibre coefficients of a linear connection are equal to the 3\ndash
index ones lifted by the projection $\pi$.
    \end{Cor}

    \begin{Proof}
    Since~\eref{3.19a} and~\eref{4.16} imply
    \begin{equation}    \label{4.21}
[X_\mu,X_b]_{\_} = (\Gamma_{b\mu}^{a}\circ\pi) X_a,
    \end{equation}
the equation~\eref{4.22} follows from~\eref{3.13a} and~\eref{3.14a}
or~\eref{3.24b} and~\eref{4.16}.
    \end{Proof}

    As the vector bundle coordinates $\{u^{I}\}$ are, by definition,
linear on the fibres of the bundle, the general change of such coordinates is
    \begin{equation}    \label{4.23}
\{u^\mu,u^a\} \mapsto
\{\tilde{u}^\mu = \tilde{x}^\mu\circ\pi ,
\tilde{u}^a=(B_b^a\circ\pi)\cdot u^b\} ,
    \end{equation}
with $B=[B_b^a]$ being a non-degenerate matrix-valued function on $\pi(U)$.
The change~\eref{4.23} entails the following transformation of the
corresponding adapted frames
    \begin{equation}    \label{4.24}
\{X_\mu,X_a\} \mapsto
\{
\tilde{X}_\mu=(B_\mu^\nu\circ\pi)\cdot X_\nu,
                \tilde{X}_a=(B_a^b\circ\pi)\cdot X_b
\} ,
    \end{equation}
where $[B_\mu^\nu]=\bigl[\frac{\pd x^\nu}{\pd\tilde{x}^\mu}\bigr]$ is a
non-degenerate matrix-valued function on the intersection of the domains of
$\{x^\mu\}$ and $\{\tilde{x}^\mu\}$. (In~\eref{4.24} we have used that
\(
\frac{\pd u^\nu}{\pd\tilde{u}^\mu}\big|_p
=
\frac{\pd (x^\nu\circ\pi)}{\pd(\tilde{x}^\mu\circ\pi)}\big|_p
=
\frac{\pd x^\nu}{\pd\tilde{x}^\mu}\big|_{\pi(p)}
\).)

    \begin{Prop}    \label{Prop4.1-1}
The change~\eref{4.23} implies the following transformations of the 3\ndash
index coefficients of the linear connection:
    \begin{equation}    \label{4.25}
\Gamma_{b\mu}^{a} \mapsto
\tilde{\Gamma}_{b\mu}^{a}
=
B_\mu^\nu
\Bigl( B_d^a \Gamma_{c\nu}^{d} - \frac{\pd B_c^a}{\pd x^\nu} \Bigr)
(B^{-1})_b^c .
    \end{equation}
    \end{Prop}

    \begin{Proof}
Apply~\eref{4.24},~\eref{3.22} and~\eref{4.16}. Alternatively, the same
transformation law follows also from equations ~\eref{3.15a} and~\eref{4.22}.
    \end{Proof}

	If we introduce the matrix\ndash valued functions
$\Gamma_\mu:=[\Gamma_{b\mu}^{a}]$ and
$\tilde{\Gamma}_\mu:=[\tilde{\Gamma}_{b\mu}^{a}]$ on $M$, we can
rewrite~\eref{4.25} as
    \begin{equation}
    \tag{\protect\ref{4.25}$^\prime$}   \label{4.25'}
    \begin{split}
\Gamma_\mu \mapsto \tilde{\Gamma}_\mu
& = B_\mu^\nu\
\Bigl( B\cdot\Gamma_\nu - \frac{\pd B}{\pd x^\nu} \Bigr) \cdot B^{-1}
\\
&= B_\mu^\nu B \cdot
\Bigl( \Gamma_\nu\cdot B^{-1} + \frac{\pd B^{-1}}{\pd x^\nu} \Bigr) .
    \end{split}
    \end{equation}
This relation correspond to~\eref{3.16} with $[A_b^a]=B^{-1}\circ\pi$ (see
also~\eref{4.22}) as the frame $\{e_a\colon M\to E\}$, relative to which the
vector fibre coordinates $\{u^a\}$ are defined
($E\ni p \mapsto u^a(p)$ with $p=u^a(p)e_a(\pi(p))$), transforms via the
matrix inverse to $B\circ\pi$.

	Let $E$ be a $C^2$ manifold and $\Delta^h$ a $C^1$ connection on
$(E,\pi,M)$.
    Substituting~\eref{4.16} into~\eref{3.24a}, we get the fibre
components of the curvature of a linear connection as
    \begin{align}   \label{4.26}
R_{\mu\nu}^{a}
&= - (R_{b\mu\nu}^{a}\circ\pi) \cdot u^b
\\\intertext{where} \label{4.27}
R_{b\mu\nu}^{a}
&:=
\frac{\pd}{\pd x^\mu} (\Gamma_{b\nu}^{a}) -
\frac{\pd}{\pd x^\nu} (\Gamma_{b\mu}^{a}) -
\Gamma_{b\mu}^{c}\Gamma_{c\nu}^{a} +
\Gamma_{b\nu}^{c}\Gamma_{c\mu}^{a},
\\\intertext{or in a matrix form}
    \tag{\protect\ref{4.27}$^\prime$}   \label{4.27'}
R_{\mu\nu}
&:=[R_{b\mu\nu}^{a}]
=
\frac{\pd \Gamma_\nu}{\pd x^\mu} -
\frac{\pd \Gamma_\mu}{\pd x^\nu} -
\Gamma_\nu\cdot\Gamma_\mu +
\Gamma_\mu\cdot\Gamma_\nu ,
    \end{align}
are the \emph{components of the curvature operator} (see below~\eref{4.44}).
As a result of~\eref{3.13b} and~\eref{4.26}, the transformation~\eref{4.23}
entails the change
    \begin{equation}    \label{4.28}
R_{b\mu\nu}^{a} \mapsto \tilde{R}_{b\mu\nu}^{a}
= B_\mu^\lambda B_\nu^\varrho (B^{-1})_c^a B_b^d R_{d\lambda\varrho}^{c} ,
    \end{equation}
or in a matrix form
    \begin{equation}
    \tag{\protect\ref{4.28}$^\prime$}   \label{4.28'}
R_{\mu\nu} \mapsto \tilde{R}_{\mu\nu}
= B_\mu^\lambda B_\nu^\varrho B^{-1} \cdot R_{\lambda\varrho}\cdot B,
    \end{equation}
which corresponds to~\eref{3.15b} with $A=B^{-1}\circ\pi$ (see
also~\eref{4.26}).



\subsection{Covariant derivatives in vector bundles}
    \label{Subsect4.4}

    A possibility for introduction of differentiation in vector bundles,
endowed with connection, comes from the vector space structure of their
fibres. This operation can be defined in many independent ways,
leading to identical results. In one of them is involved the parallel
transport induced by the connection: the idea is the values of sections
 to be parallel transported (along paths in the base) into a single fibre
(over the paths), where one can work with the `transported' sections as with
functions with values in a vector space. Other method uses the
existence of natural vertical lifts of sections of the bundle and horizontal
lifts of the vector fields on the base space; since the both lifts are vector
fields on the bundle space, their commutator (or Lie derivative relative to
each other) is well defined and can be used as a prototype of some sort of
differentiation. We shall realize below the second method mentioned, which
seems is first introduced in a rudimentary form in~\cite[p.~31]{Rahula}.~%
\footnote{~%
In~\cite[p.~31]{Rahula} is proved that, for $F=\frac{\pd }{\pd x^\mu}$ and in
our notation, the $a$-th component of the right hand sides of~\eref{4.36-1} and
of~\eref{4.37} coincide in a frame $\{E_a\}$ in $E$.%
}%
~%
\footnote{~%
An equivalent alternative approach is realized
in~\cite[sections~2.49--2.52]{Poor}.%
}
The first way, as well as the axiomatic  approach, for introduction of
covariant derivatives will be obtained as theorems in what follows.

    Let $(E,\pi,M)$ be a vector bundle on which a \emph{linear} connection
$\Delta^h$ is defined. Suppose $\{E_a\}$ is a frame in $E$ to which vector
fibre coordinates $\{u^a\}$ are associated and $\{u^{I}\}$ be the
corresponding vector bundle coordinates. The frame adapted to $\{u^{I}\}$
will be denoted by $\{X_{I}\}$ and $\{\omega^{I}\}$ will be its dual
coframe, both defined by~\eref{3.19} through the (2\ndash index) coefficients
$\Gamma_\mu^a$ of $\Delta^h$.

    Let $\hat{Z}=\hat{Z}^aX_a\in\Delta^v$ and
$\bar{Z}=\bar{Z}^\mu X_\mu\in\Delta^h$  be respectively  vertical and
horizontal vector fields on $E$. Define a mapping
$\hat{\nabla}\colon\Delta^v\oplus\Delta^h=T(E)\to\mathcal{X}(E)$ such that~%
\footnote{~%
The idea of the construction~\eref{4.30} is to drag the vertical vector field
$\hat{Z}$ along the horizontal one $\bar{Z}$, which will give a vector field
in $\mathcal{X}(E)$, and then to project the result onto the \emph{vertical}
distribution $\Delta^v$ by means of the invariant projection operator
\(
\Pi
=X_a\otimes \omega^a\colon \mathcal{X}(E)\to \mathcal{X}(E) .
\)
Evidently $\Pi^2=\Pi\circ\Pi=\Pi$ and $\Pi$ is the unit (identity) tensor in
the tensor product of vector fields and 1\ndash forms on $E$.%
}
    \begin{equation}    \label{4.30}
\hat{\nabla} \colon(\hat{Z},\bar{Z})
\mapsto \hat{\nabla}_{\bar{Z}} (\hat{Z})
:=
\Pi(\Lied_{\bar{Z}}\hat{Z}) \in \mathcal{X}(E) ,
    \end{equation}
where the (1,1) tensor field
    \begin{equation}    \label{4.31}
\Pi := \sum_a X_a\otimes \omega^a
    \end{equation}
is considered as a operator on the set of vector fields on $E$. Since
(see~\eref{2.1b} and~\eref{2.7})
\[
\Lied_{\bar{Z}}\hat{Z}
= \bar{Z}(\hat{Z}^a) X_a + \bar{Z}^\mu \hat{Z}^a [X_\mu,X_a]_{\_}
\]
and $\omega^a(X_\mu)=\delta_\mu^a=0$, form~\eref{3.23},~\eref{3.24b}
and~\eref{4.30}, we obtain
    \begin{equation}    \label{4.32}
\hat{\nabla}_{\bar{Z}} \hat{Z}
=
\bar{Z}^\mu\{ X_\mu(\hat{Z}^a) - \hat{Z}^b\pd_b(\Gamma_\mu^a) \} X_a ,
    \end{equation}
from where one can prove, via direct calculation, the independence of
 $\hat{\nabla}_{\bar{Z}} \hat{Z}$ of the particular (co)frame used. For any
particular point $p\in E$, the value of the vector field~\eref{4.32} is a
vertical vector, $(\hat{\nabla}_{\bar{Z}} \hat{Z})|_p\in\Delta_p^v$, but
generally $\hat{\nabla}_{\bar{Z}} \hat{Z}$ is not a vertical vector field.
The reason is that a  vertical vector field on $E$ is a mapping
\(
V\colon p\mapsto V_p\in\Delta^v_p
:=T_p(\pi^{-1}(\pi(p))
:=T_{\imath(p)}(\pi^{-1}(\pi(p))
=(\pi_*|_p)^{-1}(0_{\pi(p)}) ,
\)
with $\imath\colon\pi^{-1}(p)\to E$ being the inclusion mapping and
$0_{\pi(p)}\in T_{\pi(p)}(M)$ being the zero vector, due to which $V_p$, and
hence its components, must depend only on $\pi(p)\in M$. Therefore, we have
    \begin{equation}    \label{4.33}
\hat{\nabla}_{\bar{Z}}\hat{Z} \in \Delta^v
\iff
\pd_b(\Gamma_\mu^a) = - \Gamma_{b\mu}^{a}\circ\pi
\iff
\Gamma_\mu^a = - (\Gamma_{b\mu}^{a}\circ\pi) \cdot u^b + G_\mu^a\circ\pi ,
    \end{equation}
for some functions $\Gamma_{b\mu}^{a},G_\mu^a\colon M\to\field$. Thus
$\hat{\nabla}_{\bar{Z}}\hat{Z}$ is a vertical vector field if and only if the
2\ndash index coefficients $\Gamma_{\mu}^{a}$ in $\{X_I\}$ of the connection
$\Delta^h$ are of the form
    \begin{equation}    \label{4.33-1}
\Gamma_{\mu}^{a}
= - (\Gamma_{b\mu}^{a}\circ\pi)\cdot u^b + G_\mu^{a}\circ\pi .
    \end{equation}
This equality selects the set of \emph{affine connections} among all
connections (see Subsect~\ref{Subsect4.5} below);~%
\footnote{~%
Usually the affine connections are defined on affine
bundles~\cite{K&N-1,Mangiarotti&Sardanashvily}.%
}
in particular, of this type are the linear connections for which $G_\mu^{a}=0$
and $\Gamma_{b\mu}^{a}$ are their 3\ndash index coefficients (see~\eref{4.16}).
For connections with 2\ndash index coefficients~\eref{4.33-1},
equation~\eref{4.32} reduces to
    \begin{equation}    \label{4.34}
\hat{\nabla}_{\bar{Z}} \hat{Z}
=
\bar{Z}^\mu\{ X_\mu(\hat{Z}^a) + \hat{Z}^b(\Gamma_{b\mu}^a\circ\pi) \} X_a
\in\Delta^v .
    \end{equation}

    Now the idea of introduction of a covariant derivative of a section
$Y\in\Sec(E,\pi,M)$ along a vector field $F\in\mathcal{X}(M)$ is to `lower
the operator $\hat{\nabla}$ from $T(E)$ to $T(M)$.

    \begin{Defn}    \label{Defn4.2}
A \emph{covariant derivative} or \emph{covariant derivative operator},
associated to a linear (or affine) connection $\Delta^h$ on a vector bundle
$(E,\pi,M)$, is a mapping
    \begin{equation}    \label{4.35}
    \begin{split}
\nabla & \colon\mathcal{X}(M)\times\Sec^1(E,\pi,M) \to  \Sec^0(E,\pi,M)
\\
\nabla & \colon (F,Y)\mapsto\nabla_FY
    \end{split}
    \end{equation}
such that, for $F\in\mathcal{X}(M)$ and $Y\in\Sec^1(E,\pi,M)$, $\nabla_FY$ is
the unique section of $(E,\pi,M)$ whose vertical lift is
$\hat{\nabla}_{F^h}Y^v$, with $\hat{\nabla}$ defined by~\eref{4.30}
(or~\eref{4.34}), \viz
    \begin{equation}    \label{4.36}
(\nabla_FY)^v := \hat{\nabla}_{F^h} Y^v
    \end{equation}
or
    \begin{equation}    \label{4.36-1}
(\nabla_FY) = v^{-1}\circ\hat{\nabla}_{ (\pi_*|_{\Delta_h})^{-1}(F) }(v(Y)) ,
    \end{equation}
where $F^h\in\Delta^h$ and $Y^v\in\Delta^v$ are respectively the horizontal
and vertical lifts of $F$ and $Y$.
    \end{Defn}

    \begin{Rem}    \label{Rem4.0-1}
Definition~\ref{Defn4.2} and the rest of this subsection are valid also for
affine connections for which~\eref{4.33-1} holds, not only for the linear
ones. For some details, see Subsect.~\ref{Subsect4.5}.
    \end{Rem}

    \begin{Prop}    \label{Prop4.6}
Let $\{E_a\}$ be a frame in $E$ and $\{x^\mu\}$ local coordinates on $M$. If
$Y=Y^aE_a\in\Sec^1(E,\pi,M)$ and
$F=F^\mu\frac{\pd}{\pd x^\mu}\in\mathcal{X}(M)$, then we have the explicit
local expression
    \begin{equation}    \label{4.37}
\nabla_FY
= F^\mu\Bigl( \frac{\pd Y^a}{\pd x^\mu} +\Gamma_{b\mu}^{a}Y^b \Bigr) E_a .
    \end{equation}
    \end{Prop}

    \begin{Proof}
Apply~\eref{4.36},~\eref{4.6-4},~\eref{4.6-5},~\eref{4.34}, and~\eref{4.2}.
    \end{Proof}

    \begin{Prop}    \label{Prop4.2}
Let $\Delta^h$ be a linear connection on $(E,\pi,M)$ and $\Psf$ be the
generated by it parallel transport. Let $x\in M$,
$\gamma\colon[\sigma,\tau]\to M$, $\gamma(t_0)=x$ for some
$t_0\in[\sigma,\tau]$, and $\dot{\gamma}(t_0)=F_x$, \ie $\gamma$ to be the
integral path of $F\in\Fields{X}(M)$ through $x$. Then
    \begin{equation}    \label{4.38}
(\nabla_F Y)|_x
=
\lim_{s\to t_0}
\frac{P_{s\to t_0}^{\gamma} (Y_{\gamma(s)}) - Y_{\gamma(t_0)} }
     {s-t_0}
=
\lim_{\varepsilon\to 0}
\frac{P_{t_0+\varepsilon\to t_0}^{\gamma} (Y_{\gamma(t_0+\varepsilon)})
                            - Y_{\gamma(t_0)} }
     {\varepsilon} ,
    \end{equation}
where  $Y\in\Sec^1(E,\pi,M)$ and
    \begin{equation}    \label{4.39}
P_{s\to t}^{\gamma}
:=
    \begin{cases}
\Psf^{\gamma|[s,t]} & \text{\textup{for }} s\le t \\
\bigl(\Psf^{\gamma|[t,s]}\bigr)^{-1} & \text{\textup{for }} s\ge t .
    \end{cases}
    \end{equation}
    \end{Prop}

    \begin{Proof}
Use definition~\ref{Defn3.2} and apply the parallel transport
equation~\eref{4.18} with initial value
$\bar{\gamma}_{Y_{\gamma(s)}}(s)=Y_{\gamma(s)}$ at the point
$t=s\in[\sigma,\tau]$.
    \end{Proof}

    By proposition~\ref{Prop4.2}, the equation~\eref{4.38} can be used as
an equivalent definition of a covariant derivative associated with a linear
connection.

    \begin{Prop}    \label{Prop4.3}
Let $F,G\in\mathcal{X}(M)$, $Y,Z\in\Sec^1(E,\pi,M)$, and $f\colon M\to\field$
be a $C^1$ function. Then:
    \begin{subequations}    \label{4.40}
    \begin{align}   \label{4.40a}
\nabla_{F+G}Y &= \nabla_{F}Y + \nabla_{G}Y
\\          \label{4.40b}
\nabla_{fF}Y &= f\nabla_{F}Y
\\          \label{4.40c}
\nabla_{F}(Y+Z) &= \nabla_{F}Y + \nabla_{F}Z
\\          \label{4.40d}
\nabla_{F}(fY) &= F(f)\cdot Y + f\cdot \nabla_{F}Y .
    \end{align}
    \end{subequations}
    \end{Prop}

    \begin{Proof}
Apply~\eref{4.37}.
    \end{Proof}

    \begin{Prop}    \label{Prop4.4}
If a mapping~\eref{4.35} satisfies~\eref{4.40}, there exists a unique linear
connection $\Delta^h$, the assigned to which covariant derivative is exactly
$\nabla.$
    \end{Prop}

    \begin{Proof}
Define local functions $\Gamma_{b\mu}^{a}$ on $M$, called \emph{components}
of $\nabla$, by the decomposition
    \begin{equation}    \label{4.41}
\nabla_{\frac{\pd}{\pd x^\mu}} E_b =: \Gamma_{b\mu}^{a} E_a .
    \end{equation}
A simple verification proves that they transform according to~\eref{4.25} and
hence the quantities~\eref{4.16} transform by~\eref{3.22}.
Proposition~\ref{Prop3.1} ensures the existence of a unique linear connection
whose 2\ndash index (3\ndash index) coefficients are $\Gamma_{\mu}^{a}$
($\Gamma_{b\mu}^{a}$). Thus the covariant derivative of $Y\in\Sec(E,\pi,M)$
relative to $F\in\mathcal{X}(M)$ is given by the r.h.s.\ of~\eref{4.37}. On
another hand,~\eref{4.40} entail~\eref{4.37}, with $\Gamma_{b\mu}^{a}$
defined by~\eref{4.41}, so that $\nabla$ is exactly the covariant derivative
operator assigned to the connection with 3\ndash index coefficients
$\Gamma_{b\mu}^{a}$.
    \end{Proof}

    Consequently, equations~\eref{4.40} and~\eref{4.41} provide a third
equivalent definition of a covariant derivative (covariant derivative
operator). Moreover, since proposition~\ref{Prop4.4} establishes a bijective
correspondence between linear connections and operators~\eref{4.35}
satisfying~\eref{4.40}, quite often such operators are called linear
connections.~%
\footnote{~%
See also~\cite[sections~2.15 and~2.52]{Poor}.%
}
As it is clear from the proof of proposition~\ref{Prop4.4}, the bijection
between linear connection and covariant derivative operators is locally given
by the coincidence of their (3\ndash index) coefficients and components,
respectively.

    \begin{Exrc}    \label{Exrc4.1}
    A  $C^1$ section $\omega=\omega_a E^a$ of the bundle dual to
$(E,\pi,M)$ can be differentiated covariantly similarly to the sections of
$(E,\pi,M)$. Show that the corresponding operator, say $\nabla^*$, can
equivalently be defined by (the `Leibnitz rule')
    \begin{equation}    \label{4.42}
(\nabla^*_F\omega)(Y) = F(\omega(Y)) - \omega(\nabla_FY)
    \end{equation}
and locally is valid the equation
    \begin{equation}    \label{4.43}
\nabla^*_F\omega
= F^\mu
\Bigl( \frac{\pd\omega_a}{\pd x^\mu} - \Gamma_{a\mu}^{b} \omega_b \Bigr) E^a.
    \end{equation}
    \end{Exrc}

    Equipped with the covariant derivative $\nabla$ assigned to a $C^1$
linear connection $\Delta^h$, we define the \emph{curvature operator} of
$\Delta^h$ (or $\nabla$) by
    \begin{align}   \label{4.44}
    \begin{split}
R & \colon \mathcal{X}(M)\times\mathcal{X}(M)\to \End(\Sec(E,\pi,M))   \\
R & \colon (F,G)\mapsto R(F,G)
:=
\nabla_F\circ\nabla_G - \nabla_G\circ\nabla_F - \nabla_{[F,G]_{\_}} ,
    \end{split}
    \end{align}
with $\End(\dots)$ denoting the set of endomorphisms of $(\dots)$.

    \begin{Exrc}    \label{Exrc4.2}
    Prove that locally
    \begin{equation}    \label{4.45}
(R(F,G))(Y) = (R_{b\mu\nu}^{a} Y^b F^\mu G^\nu) E_a ,
    \end{equation}
where the functions $R_{b\mu\nu}^{a}\colon M\to\field$, called the
\emph{components} of the curvature operator $R$ in the pair of frames
$\bigl(\bigl\{\frac{\pd}{\pd x^\mu}\bigr\},\{E_a\}\bigr)$,
are defined by
    \begin{equation}    \label{4.46}
R\Bigl( \frac{\pd}{\pd x^\mu},\frac{\pd}{\pd x^\nu} \Bigr)(E_b)
=:R_{b\mu\nu}^{a} E_a
    \end{equation}
and are explicitly expressed through the coefficients of $\nabla$(= 3\ndash
index coefficients of $\Delta^h$) via~\eref{4.27}.
    \end{Exrc}

    A linear connection or covariant derivative operator is called
\emph{flat} or \emph{curvature free} if
    \begin{equation}    \label{4.47}
R=0 \qquad ( \iff R_{b\mu\nu}^{a}=0 ).
    \end{equation}
Obviously, the flatness of $\Delta^h$ or $\nabla$ is a necessary and
sufficient condition for the (local) integrability of the horizontal
distribution $\Delta^h\colon p\mapsto\Delta_p^h\subseteq T_p(E)$, $p\in E$
(see~\eref{3.14b} and~\eref{4.26}).

    \begin{Thm} \label{Thm4.2}
Let $Y$ be a $C^1$ section of a vector bundle $(E,\pi,M)$ endowed with a
linear connection $\Delta^h$. The following three conditions are
equivalent:
    \\\indent\textup{(i)}
    $Y$ is covariantly constant, viz., if $F\in\mathcal{X}(M)$, then
    \begin{equation}    \label{4.48}
\nabla_FY = 0.
    \end{equation}
    \\\indent\textup{(ii)}
    $Y$ is a solution of $\Delta^h$, \ie
    \begin{equation}    \label{4.49}
\Image Y_* \subset \Delta^h
\quad  (\iff Y_*|_x(T_x(M))\subseteq\Delta_{Y_x}^h \text{ for } x\in M ) .
    \end{equation}
    \\\indent\textup{(iii)}
    $Y$ is parallelly transported along any path
$\gamma\colon[\sigma,\tau]\to M$,
    \begin{equation}    \label{4.50}
\Psf^\gamma(Y_{\gamma(\sigma)}) = Y_{\gamma(\tau)} .
    \end{equation}
    \end{Thm}

    \begin{Proof}
Since $Y=u^a(Y)e_a$, $\pi\circ Y=\id_M$, and
$\omega^a=\od u^a - \Gamma_\mu^a\od u^\mu$, we have for $x\in M$:
    \begin{multline*}
\omega^a\circ Y \Bigl(\frac{\pd}{\pd x^\mu}\Bigr)
=
\omega^a
\Bigl(
\frac{\pd(u^\nu\circ Y)}{\pd x^\mu}\Big|_x \frac{\pd}{\pd u^\nu} \Big|_{Y_x}
+
\frac{\pd(u^a\circ Y)}{\pd x^\mu}\Big|_x \frac{\pd}{\pd u^a} \Big|_{Y_x}
\Bigr)
\\
=
- \frac{\pd(x^\nu\circ\pi\circ Y)}{\pd x^\mu}\Big|_x  \Gamma_\nu^a(Y_x)
+ \frac{\pd Y^a}{\pd x^\mu}\Big|_x
=\Bigl(
 \frac{\pd Y^a}{\pd x^\mu}
- \Gamma_\mu^a \circ Y
\Bigr)(x).
    \end{multline*}
The equivalence of (i) and (ii) follows from here,~\eref{4.16},~\eref{4.37},
and that $\Delta^h$ annihilates the 1\ndash forms $\omega^a$,
$\omega^a(Z) \iff Z\in\Delta^h$.

    If we rewrite the parallel transport equation~\eref{4.18} as
(see~\eref{4.37})
    \begin{equation}    \label{4.51}
(\nabla_{\dot{\gamma}(t)} \bar{\gamma})|_{\gamma(t)} = 0 ,
    \end{equation}
the equivalence of (i) and (iii) follows from definition~\ref{Defn3.2} of a
parallel transport and the arbitrariness of $\gamma$ in~\eref{4.51}.
    \end{Proof}

    \begin{Exrc}    \label{Exrc4.3}
    Formulate and prove a theorem dual to theorem~\ref{Thm4.2}; \eg a section
$\varphi=\varphi_au^a$ of the bundle dual to $(E,\pi,M)$ is a first integral of
$\Delta^h$, \ie $\Ker\varphi_*\supseteq\Delta^h$
($\iff \varphi_*|_p(\Delta_p^h)=0_{\varphi(p)}\in T_{\varphi(p)}(\field)$ for
$p\in E$), if and only if
    \begin{equation}    \label{4.52}
\nabla^*\varphi = 0 .
    \end{equation}
    \end{Exrc}

    \begin{Prop}[\normalfont
			cf.~\protect{\cite[p.~32]{Rahula}} \bfseries]

    \label{Prop4.5}
Let a linear connection $\Delta^h$ on a vector bundle be given and
$\Gamma_{b\mu}^{a}$ be its (3\ndash index) coefficients. The following
conditions are (locally) equivalent:
    \\\indent\textup{(a)}
    $\Delta^h$ is integrable.
    \\\indent\textup{(b)}
    $\Delta^h$ is flat.
    \\\indent\textup{(c)}
    There exists a solution of the system of partial
differential equations
    \begin{equation}    \label{4.53}
\frac{\pd U^a}{\pd x^\mu} +\Gamma_{b\mu}^{a} U^b = 0
    \end{equation}
relative to $U^a$ and the solution of~\eref{4.53} satisfying
$U^a|_{x=x_0}=U_0^a$ is $U^a=B_b^a U_0^b$, where $B=[B_b^a]$ is the
fundamental solution of~\eref{4.53}, \viz
    \begin{equation}    \label{4.54}
\frac{\pd B_b^a}{\pd x^\mu} +\Gamma_{c\mu}^{a} B_b^c = 0
\qquad B_b^a|_{x=x_0} = \delta_b^a .
    \end{equation}
    \\\indent\textup{(d)}
 There is an integrating matrix $B^{-1}$ for the 1-forms $\omega^a$, that is
$(B^{-1}\circ\pi)_b^a \omega^b=\od f^a$, where the functions
$f^a\colon E\to\field$ are first integrals of $\Delta^h$, \ie $\Ker
f^a\supset\Delta^h$. %
    \\\indent\textup{(e)}
    The coefficients of $\Delta^h$ have the form
    \begin{equation}    \label{4.55}
\Gamma_\mu
:=[\Gamma_{b\mu}^{a}]
 = B\cdot \frac{\pd B^{-1}}{\pd x^\mu}
 = -\frac{\pd B}{\pd x^\mu}\cdot B^{-1}
    \end{equation}
for some matrix-valued function $B$ on $M$.
    \end{Prop}

    \begin{Proof}
    (a)$\iff$(b):
    See~\eref{4.47} and the comment after it.

    (c)$\iff$(e):
    The matrix form of the equation in~\eref{4.54}, \ie
    \begin{equation}
    \tag{\protect\ref{4.54}$^\prime$}   \label{4.54'}
\frac{\pd B}{\pd x^\mu} + \Gamma_\mu\cdot B = 0 ,
    \end{equation}
is tantamount to~\eref{4.55}.

    (b)$\iff$(c):
    The flatness of $\Delta^h$, \ie $R_{\mu\nu}=0$ (see~\eref{4.27'}), is
the integrability condition for~\eref{4.54'} as an equation relative to $B$
-- see~\cite[lemma~2.1]{bp-NF-D+EP}.

    (c)$\iff$(d):
    Since~\eref{3.23} and the first equality in~\eref{2.1c} entail
    \begin{equation}    \label{4.56}
\Lied_{X_\mu}(\varphi_a\omega^a)
=
- \varphi_a R_{\mu\nu}^{a} \omega^\nu
+ \{ X_\mu(\varphi_a)
           - \lindex[\Gamma]{}{\circ}{}_{a\mu}^{b} \varphi_b \} \omega^a ,
    \end{equation}
we have (see also~\eref{4.22}) for a flat linear connection:
    \begin{multline*}
\Lied_{X_\mu} ((B^{-1})_b^a \omega^b)
= \Bigl\{\Bigl(\frac{\pd B^{-1}}{\pd x^\mu} - B^{-1}\cdot\Gamma_\mu
     \Bigr)\circ\pi\Bigr\}_b^a \; \omega^b
\\
= \Bigl\{\Bigl[ - B^{-1} \cdot
\Bigl( \frac{\pd B}{\pd x^\mu} + \Gamma_\mu \cdot B \Bigr) B^{-1}
     \Bigr]\circ\pi\Bigr\}_b^a \; \omega^b.
    \end{multline*}
Thus~\eref{4.54}, which entails~(c), is equivalent to
 $\Lied_{X_\mu}((B^{-1}\circ\pi)_b^a\:\omega^b)=0$, which is equivalent to
 $\od((B^{-1}\circ\pi)_b^a\;\omega^b)=0$, due to
$\omega^a(X_\mu)=\delta_\mu^a=0$ and the second equality
in~\eref{2.1a} (applied, e.g., for $Y=X_\nu$). Now the
Poincar\'e's lemma (see~\cite[sec.~6.3]{Nash&Sen}
or~\cite[pp.~21,~106]{Gockeler&Schucker}) tells us that locally
(on a contractible region in $E$) there are functions $f^a$ on $E$
such that the last equality is tantamount to
 $\od f^a = (B^{-1}\circ\pi)_b^a\; \omega^b$.

    It remains to be proved that $f^a\colon E\to\field$ are first
integrals of $\Delta^h$, \ie $\Ker f^a\supset\Delta^h$ which means
 $(f^a)_*|_p(\Delta_p^h) = 0$, $p\in E$, or
 $(f^a)_*|_p(X_\mu)=0$ as $\Delta^h$
is spanned by $\{X_\mu\}$. Using the global chart $(\field,\id_{\field})$ on
$\field$, which induces the one\ndash vector frame $\bigl\{\frac{\pd}{\pd
r}\bigr\}$ for $r\in\field$ on $\field$, we have (see~\eref{3.19})
\[
(f^a)_*|_p(X_\mu)
= (f^a)_*|_p\Bigl(\frac{\pd}{\pd u^\mu}
        + \Gamma_\mu^b\frac{\pd}{\pd u^b}\Bigr)\Big|_p
= \Bigl( \frac{\pd f^a}{\pd u^\mu}\Big|_p
          + \Gamma_\mu^b(p) \frac{\pd f^a}{\pd u^b}\Bigr|_p \Bigr)
        \frac{\od}{\od r}\Big|_{f^a(p)}
\equiv 0
\]
as
\(
\od f^a
= (B^{-1}\circ\pi)_b^a\; \omega^a
= (B^{-1}\circ\pi)_b^a (\od u^a - \Gamma_\mu^b\circ \od u^\mu)
\equiv
\frac{\pd f^a}{\pd u^b} \od u^b + \frac{\pd f^a}{\pd u^\mu} \od u^\mu
= \od f^a.
\)
    \end{Proof}




\subsection{Affine connections}
    \label{Subsect4.5}

	In Subsect.~\ref{Subsect4.4}, we met a class of connections on a
vector bundle whose local 2\ndash index coefficients have the form
(see~\eref{4.33-1})
    \begin{equation}    \label{4.61}
\Gamma_{\mu}^{a}
= - (\Gamma_{b\mu}^{a}\circ\pi)\cdot u^b + G_\mu^{a}\circ\pi .
    \end{equation}
in the frame $\{X_I\}$ adapted to vector bundle coordinates $\{u^I\}$. From
$\pd_b\Gamma_\mu^a=-\Gamma_{b\mu}^{a}$ and~\eref{3.22}, one derives that the
functions $\Gamma_{b\mu}^{a}$ in~\eref{4.61} transform according
to~\eref{4.25}, \viz
    \begin{equation}    \label{4.62}
\Gamma_{b\mu}^{a} \mapsto
\tilde{\Gamma}_{b\mu}^{a}
=
B_\mu^\nu
\Bigl( B_d^a \Gamma_{c\nu}^{d} - \frac{\pd B_c^a}{\pd x^\nu} \Bigr)
(B^{-1})_b^c .
    \end{equation}
when the vector bundle coordinates or adapted frames undergo the
change~\eref{4.23} or~\eref{4.24}, respectively. Thus, combining~\eref{3.22},
\eref{4.62} and~\eref{4.61}, we see that~\eref{4.23} or~\eref{4.24} implies the
transition
    \begin{equation}    \label{4.63}
G_\mu^a \mapsto \tilde{G}_\mu^a = B_b^a G_\nu^b B_\mu^\nu .
    \end{equation}
Consequently, the functions $\Gamma_{b\mu}^{a}$ in~\eref{4.61} are 3-index
coefficients of a linear connection, while $G_\mu^a$ in it are the components
of a linear mapping
$G \colon \mathcal{X}(M)\to \End(\Sec((E,\pi,M)^*))$
such that $G \colon F\mapsto G(F) \colon \omega \mapsto (G(F))(\omega)$, for
$F\in\mathcal{X}(M)$ and a section $\omega$ of the bundle $(E,\pi,M)^*$ dual to
$(E,\pi,M)$, and
$\bigl( G\bigl( \frac{\pd }{\pd x^\mu} \bigr) \bigr) (E^a) = G_\mu^a$.
The invariant description of the connections with local 2\ndash
index coefficients of the type~\eref{4.61} is as follows.

    \begin{Defn}    \label{Defn4.3}
A connection on a vector bundle is termed \emph{affine connection}
if the assigned to it parallel transport $\Psf$ is an affine mapping along all
paths $\gamma \colon [\sigma,\tau]\to M$ in the base space, \ie
    \begin{subequations}    \label{4.64}
    \begin{align}    \label{4.64a}
\Psf^\gamma(\rho X) &= \rho\Psf^\gamma(X) + (1-\rho) \Psf^\gamma(\bs{0})
\\			    \label{4.64b}
\Psf^\gamma(X+Y)
&= \Psf^\gamma(X) + \Psf^\gamma(Y) -  \Psf^\gamma(\bs{0}),
    \end{align}
    \end{subequations}
where $\rho\in\field$, $X,Y\in\pi^{-1}(\gamma(\sigma))$, and $\bs{0}$ is the
zero vector in the fibre $\pi^{-1}(\gamma(\sigma))$, which is a \field\ndash
vector space.
    \end{Defn}

    \begin{Thm}    \label{Thm4.3}
Let $(E,\pi,M)$ be a vector bundle, $\{u^I\}$ be vector bundle coordinates over
an open set $U\subseteq E$, and $\Delta^h$ be a connection on it with 2\ndash
index coefficients $\Gamma_\mu^a$ in the frame $\{X_I\}$ adapted to $\{u^I\}$.
The connection $\Delta^h$ is an affine connection if and only if
equation~\eref{4.61} holds for some functions
$\Gamma_{b\mu}^{a},G_\mu^a \colon \pi(U)\to \field$.
    \end{Thm}

    \begin{Proof}[Proof (cf.\ the proof of theorem~\protect{\ref{Thm4.1}}).]
Take a $C^1$ path $\gamma\colon[\sigma,\tau]\to\pi(U)$ and consider the
parallel transport equation~\eref{3.26''a}, \viz
    \begin{equation}    \label{4.65}
\frac{\od\bar{\gamma}_p^a(t)}{\od t}
=
\Gamma_\mu^a(\bar{\gamma}_p(t)) \dot\gamma^\mu(t),
    \end{equation}
where $\bar{\gamma}_p\colon[\sigma,\tau]\to U$ is the horizontal
lift of $\gamma$ through $p\in\pi^{-1}(\gamma(\sigma))$,
$\bar{\gamma}^a:=u^a\circ\bar{\gamma}$, and \( \dot\gamma^\mu(t)
=\frac{\od(x^\mu\circ\gamma(t))}{\od t}
=\frac{\od(u^\mu\circ\bar{\gamma}(t))}{\od t} \) as
$u^\mu=x^\mu\circ\pi$ for some coordinates $\{x^\mu\}$ on
$\pi(U)$.

    SUFFICIENCY.
    If~\eref{4.61} holds,~\eref{4.65} is transformed into
    \begin{equation}    \label{4.66}
\frac{\od\bar{\gamma}_p^a(t)}{\od t}
=
- \Gamma_{b\mu}^a(\gamma(t)) \bar{\gamma}_p^b(t) \dot\gamma^\mu(t)
+ G_\mu^a(\gamma(t)) \dot\gamma^\mu(t) ,
    \end{equation}
which is a system of $r$ linear inhomogeneous first order ordinary differential
equations for the $r$ functions
$\bar{\gamma}_p^{n+1},\dots,\bar{\gamma}_p^{n+r}$. According to the general
theorems of existence and uniqueness of the solutions of such
systems~\cite{Hartman}, it has a unique solution
    \begin{equation}    \label{4.67}
\bar{\gamma}_p^{a}(t) = Y_b^a(t) p^b + y^a(t)
    \end{equation}
satisfying the initial condition $\bar{\gamma}_p^{a}(\sigma)=u^a(p)=:p^a$,
where $Y=[Y_b^a]$ is the fundamental solution of~\eref{4.18} (see~\eref{4.20})
and $y^a(t)$ is the solution of~\eref{4.66} with $y^a(t)$ for
$\bar{\gamma}_p^a(t)$ satisfying the initial condition $y^a(\sigma)=0$.
The affinity of~\eref{3.9-4} in $p$, \ie~\eref{4.64}, follows from~\eref{4.67}
for $t=\tau$.

    NECESSITY.
    Suppose~\eref{3.9-4} is affine in $p$ for all paths
$\gamma\colon[\sigma,\tau]\to\pi(U)$. Then
$\bar{\gamma}_p(t):=\Psf^{\gamma|[\sigma,t]}(p)$ is the horizontal lift of
$\gamma|[\sigma,t]$ through $p$ and (cf.~\eref{4.67})
 $\bar{\gamma}_p^{a}(t)=A_b^a(\gamma(t))p^b + A^a(\gamma(t))$
for some $C^1$ functions $A_b^a,A^a\colon\pi(U)\to\field$. The substitution of
this equation in~\eref{4.65} results into
\[
\frac{\pd A_b^a(x)}{\pd x^\mu} \Big|_{x=\gamma(t) = \pi(\bar{\gamma}_p(t))}
		\cdot \dot{\gamma}^\mu  p^b
+
  \frac{\pd A^a(x)}{\pd x^\mu} \Big|_{x=\gamma(t) = \pi(\bar{\gamma}_p(t))}
		\cdot \dot{\gamma}^\mu(t) .
=
   \Gamma_{\mu}^{a}(\bar{\gamma}_p(t)) \dot\gamma^\mu(t)
\]
Since $\gamma\colon[\sigma,\tau]\to M$, we get equation~\eref{4.61} from here,
for $t=\sigma$, with
 $\Gamma_{b\mu}^{a}(x) = - \frac{\pd A_b^a(x)}{\pd x^\mu}$ and
 $G_{\mu}^{a}(x) =  \frac{\pd A^a(x)}{\pd x^\mu}$
for $x\in\pi(U)$.
    \end{Proof}

    \begin{Prop}    \label{Prop4.7}
There is a bijective mapping $\alpha$ between the sets of affine connections
and of pairs of a linear connection and a linear mapping $G \colon
\mathcal{X}(M)\to \End(\Sec((E,\pi,M)^*))$.
    \end{Prop}

    \begin{Proof}
If $\lindex{}{A\mspace{-2mu}}\Delta^h$ is an affine connection with 2\ndash
index coefficients give by~\eref{4.61} (see theorem~\ref{Thm4.3}), then (see the
discussion after equation~\eref{4.61}) to it corresponds the pair
\(
\alpha(\lindex{}{A\mspace{-2mu}}\Delta^h)
:=
(\lindex{}{L\mspace{-2mu}}\Delta^h,G)
\)
of a linear connection, with 3\ndash index coefficients $\Gamma_{b\mu}^{a}$ and
linear mapping $G \colon \mathcal{X}(M)\to \End(\Sec((E,\pi,M)^*))$, with
components $G_\mu^a$. Conversely, to a pair
$(\lindex{}{L\mspace{-2mu}}\Delta^h,G)$, locally described via the 3\ndash
index coefficients $\Gamma_{b\mu}^{a}$  of $\lindex{}{L\mspace{-2mu}}\Delta^h$
and components $G_\mu^a$ of $G$, there corresponds an affine connection
\(
\lindex{}{A\mspace{-2mu}}\Delta^h
=
\alpha^{-1}(\lindex{}{L\mspace{-2mu}}\Delta^h,G)
\)
with 2\ndash index coefficients given by~\eref{4.61}.
    \end{Proof}

	In Subsect.~\ref{Subsect4.4}, it was demonstrated that covariant
derivatives can be introduced for affine connections, not only for linear ones.

    \begin{Prop}    \label{Prop4.8}
The covariant derivative for an affine connection
$\lindex{}{A\mspace{-2mu}}\Delta^h$ coincides with the one for the linear
connection $\lindex{}{L\mspace{-2mu}}\Delta^h$ given via
\(
\alpha(\lindex{}{A\mspace{-2mu}}\Delta^h)
=
(\lindex{}{L\mspace{-2mu}}\Delta^h, G)
\)
with $\alpha$ defined in the proof of proposition~\ref{Prop4.7}.
    \end{Prop}

    \begin{Proof}
Apply~\eref{4.32}--\eref{4.37}.
    \end{Proof}

If a linear connection $\lindex{}{L\mspace{-2mu}}\Delta^h$ and an affine one
$\lindex{}{A\mspace{-2mu}}\Delta^h$ are connected by
\(
\alpha(\lindex{}{A\mspace{-2mu}}\Delta^h)
=
(\lindex{}{L\mspace{-2mu}}\Delta^h,G)
\)
for some $G$, then some of their characteristics coincide; \eg such are
their fibre coefficients (see~\eref{3.24b}, \eref{4.61} and~\eref{4.16}) and
all quantities expressed via the corresponding to them (identical) covariant
derivatives. However, quantities, containing (depending on) partial derivatives
relative to the basic coordinates $\{u^\mu\}$, are generally different for
those connections. For instance, if
 $\lindex{}{A}R_{\mu\nu}^a$ and $\lindex{}{L}R_{\mu\nu}^a$ are the fibre
components of the curvatures  of
 $\lindex{}{A\mspace{-2mu}}\Delta^h$ and $\lindex{}{L\mspace{-2mu}}\Delta^h$,
respectively, then, by~\eref{3.24a} and~\eref{4.61}, we have
    \begin{align}    \label{4.68}
\lindex{}{A}R_{\mu\nu}^a
& =
  - (\lindex{}{L}R_{b\mu\nu}^a\circ \pi) \cdot u^b
  - T_{\mu\nu}^a \circ \pi
\\		    \label{4.69}
\lindex{}{L}R_{\mu\nu}^a
& =
  - (\lindex{}{L}R_{b\mu\nu}^a\circ \pi) \cdot u^b
    \end{align}
where (see~\eref{4.27})
    \begin{align}    \label{4.70}
\lindex{}{L}R_{b\mu\nu}^{a}
&:=
\frac{\pd}{\pd x^\mu} (\Gamma_{b\nu}^{a}) -
\frac{\pd}{\pd x^\nu} (\Gamma_{b\mu}^{a}) -
\Gamma_{b\mu}^{c}\Gamma_{c\nu}^{a} +
\Gamma_{b\nu}^{c}\Gamma_{c\mu}^{a} ,
\\		    \label{4.71}
T_{\mu\nu}^{a}
&:=
- \frac{\pd}{\pd x^\mu} (G_{\nu}^{a}) +
  \frac{\pd}{\pd x^\nu} (G_{\mu}^{a}) +
 \Gamma_{c\nu}^{a} G_{\mu}^{c} -
 \Gamma_{c\mu}^{a} G_{\nu}^{c}
    \end{align}
and the functions $T_{\mu\nu}^{a}$ have a sense of components of the torsion of
$\lindex{}{L\mspace{-2mu}}\Delta^h$ relative to
$G$~\cite[pp.~42,~46]{Mangiarotti&Sardanashvily}.

	Thus, in general, the affine connections and linear connections are
essentially different. However, they imply identical theories of covariant
derivatives.

	If, for some reason, the linear mapping $G$ is fixed, then the set of
linear connections $\{\lindex{}{L\mspace{-2mu}}\Delta^h\}$ can be identified
with the subset $\{\alpha^{-1}(\lindex{}{L\mspace{-2mu}}\Delta^h,G)\}$ of the
set of affine connections $\{\lindex{}{A\mspace{-2mu}}\Delta^h\}$. We shall
exemplify this situation on the tangent bundle $(T(M),\pi_T,M)$ over a manifold
$M$. Using the base indices $\mu,\nu,\dots$ for the fibre ones $a,b,\dots$
according to the rule $a \mapsto \mu=a-\dim M$ (see Subsect.~\ref{Subsect4.2}),
we rewrite~\eref{4.61} as
    \begin{equation}    \label{4.72}
\Gamma_\nu^\mu
=
- (\Gamma_{\lambda\nu}^{\mu} \circ \pi_T) \cdot u_1^\lambda
+ G_\nu^\mu\circ\pi_T .
    \end{equation}
Now the affine  connections on $(T(M),\pi_T,M)$ are the
\emph{generalized affine connections on} $M$~\cite[ch.~III, \S~3]{K&N-1}. The
choice of $G$ via
    \begin{equation}    \label{4.73}
G_\nu^\mu \colon M\to \delta_\nu^\mu ,
    \end{equation}
which corresponds to the identical transformation of the spaces tangent to $M$,
singles out the set of
\emph{affine connections on} $M$
-- see ~\cite[ch.~III, \S~3]{K&N-1} or \cite[pp.~103--105]{Poor} --
(known also as \emph{Cartan connections on}
$M$~\cite[p.~46]{Mangiarotti&Sardanashvily}) whose 2\ndash index coefficients
have the form (see~\eref{4.72}, \eref{4.6a} and~\eref{4.73})
    \begin{equation}    \label{4.74}
\Gamma_\nu^\mu
=
- (\Gamma_{\lambda\nu}^{\mu} \circ \pi_T) \cdot \od x^\lambda
+ \delta_\nu^\mu .
    \end{equation}

	Combining this with proposition~\ref{Prop4.7}, we derive

    \begin{Prop}%
[\normalfont cf.~\protect{\cite[ch.~III, \S~3, theorem~3.3]{K&N-1}}\bfseries]
			\label{Prop4.9}
There is a bijective correspondence between the sets of linear connections and
of affine ones on a manifold.
    \end{Prop}

	Often the terms ``linear connection'' and ``affine connection'' on a
manifold are used as synonyms, due to the last result.



\section
[Morphisms of bundles with connection]
{Morphisms of bundles with connection\footnotesize~%
\footnote{~%
Some ideas in this section are borrowed
from~\protect{\cite[ch.~I, \S~6]{Rahula}}%
}
}
\label{Sect5}

    A \emph{morphism} between two bundles $(E,\pi,M)$ and $(E',\pi',M')$
is a pair of mappings $(F,f)$ such that $F\colon E\to E'$, $f\colon M\to M'$,
and $\pi'\circ F=f\circ\pi$. If $(U,u)$ and $(U',u')$ are charts in $E$ and
$E'$, respectively, and $F(U)\subseteq U'$, we have the following local
representation of $(F,f)$
    \begin{subequations}    \label{5.1}
    \begin{align}   \label{5.1a}
\bar{F} & =u'\circ F\circ u^{-1} \colon u(U)\to u'(U')
\\              \label{5.1b}
\bar{f} & =x'\circ f\circ x^{-1} \colon x(V)\to x'(V') ,
    \end{align}
    \end{subequations}
where $(V,x)$ and $(V',x')$ are local charts respectively on $M$ and $M'$.
Further, we assume that $U'=F(U)$ and that the charts in the base and bundle
spaces respect the fibre structure, $V=\pi(U)$ and $V'=\pi'(U')$ so that
$V'=f(V)$, and that the basic coordinates are $u^\mu=x^\mu\circ\pi$ and
$u^{\prime\mu'}=x^{\prime\mu'}\circ\pi'$. Here and henceforth the quantities
referring to $(E',\pi',M')$ will inherit the same notation as the similar
ones with respect to $(E,\pi,M)$ with exception of the prime symbol added to
the latter ones; in particular, the primed indices
$\lambda',\mu',\nu',\dots$ and $a',b',c',\dots$ run respectively over the
ranges $1,\dots,n'=\dim M'$ and $n'+1,\dots,n'+r'=\dim E'$ with $r'$ being
the fibre dimension of $(E',\pi',M')$, \ie $r'=\dim((\pi')^{-1}(p')$ for
$p'\in M'$.

    Using the local coordinates $\{x^\mu\}$ on $M$ and
$\{u^\mu=x^\mu\circ\pi,u^a\} $ on $E$, we rewrite~\eref{5.1} as
(cf.~\eref{3.1})
    \begin{subequations}    \label{5.1'}
    \begin{align}   \label{5.1'a}
    \tag{\protect\ref{5.1}$^\prime$a}
\bar{F}^{{I}'}&=u^{\prime{I}'}\circ F\circ u^{-1} \colon u(U)\to \field
\\              \label{5.1'b}
    \tag{\protect\ref{5.1}$^\prime$b}
\bar{f}^{\mu'}&=x^{\prime\mu'}\circ f\circ x^{-1} \colon x(\pi(U))\to \field ,
    \end{align}
    \end{subequations}
\ie one can simply write
$u^{\prime{I}'}=\bar{F}^{{I}'}(u^1,\dots,u^{n+r)}$ and
$x^{\prime\mu'}=\bar{f}^{\mu'}(x^1,\dots,x^{n})$.
However, in what follows, the mappings
    \begin{subequations}    \label{5.2}
    \begin{align}   \label{5.2a}
F^{\mu'}
&:= u^{\prime\mu'}\circ F
 = x^{\prime\mu'}\circ\pi\circ F
 = x^{\prime\mu'}\circ f \circ\pi \colon U\to \field
\quad
F^{a'} := u^{\prime a'}\circ F \colon U\to\field
\\              \label{5.2b}
f^{\mu'}
&:= x^{\prime\mu'}\circ f \colon \pi(U)\to\field
    \end{align}
    \end{subequations}
will be employed. The reason is that the derivatives
\[
F_{,{J}}^{{I}'}
:= \frac{\pd}{\pd u^{J}}  F^{{I}'}
    \colon p\mapsto \frac{\pd}{\pd u^{J}}\Big|_p  F^{{I}'}
=\frac{\pd(F^{{I}'}\circ u^{-1})}{\pd (u^{J}\circ u^{-1})}\Big|_{u(p)}
=\frac{\pd\bar{F}^{{I}'}}{\pd (u^{J}\circ u^{-1})}\Big|_{u(p)}
\qquad p\in U
\]
(note, $\{u^{J}\circ u^{-1}\}$ are Cartesian coordinates on
$u(U)\subseteq\field^{n+r}$) are the elements of the matrix of the tangent
mapping $F_*\colon T(E)\to T(E')$ in the charts $(U,u)$ and $(U',u')$.
Indeed, since this matrix, known as the Jacobi matrix of $F$, is defined
by~\cite[sec.~1.23(a)]{Warner}
    \begin{equation}    \label{5.3}
F_*(\pd_{J}|_p) = F_{J}^{{I}'}|_p (\pd'_{{I}'}|_{F(p)})
\qquad p\in U ,
    \end{equation}
we have
    \begin{equation}    \label{5.4}
[F_{J}^{{I}'}]_{ \text{in } (\{\pd_{{K}}\},\{\pd_{{K}'}\}) }
=
\Bigl[\frac{\pd F^{{I}'}}{\pd u^{J}}
\Bigr]_{
    \begin{subarray}{l}
{I}'=1,\dots,n'+r'\\
{J}=1,\dots,n+r
    \end{subarray}
}
=
    \begin{pmatrix}
[F_{,\mu}^{\nu'}]   & 0_{n'\times r} \\
[F_{,\mu}^{a'}]     & [F_{,b}^{a'}]
    \end{pmatrix}
=
    \begin{bmatrix}
F_{,\mu}^{\nu'} & 0_{n'\times r} \\
F_{,\mu}^{a'}   & F_{,b}^{a'}
    \end{bmatrix}
\ .
    \end{equation}

    Let connections $\Delta^{h}$ and $\Delta^{\prime h}$ on $(E,\pi,M)$
and $(E,'\pi',M')$, respectively, be given. To the local coordinates
$\{u^{I}\}$ and $\{u^{\prime{I}}\}$ correspond the adapted frames
(see~\eref{3.18}--\eref{3.20})
    \begin{equation}    \label{5.5}
    \begin{split}
(X_\mu,X_a)
&= (\pd_\nu,\pd_b) \cdot
    \begin{bmatrix}
\delta_\mu^\nu  & 0 \\
+\Gamma_\mu^b   & \delta_a^b
    \end{bmatrix}
=(\pd_\mu+\Gamma_\mu^b\pd_b,\pd_a)
\\
(X'_{\mu'},X'_{a'})
&= (\pd'_{\nu'},\pd'_{b'}) \cdot
    \begin{bmatrix}
\delta_{\mu'}^{\nu'}    & 0 \\
+\Gamma_{\mu'}^{b'} & \delta_{a'}^{b'}
    \end{bmatrix}
=(\pd'_{\mu'}+\Gamma_{\mu'}^{\prime b'}\pd_{b'},\pd_{a'}) ,
    \end{split}
    \end{equation}
where $\pd_{I}:=\frac{\pd}{\pd u^{I}}$, and adapted coframes
    \begin{equation}    \label{5.6}
\begin{pmatrix} \omega^\mu \\ \omega^a  \end{pmatrix}
=
    \begin{bmatrix}
\delta^\mu_\nu  & 0 \\
- \Gamma_\nu^a   & \delta_b^a
    \end{bmatrix}
\cdot
\begin{pmatrix} \od u^\nu \\ \od u^b    \end{pmatrix}
=\dots
\qquad
\begin{pmatrix} \omega^{\prime\mu'} \\ \omega^{\prime a'} \end{pmatrix}
=
    \begin{bmatrix}
\delta^{\mu'}_{\nu'}    & 0 \\
- \Gamma_{\nu'}^{\prime a'}  & \delta_{b'}^{a'}
    \end{bmatrix}
\cdot
\begin{pmatrix} \od u^{\prime\nu'} \\ \od u^{\prime b'} \end{pmatrix}
=\dots
\ .
    \end{equation}
The symbols $\Gamma_{\mu}^{a}$ and $\Gamma_{\mu'}^{\prime a'}$ in~\eref{5.5}
and~\eref{5.6} denote the 2\ndash index coefficients of respectively
$\Delta^h$ and $\Delta^{\prime h}$ in the respective adapted frames.

    If $\{e_{I}\}$ and $\{e'_{{I}'}\}$ are arbitrary frames over
$U$ in $T(E)$ and over $U'=F(U)$ in $T(E')$, respectively, the (Jacobi)
matrix of $F_*$ in them is defined via (cf.~\eref{5.3})
    \begin{equation}    \label{5.7}
F_*(e_{I}|_p) = (F_{{I}}^{{I}'}|_p) (e'_{{I}'}|_{F(p)}) .
    \end{equation}
In particular, in the adapted frames~\eref{5.5}, we have
 $F_*(X_{I}|_p)=(F_{{I}}^{{I}'}|_p) (X'_{{I}'}|_{F(p)})$
and therefore the Jacobi matrix of $F_*$ in the adapted frames~\eref{5.5} is~%
\footnote{~%
The changes
$e:=\{e_I\} \mapsto \{B_I^J e_J\}$ and
$e':=\{e'_{I'}\} \mapsto \{B_{I'}^{\prime\, J'} e'_{J'}\}$, with non-degenerate
matrix\ndash valued functions $B:=[B_I^J]$ and $B':=[B_{I'}^{\prime\, J'}]$,
imply the transformation
 $F_{(e,e')}:=[F_{J}^{I'}] \mapsto (B')^{-1}\cdot F_{(e,e')}\cdot B$
of the Jacobi matrix of $F_*$. From here,~\eref{5.8} follows immediately.%
}
    \begin{multline}    \label{5.8}
[F_{{J}}^{{I}'}]
= [F_{{J}}^{{I}'}]_{\text{in } ( \{X_{K}\} , \{X'_{{K}'}\} ) }
=
    \begin{bmatrix}
F_{\nu}^{\mu'}  & F_{a}^{\mu'} \\
F_{\nu}^{b'}    & F_{a}^{b'}
    \end{bmatrix}
=
    \begin{bmatrix}
\delta_{\lambda'}^{\mu'}    &0\\
- \Gamma_{\lambda'}^{\prime b'} \circ F & \delta_{c'}^{b'}
    \end{bmatrix}
\cdot
    \begin{bmatrix}
F_{,\varrho}^{\lambda'} & 0 \\
F_{,\varrho}^{c'}   & F_{,d}^{c'}
    \end{bmatrix}
\cdot
    \begin{bmatrix}
\delta_{\nu}^{\varrho}  & 0 \\
+\Gamma_{\nu}^{d}   & \delta_{a}^{d}
    \end{bmatrix}
\\
=
    \begin{bmatrix}
F_{,\nu}^{\mu'} & 0\\
X_\nu(F^{b'}) - (\Gamma_{\lambda'}^{\prime b'}\circ F) F_{,\nu}^{\lambda'}
        & F_{,a}^{b'}
    \end{bmatrix}
    \end{multline}
with $F_{,{J}}^{{I}'}:=\frac{\pd F^{{I}'}}{\pd u^{J}}$
defining the matrix of $F_*$ in $(\{\pd_{K}\},\{\pd_{{K}'}^{\prime}\})$
via~\eref{5.4}. Thus, the general formula~\eref{5.7} now reads
    \begin{equation}    \label{5.9}
F_*(X_\mu,X_a)
=
(X_{\nu'}^{\prime},X_{b'}^{\prime})
\cdot
    \begin{bmatrix}
F_{\mu}^{\nu'}  & 0 \\
F_{\mu}^{b'}    & F_{a}^{b'}
    \end{bmatrix}
=
( F_{,\mu}^{\nu'}X_{\nu'}^{\prime}+ F_{\mu}^{b'} X_{b'}^{\prime} ,
            F_{,a}^{b'} X_{b'}^{\prime}
)
    \end{equation}
with
    \begin{equation}    \label{5.10}
F_{\mu}^{b'}
= X_\mu(F^{b'})
- (\Gamma_{\lambda'}^{\prime b'}\circ F)\cdot F_{,\mu}^{\lambda'} .
    \end{equation}
From~\eref{5.7}, it is clear that the elements $F_{{J}}^{{I}'}|_p$ of the
Jacobi matrix of $F_*$ at $p\in U$ are elements of a $(1,1)$ (mixed) tensor
from $T^*_p(E)\otimes T_{F(p)}(E')$; in particular, if the adapted frames are
changed (see~\eref{3.21}), the block structure of~\eref{5.8} is preserved and
the elements of its blocks are transformed as elements of the corresponding to
them tensors~%
\footnote{~%
E.g.  $F_{\nu}^{b'}(p)$ are elements of a tensor from the tensor space
spanned by $\{\omega^\nu|_p\otimes X_{b'}^{\prime}|_{F(p)}\}$.%
}.
An important corollary from~\eref{5.9} is
    \begin{equation}    \label{5.11}
F_*(\Delta^h)\subseteq \Delta^{\prime h}
\iff
F_{\mu}^{b'} = 0
    \end{equation}
in any pair $(\{X_{I}\},\{X_{{I}'}\})$ of adapted frames. If it happens
that $F_*(\Delta^h)=\Delta^{\prime h}$, we say that $F$ \emph{preserves the
connection} $\Delta^h$, \ie $F$ is a \emph{connection preserving mapping}; in
particular, if $(E',\pi',M')=(E,\pi,M)$ and $F_*(\Delta^h)=\Delta^{\prime
h}$, the mapping $F$ is called a \emph{symmetry} of $\Delta^h$.

    If the bundles considered are vectorial ones, the fibre coordinates,
morphisms, and connections which are compatible with the vector structure
must be linear on the fibres, \viz
    \begin{equation}    \label{5.12}
F^{a'} = (\mathcal{F}_{b}^{a'}\circ\pi) u^b \quad
\Gamma_\mu^a = - (\Gamma_{b\mu}^{a}\circ\pi) u^b \quad
\Gamma_{\mu'}^{a'} = - (\Gamma_{b'\mu'}^{a'}\circ\pi') u^{b'} ,
    \end{equation}
where the functions $\mathcal{F}_{b}^{a'}\colon\pi(U)\to\field$ are of class
$C^1$ and $\Gamma_{b\mu}^{a}$ (resp.\ $\Gamma_{b'\mu'}^{a'}$) are the
3\ndash index coefficients of the linear connection $\Delta^h$ (resp.\
$\Delta^{\prime h}$). Consequently, in a case of vector bundles, the Jacobi
matrix~\eref{5.8} takes the form
    \begin{equation}    \label{5.13}
    \begin{bmatrix}
F_{\nu}^{\mu'}  & F_{a}^{\mu'}  \\
F_{\nu}^{b'}    & F_{a}^{b'}
    \end{bmatrix}
=
    \begin{bmatrix}
F_{,\nu}^{\mu'} & 0 \\
(F_{c\nu}^{b'}\circ\pi) u^c & \mathcal{F}_{a}^{b'}\circ \pi
    \end{bmatrix}
    \end{equation}
with
    \begin{equation}    \label{5.14}
F_{a\mu}^{b'}
:= \pd_\mu(\mathcal{F}_{a}^{b'}) - \Gamma_{a\mu}^{c} \mathcal{F}_{c}^{b'}
   + (\Gamma_{c'\lambda'}^{\prime b'}\circ f)
   \cdot \mathcal{F}_{a}^{c'}\cdot  f_{\mu}^{\lambda'} ,
    \end{equation}
where we have used that $\pi'\circ F=f\circ\pi$ and
$ u^{\prime c'}\circ F= F^{c'}$
and we have set
$f_{\mu}^{\lambda'}:=\frac{\pd(x^{\prime\lambda'}\circ f)}{\pd x^\mu}$, so
that $F_{,\mu}^{\lambda'}= f_{\mu}^{\lambda'}\circ\pi$. Therefore~\eref{5.10}
now reads
    \begin{equation}    \label{5.15}
F_{\mu}^{b'} = (F_{a\mu}^{b'}\circ\pi)\cdot u^a.
            \end{equation}

    If $M$ and $M'$ are manifolds and $f\colon M\to M'$ is of class
$C^1$, the above general considerations are valid for the morphism
$(f_*,f)$ of the tangent bundles $(T(M),\pi_T,M)$ and $(T(M'),\pi'_T,M')$. A
peculiarity of a tangent bundle is that the fibre dimension of the bundle
equals to the dimension of its base. Due to that fact, the base indices
$\lambda,\mu,\nu,\dots=1,\dots,n$ is convenient to be used for the fibre ones
$a,b,c,\dots=n+1,\dots,n+r$ according to the rule
    \begin{subequations}    \label{5.16}
    \begin{gather}  \label{5.16a}
a\mapsto \mu=a-\dim M,
\\\intertext{which must be combined with a change of the notation for the
fibre coordinates, like}
            \label{5.16b}
u^a\mapsto u_1^\mu,
    \end{gather}
    \end{subequations}
as otherwise the change~\eref{5.16a} will entail $u^a\mapsto u^\mu$, the
result of which coincides with the notation for the basic coordinates.~%
\footnote{~%
The subscript $1$ in~\eref{5.16b} indicates that $u_1^\mu$ are fibre
coordinates in the \emph{first} order tangent bundle $(T(M),\pi_T,M)$ over
$M$.%
}
Since
\(
f_*\bigl(\frac{\pd}{\pd x^\mu}\big|_z\bigr)
=\frac{\pd(x^{\prime\mu'}\circ f)}{\pd x^\mu} \big|_{z}
        \frac{\pd}{\pd x^{\prime \mu'}} \big|_{f(z)}
\)
for $z\in\pi(U)$~\cite[sec.~1.23(a)]{Warner}, the Jacobi matrix of $f$
relative to the charts $(\pi(U),x)$ and $(\pi'(U'),x')=(f(\pi(U)),x')$ has
the elements
    \begin{equation}    \label{5.17}
f_{\nu}^{\mu'}
:= \frac{\pd(x^{\prime\mu'}\circ f)}{\pd x^\nu} \colon\pi(U)\to\field .
    \end{equation}
Combining this with the definition of the vector fibre coordinates
$u_1^\mu$,
$u_1^\mu\big( p^\nu\frac{\pd}{\pd x^\nu}\big|_{\pi(p)} \bigr) = p^\mu$,
we see that~\eref{5.2}, with $f_*$ for $F$, reads
    \begin{subequations}    \label{5.18}
    \begin{gather}  \label{5.18a}
u^{\prime\mu'}
= f_*^{\mu'}(u^1,\dots,u^n)
= x^{\prime\mu'} \circ f \circ \pi
\quad
u_1^{\prime\mu'}
= f_{*}^{\prime\mu'}( u^1,\dots,u^n, u_1^1,\dots,u_1^n)
= (f_\nu^{\mu'}\circ\pi)\cdot u_1^\nu
\\          \label{5.18b}
x^{\prime\mu'} = f^{\mu'}(x^1,\dots,x^n) = x^{\mu'}\circ f .
    \end{gather}
    \end{subequations}
Therefore the derivatives in~\eref{5.4} and~\eref{5.8}--\eref{5.10} should be
replace according to ($u^\mu=x^\mu\circ\pi)$
    \begin{equation}    \label{5.19}
F_{,\mu}^{\nu'}\mapsto \frac{\pd f_*^{\mu'}}{\pd x^\mu }
= f_{\mu}^{\nu'} \circ \pi
\quad
F_{,\nu}^{a'}\mapsto \frac{\pd f_*^{\mu'}}{\pd x^\nu}
=\Bigl( \frac{\pd f_{\lambda}^{\mu'}}{\pd x^\nu} \circ\pi \Bigr) u_1^\lambda
\quad
F_{,b}^{a'}\mapsto \frac{\pd f_*^{\mu'}}{\pd u_1^\nu}
= f_{\nu}^{\mu'}\circ\pi .
    \end{equation}

    If $\Delta^h$ and $\Delta^{\prime h}$ are linear connections on $M$
and $M'$, respectively, their 2- and 3-index coefficients are connected
through (cf.~\eref{5.12})
    \begin{equation}    \label{5.20}
\Gamma_{\nu}^{\lambda}
= - (\Gamma_{\mu\nu}^{\lambda} \circ\pi) \cdot u_1^\lambda
\quad
\Gamma_{\prime\nu'}^{\lambda'}
= - (\Gamma_{\mu'\nu'}^{\prime\lambda'} \circ\pi') \cdot u_1^{\lambda'} .
    \end{equation}
Thus the Jacobi matrix of $(f_*)_*=:f_{**}$ in the pair of frames
\(
\bigl(
\bigl\{
X_\mu
=\frac{\pd}{\pd x^\mu} + \Gamma_{\mu}^{\lambda}\frac{\pd}{\pd u_1^\lambda}
, X_\nu^1=\frac{\pd}{\pd u_1^\nu}
\bigr\}
,
\bigl\{ X_{\mu'}^{\prime},X_{\nu'}^{\prime1} \bigr\}
\bigr)
\)
is (cf.~\eref{5.13} and~\eref{5.14})
    \begin{gather}  \label{5.21}
    \begin{bmatrix}
f_{\nu}^{\mu'}\circ\pi  & 0 \\
- f_{\lambda\nu}^{\varrho'}\circ\pi) \cdot u_1^\lambda
            & f_{\tau}^{\varrho'}\circ \pi
    \end{bmatrix}
\\\intertext{where}          \label{5.22}
f_{\mu\nu}^{\lambda'}
:= f_{*\mu\nu}^{\lambda'}
:= \pd_\nu(f_{\mu}^{\lambda'})
   - f_{\sigma}^{\lambda'} \Gamma_{\mu\nu}^{\sigma}
   + (\Gamma_{\sigma'\tau'}^{\prime\lambda'} \circ f)
      f_{\mu}^{\sigma'} f_{\nu}^{\tau'}.
    \end{gather}
The quantities~\eref{5.22} are components of a $T(M')$-valued 2-form on $M$,
\ie of an element in $T(M')\otimes\Lambda^2(M)$.~%
\footnote{~%
Moreover, if we consider $f_{\nu}^{\mu'}$, defined via~\eref{5.17}, as
components of an element in $T_{f(p)}(M')\otimes\Lambda^1_p(M)$,
then~\eref{5.22} are the components of the \emph{mixed} covariant derivative
(along $\frac{\pd}{\pd x^\nu}$) of
 $f_{\mu}^{\mu'} (\pd'_{\mu'}|_{f(\cdot)})\otimes \od u^\mu$ relative to the
connection $\Delta^h\times\Delta^{\prime h}$ on $M\times M$.%
}


\section {General (co)frames}
\label{Sect6}

    Until now two special kinds of local (co)frames in the (co)tangent
bundle to the bundle space of a bundle were employed, \viz the natural
holonomic ones, induced by some local coordinates, and the adapted (co)frames
determined by local coordinates and a connection on the bundle. The present
section is devoted to (re)formulation of some important results and formulae in
arbitrary (co)frames, which in particular can be natural or adapted (if a
connection is presented) ones.

    Let $(E,\pi,M)$ be a $C^2$ bundle and $\{e_{I}\}$ a (local) frame
in $T(E)$. The components $C_{{I}{J}}^{{K}}$ of the anholonomy object
of $\{e_{I}\}$ are defined by~\eref{3.11} and a change
    \begin{equation}    \label{6.1}
\{e_{I}\} \mapsto \{ \bar{e}_{I} = B_{I}^{J} e_{J} \}
    \end{equation}
with a non-degenerate matrix-valued function
$B=[B_{I}^{J}]_{{I},{J}=1}^{n+r}$ entails (see~\eref{2.7-1})
    \begin{equation}    \label{6.2}
C_{{I}{J}}^{{K}}
\mapsto \bar{C}_{{I}{J}}^{{K}}
=
(B^{-1})_{L}^{K}
  \bigl( B_{I}^{M} e_{M}(B_{J}^{L}) -
     B_{J}^{M} e_{M}(B_{I}^{L}) +
    B_{I}^{M} B_{J}^N C_{{M}N}^{{L}}
  \bigr) .
    \end{equation}

    Let a connection $\Delta^h$ on $(E,\pi,M)$ be given. If
$\{e_{I}\}$ is a specialized frame for $\Delta^h$ (see
Sebsect.~\ref{Subsect3.2}), then the set $\{C_{{I}{J}}^{{K}}\}$ is
naturally divided into the six groups~\eref{3.12}. The value of that division
is in its invariance with respect to the class of specialized frames, which
means that, if $\{\bar{e}_{I}\}$ is also a specialized   frame, then the
transformed components of the elements of each group are functions only in the
elements of the non\ndash transformed components of the same group ---
see~\eref{3.15}, \eref{3.12-1}, and~\eref{2.7-1}. By means of~\eref{6.1}, one
can prove that, if such a division holds in a frame $\{e_{I}\}$, then it
holds in $\{\bar{e}_{I}\}$ if and only if the matrix\ndash valued
function $B$ is of the form~\eref{3.10}. In particular, we cannot talk about
fibre coefficients of $\Delta^h$ and of fibre components of the curvature of
$\Delta^h$ in frames more general than the specialized ones as in that case
the  transformation~\eref{6.1}, with $\{e_{I}\}$ (resp.\
$\{\bar{e}_{I}\}$) being a specialized (resp.\ non\ndash specialized)
frame, will mix, for instance, the fibre coefficients and the curvature's fibre
components of $\Delta^h$ in $\{\bar{e}_I\}$ --- see~\eref{6.2}.

    It is a simple, but important, fact that the specialized frames are (up to
renumbering) the most general ones which respect the splitting of $T(E)$ into
vertical and horizontal components. Suppose $\{e_{I}\}$ is a specialized frame.
Then the general element of the set of all specialized frames is
(see~\eref{3.4a} and~\eref{3.10})
    \begin{subequations}    \label{6.3}
    \begin{equation}    \label{6.3a}
(\bar{e}_\mu,\bar{e}_a)
=
({e}_\nu,{e}_b) \cdot
    \begin{bmatrix}
A_\mu^\nu   & 0 \\
0       & A_a^b
    \end{bmatrix}
=
(A_\mu^\nu e_\nu, A_a^b e_b),
    \end{equation}
where $[A_\mu^\nu]_{\mu,\nu=1}^{n}$ and
$[A_a^b]_{a,b=n+1}^{n+r}$ are non-degenerate matrix-valued functions on $E$,
which are constant on the fibres of $(E,\pi,M)$, \ie we can set
 $A_\mu^\nu=B_\mu^\nu\circ\pi$ and $A_a^b=B_a^b\circ\pi$ for some
non\ndash degenerate matrix\ndash valued functions $[B_\mu^\nu]$ and
$[B_a^b]$ on $M$. Respectively, the general specialized coframe dual to
$\{\bar{e}_{I}\}$ is (see~\eref{3.4b} and~\eref{3.10})
    \begin{equation}    \label{6.3b}
\begin{pmatrix} \bar{e}^\mu \\ \bar{e}^a  \end{pmatrix}
 =
    \begin{bmatrix}
[A_\rho^\lambda]^{-1}   & 0\\
0           & [A_d^c]^{-1}
    \end{bmatrix}
\cdot \begin{pmatrix} {e}^\nu \\ {e}^b  \end{pmatrix}
 =
    \begin{bmatrix}
\bigl([A_\rho^\lambda]^{-1}\bigr)_\nu^\mu e^\nu \\
\bigl([A_d^c]^{-1}\bigr)_b^a e^b
    \end{bmatrix} \ ,
    \end{equation}
    \end{subequations}
where $\{e^{I}\}$ is the specialized coframe dual to $\{e_{I}\}$,
$e^{I}(e_{J})=\delta_{J}^{I}$.

    Since $\pi_*|_{\Delta^h}\colon\{X\in\Delta^h\}\to \mathcal{X}(M)$ is
an isomorphism, any basis $\{\varepsilon_\mu\}$ \emph{for} $\Delta^h$
defines a basis $\{E_\mu\}$ of $\mathcal{X}(M)$ such that
    \begin{equation}    \label{6.4}
E_\mu = \pi_*|_{\Delta^h}(\varepsilon_\mu)
    \end{equation}
and v.v., a basis $\{E_\mu\}$ for $\mathcal{X}(M)$ induces a basis
$\{\varepsilon_\mu\}$ for $\Delta^h$ via
    \begin{equation}    \label{6.5}
\varepsilon_\mu = (\pi_*|_{\Delta^h})^{-1} (E_\mu) .
    \end{equation}
Similarly, there is a bijection $\{\varepsilon^\mu\}\mapsto\{E^\mu\}$ between
the `horizontal' coframes $\{\varepsilon^\mu\}$ and the coframes $\{E^\mu\}$
dual to the frames in $T(M)$
($E^\mu\in\Lambda^1(M)$, $E^\mu(E_\nu)=\delta_\nu^\mu$).
Thus a `horizontal' change
    \begin{equation}    \label{6.6}
\varepsilon_\mu\mapsto
\bar{\varepsilon}_\mu = (B_\mu^\nu\circ\pi) \varepsilon_\nu ,
    \end{equation}
which is independent of a `vertical' one given by
    \begin{equation}    \label{6.7}
\varepsilon_a\mapsto
\bar{\varepsilon}_a = (B_a^b\circ\pi) \varepsilon_b
    \end{equation}
with $\{\varepsilon_a\}$ being a basis for $\Delta^v$, is
equivalent to the transformation
    \begin{equation}    \label{6.8}
E_\mu \mapsto \bar{E}_\mu = B_\mu^\nu E_\nu
    \end{equation}
of the basis $\{E_\mu\}$ for $\mathcal{X}(M)$, related via~\eref{6.4} to the
basis $\{\varepsilon_\mu\}$ for $\Delta^h$. Here $[B_\mu^\nu]$ and $[B_a^b]$
are non\ndash degenerate matrix\ndash valued functions on $M$.

    As $\pi_*(\varepsilon_a)=0\in\mathcal{X}(M)$, the `vertical'
transformations~\eref{6.7} do not admit interpretation analogous to the
`horizontal' ones~\eref{6.6}. However, in a case of a \emph{vector} bundle
$(E,\pi,M)$, they are tantamount to changes of frames in the bundle space
$E$, \ie of the bases for $\Sec(E,\pi,M)$. Indeed, if $v$ is the mapping
defined by~\eref{4.1-1}, the sections
    \begin{equation}    \label{6.9}
E_a = v^{-1}(\varepsilon_a)
    \end{equation}
form a basis for $\Sec(E,\pi,M)$ as  the vertical vector fields
$\varepsilon_a$ form a basis for $\Delta^v$. Conversely, any basis $\{E_a\}$
for the sections of $(E,\pi,M)$ induces a basis $\{\varepsilon_a\}$ for
$\Delta^v$ such that
    \begin{equation}    \label{6.10}
\varepsilon_a = v(E_a) .
    \end{equation}
As $v$ and $v^{-1}$ are linear, the change~\eref{6.7} is equivalent to the
transformation
    \begin{equation}    \label{6.11}
E_a\mapsto \bar{E}_a = B_a^b E_b
    \end{equation}
of the frame $\{E_a\}$ in $E$ related to $\{\varepsilon_a\}$ via~\eref{6.9}.
In this way, we see that
\emph{any specialized frame
$\{\varepsilon_{I}\}=\{\varepsilon_\mu,\varepsilon_a\}$
for a connection on a vector bundle $(E,\pi,M)$ is equivalent to a pair of
frames $(\{E_\mu\},\{E_a\})$%
}
such that $\{E_\mu\}$ is a basis for the set $\mathcal{X}(M)$ of vector
fields on the base $M$, \ie for the sections of the tangent bundle
$(T(M),\pi_T,M)$ (and hence is a frame in $T(M)$ over $M$), and $\{E_a\}$ is
a basis for the set $\Sec(E,\pi,M)$ of sections of the initial bundle (and
hence is a frame in $E$ over $M$). Since conceptually the frames in $T(M)$ and
$E$ are easier to be understood and in some cases have a direct physical
interpretation, one often works with the pair of frames
\(
( \{E_\mu=\pi_*|_{\Delta^h}(\varepsilon_\mu)\} ,
\{E_a=v^{-1}(\varepsilon_a)\} )
\)
instead with a specialized frame
$\{\varepsilon_{I}\}=\{\varepsilon_\mu,\varepsilon_a\}$;
for instance $\{E_\mu\}$ and $\{E_a\}$ can be completely arbitrary frames in
$T(M)$ and $E$, respectively, while the specialized frames represent only a
particular class  of frames in $T(E)$.

    One can \emph{mutatis mutandis} localize the above considerations
when $M$ is replaced with an open subset $U_M$ in $M$ and $E$ is replaced with
$U=\pi^{-1}(U_M)$. Such a localization is
important when the bases/frames considered are connected with some local
coordinates or when they should be smooth.%
\footnote{~%
Recall, not every manifold admits a \emph{global} nowhere vanishing $C^m$,
$m\ge0$, vector field (see~\cite{Spivak-1} or~\cite[sec.~4.24]{Schutz}); \eg
such are the even\ndash dimensional spheres $\mathbb{S}^{2k}$,
$k\in\field[N]$, in Euclidean space.%
}

    Let us turn now our attention to frames adapted to local coordinates
$\{u^{I}\}$ on an open set $U\subseteq E$ for a given connection
$\Delta^h$ on a general  $C^1$ bundle $(E,\pi,M)$
(see~\eref{3.18}--\eref{3.20}). Since in their definition the local
coordinates $\{u^{I}\}$ enter only via the vector fields
$\pd_{I}:=\frac{\pd}{\pd u^{I}}\in\mathcal{X}(E)$, we can generalize
this definition by replacing $\{\pd_{I}\}$ with an arbitrary frame
$\{e_{I}\}$ defined in $T(E)$ over an open set $U\subseteq E$ and such that
$\{e_a|_p\}$ is a
\emph{basis for the space $T_p(\pi^{-1}(\pi(p)))$ tangent to the fibre
through} $p\in U$.
So, using $\{e_{I}\}$ for $\{\pd_{I}\}$, we have
    \begin{equation}    \label{6.12}
(e_\mu^U,e_a^U )
= (D_\mu^\nu e_\mu + D_\mu^a e_a , D_a^b e_b )
=
(e_\nu,e_b) \cdot
    \begin{pmatrix}
[D_\mu^\nu] & 0 \\
[D_\mu^b] & [D_a^b]
    \end{pmatrix} \ ,
    \end{equation}
where $\{e_{I}^U\}$ is a \emph{specialized} frame in $T(U)$, $[D_\mu^\nu]$
and $[D_a^b]$ are non\ndash degenerate matrix\ndash valued functions on $U$,
and $D_\mu^a\colon U\to\field$.

    \begin{Defn}    \label{Defn6.1}
    The specialized frame $\{X_{I}\}$ over $U$ in $T(U)$, obtained
from~\eref{6.12} via an admissible transformation~\eref{3.4a} with matrix
\(
A = \Bigl(
    \begin{smallmatrix}
[D_\nu^\mu]^{-1}    & 0 \\
0           & [D_b^a]^{-1}
    \end{smallmatrix}
\Bigr) ,
\)
is called \emph{adapted to the frame $\{e_{I}\}$ for} $\Delta^h$.~%
\footnote{~%
Recall, here and below the adapted frames are defined only with respect to
frames $\{e_I\}=\{e_\mu,e_a\}$ such that $\{e_a\}$ is a basis for the
vertical distribution $\Delta^v$ over $U$, \ie $\{e_a|_p\}$ is a basis for
$\Delta_p^v$ for all $p\in U$. Since $\Delta^v$ is integrable, the relation
$e_a\in\Delta^v$ for all $a=n+1,\dots,n+r$ implies
$[e_a,e_b]_{\_}\in\Delta^v$ for all $a,b=n+1,\dots,n+r$.%
}
    \end{Defn}

    \begin{Exrc}    \label{Exrc6.0}
	Using~\eref{3.4} and~\eref{3.10}, verify that the adapted frame
$\{X_I\}$ and coframe $\{\omega^I\}$ dual to it are independent of the
particular specialized frame $\{e_I^U\}$ entering into their definitions
via~\eref{6.12}. The equalities~\eref{6.13a} and~\eref{6.17-4} derived below
are indirect proof of that fact too.
    \end{Exrc}

    According to~\eref{3.4}, the adapted frame
$\{X_{I}\}=\{X_\mu,X_a\}$ and the dual to it coframe
$\{\omega^{I}\}=\{\omega^\mu,\omega^a\}$ are given by the equations
    \begin{subequations}    \label{6.13}
    \begin{align}   \label{6.13a}
(X_\mu,X_a)
& =
(e_\nu,e_b) \cdot
    \begin{bmatrix}
\delta_\mu^\nu  & 0 \\
+\Gamma_\mu^b   &\delta_a^b
    \end{bmatrix}
=(e_\mu + \Gamma_\mu^b e_b , e_a)
\\                            \label{6.13b}
\begin{pmatrix} \omega^\mu \\ \omega^a \end{pmatrix}
& =
    \begin{bmatrix}
\delta_\nu^\mu  & 0 \\
- \Gamma_\nu^a   &\delta_b^a
    \end{bmatrix}
\cdot
\begin{pmatrix} e^\nu \\ e^b \end{pmatrix}
=
\begin{pmatrix}
e^\mu   \\
e^a - \Gamma_\nu^a e^\nu
\end{pmatrix} \ ,
    \end{align}
    \end{subequations}
where $\{e^{I}\}$ is the coframe dual to $\{e_{I}\}$,
$e^{I}(e_{J})=\delta_{J}^{I}$, and the functions
$\Gamma_\mu^a\colon U\to\field$, called
\emph{(2\ndash index) coefficients of} $\Delta^h$ in $\{X_{I}\}$, are
defined by
    \begin{equation}    \label{6.14}
[\Gamma_\mu^a] := + [D_\nu^a]\cdot [D_\mu^\nu]^{-1} .
    \end{equation}

    \begin{Prop}    \label{Prop6.0-1}
	A change $\{e_{I}\}\mapsto\{\tilde{e}_{I}\}$ with
    \begin{equation}    \label{6.15}
(\tilde{e}_\mu,\tilde{e}_a)
= (e_\nu,e_b) \cdot
    \begin{pmatrix}
[A_\mu^\nu] & 0 \\
[A_\mu^b]   & [A_a^b]
    \end{pmatrix}
=
(A_\mu^\nu e_\nu + A_\mu^b e_b , A_a^b e_b) ,
    \end{equation}
where $[A_\mu^\nu]$ and $[A_a^b]$ are non-degenerate matrix-valued functions
on $U$, which are constant on the fibres of $(E,\pi,M)$, and
$A_\mu^b\colon U\to\field$, entails the transformations
    \begin{align}   \label{6.16}
(X_\mu,X_a) & \mapsto (\tilde{X}_\mu , \tilde{X}_a)
=
( \tilde{e}_\mu + \tilde{\Gamma}_\mu^b\tilde{e}_b,\tilde{e}_a )
=
( A_\mu^\nu X_\nu, A_a^b X_b)
=
(X_\nu,X_b) \cdot
    \begin{bmatrix}
A_\mu^\nu   & 0 \\
0   & A_a^b
    \end{bmatrix}
\\          \label{6.17}
\Gamma_\mu^a & \mapsto \tilde{\Gamma}_\mu^a
= \bigl([A_d^c]^{-1}\bigr)_b^a ( \Gamma_\nu^b A_\mu^\nu - A_\mu^b )
    \end{align}
of the frame $\{X_{I}\}$ adapted to $\{e_{I}\}$ and of the coefficients
$\Gamma_\mu^a$ of $\Delta^h$ in $\{X_{I}\}$, \ie  $\{\tilde{X}_{I}\}$
is the frame adapted to $\{\tilde{e}_{I}\}$ and $\tilde{\Gamma}_\mu^a$
are the coefficients of $\Delta^h$ in $\{\tilde{X}_{I}\}$.
    \end{Prop}

    \begin{Proof}
Apply~\eref{6.12}--\eref{6.14}.
    \end{Proof}

	\begin{Note}	\label{Note6.1}
If $\{e_{I}\}$ and $\{\tilde{e}_{I}\}$ are adapted, then $A_\mu^b=0$.
	If $\{Y_I\}$ is a specialized frame, it is adapted to any frame
 $\{e_\mu=A_\mu^\nu Y_\nu,e_a=A_a^b Y_b\}$ and hence any specialized frame
can be considered as an adapted one; in particular, any specialized frame is
a frame adapted to itself. Obviously (see~\eref{6.14}), the coefficients of a
connection identically vanish in a given specialized frame considered as an
adapted one. This leads to the concept of a \emph{normal frame},
which will be studied on this context in a forthcoming paper.
Besides, from the above observation follows that the set of  adapted frames
coincides with the one of specialized frames.
	\end{Note}

    \begin{Exrc}    \label{Exrc6.1}
    Verify that the formulae dual to~\eref{6.15}
and~\eref{6.16} are (see~\eref{3.4b} and~\eref{3.5b})
    \begin{align}   \label{6.17-1}
\begin{pmatrix} \tilde{e}^\mu\\ \tilde{e}^a   \end{pmatrix}
& =
    \begin{pmatrix}
[A_\tau^\varrho]^{-1}   & 0 \\
-[A_d^c]^{-1} [A_\tau^c] [A_\tau^\varrho]^{-1}  & [A_d^c]^{-1}
    \end{pmatrix}
\cdot
\begin{pmatrix} e^\nu\\ e^b   \end{pmatrix}
=
\begin{pmatrix}
([A_\tau^\varrho]^{-1})_\nu^\mu \; e^\nu   \\
\bigl( [A_d^c]^{-1} \bigr )_b^a \; e^b
-\bigl(  [A_d^c]^{-1} [A_\tau^c] [A_\tau^\varrho]^{-1} \bigr)_\nu^a\; e^\nu
\end{pmatrix}
\\          \label{6.17-2}
\begin{pmatrix} \omega^\mu\\ \omega^a
  \end{pmatrix}
& \mapsto
\begin{pmatrix} \tilde{\omega}^\mu\\ \tilde{\omega}^a
  \end{pmatrix}
 =
    \begin{pmatrix}
([A_\tau^\varrho]^{-1}))_\nu^\mu e^\nu  \\
([A_d^c]^{-1})_b^a e^b
    \end{pmatrix} \ .
    \end{align}
    \end{Exrc}

    \begin{Exmp}    \label{Exmp6.2}
    If $\{e_{I}\}$ and $\{\tilde{e}_{I}\}$
are the frames generated by local coordinates $\{u^{I}\}$ and
$\{\tilde{u}^{I}\}$, \viz $e_{I}=\frac{\pd}{\pd u^{I}}$ and
$\tilde{e}_{I}=\frac{\pd}{\pd \tilde{u}^{I}}$, the changes~\eref{6.16}
and~\eref{6.17} reduce to~\eref{3.21} and~\eref{3.22}, respectively.
The choice $e_{I}=\frac{\pd}{\pd u^{I}}$ also reduces
definition~\ref{Defn6.1} to definition~\ref{Defn3.5}.
    \end{Exmp}

	A result similar to proposition~\ref{Prop3.1} is valid too provided in
its formulation equation~\eref{3.22} is replaced with~\eref{6.17}.

	If $e_\mu$ has an expansion
$e_\mu = e_\mu^\nu\frac{\pd}{\pd u^\nu} + e_\mu^b\frac{\pd}{\pd u^b}$ in the
domain $U$ of $\{u^I\}=\{u^\mu=x^\mu\circ\pi,u^a\}$,
where
 $e_\mu^b\colon U\to\field$ and
$e_\mu^\nu=x_\mu^\nu\circ\pi$ for some $x_\mu^\nu\colon\pi(U)\to\field$
such that $\det[x_\mu^\nu]\not=0,\infty$, and we define a frame
$\{x_\mu\}$ in $T(\pi(U))\subseteq T(M)$ by
$\{x_\mu:=x_\mu^\nu\frac{\pd}{\pd x^\nu}\}$, then
	\begin{equation}	\label{6.17-3}
\pi_*(X_\mu) = x_\mu,
	\end{equation}
by virtue of~\eref{3.20-1} and~\eref{3.20-2}. Thus, we have
(cf.~\eref{3.20-3})
	\begin{equation}	\label{6.17-4}
X_\mu = (\pi_*|_{\Delta^h})^{-1}(x_\mu)
      = (\pi_*|_{\Delta^h})^{-1}\circ \pi_* (e_\mu)
	\end{equation}
which can be used in an equivalent definition of a frame $\{X_I\}$ adapted to
$\{e_I\}$ (with $\{e_a\}$ being a basis for $\Delta^v$): $X_\mu$ should be
defined by~\eref{6.17-4} and $X_a=e_a$. If one accepts such a definition of
an adapted frame, the 2\ndash index coefficients of a connection should be
defined via the equation~\eref{6.13a}, not by~\eref{6.14}, and the proofs of
some results, like~\eref{6.16} and~\eref{6.17}, should be modified.

    \begin{Prop}    \label{Prop6.0}
If $\{X_{I}\}$ is a frame adapted to a frame $\{e_{I}\}$, with $\{e_a\}$
being a basis for $\Delta^v$, for a $C^1$ connection $\Delta^h$, then
(cf.~\eref{3.23})
    \begin{subequations}    \label{6.18}
    \begin{align}   \label{6.18a}
[X_\mu,X_\nu]_{\_} &= R_{\mu\nu}^{a} X_a + S_{\mu\nu}^{\lambda} X_\lambda
\\          \label{6.18b}
[X_\mu,X_b]_{\_} &= \lindex[\Gamma]{}{\circ}{}_{b\mu}^{a} X_a
           				+ C_{\mu b}^{\lambda} X_\lambda
\\          \label{6.18c}
[X_a,X_b]_{\_} &= C_{ab}^{d} X_d ,
    \end{align}
    \end{subequations}
where (cf.~\eref{3.24})
    \begin{subequations}    \label{6.19}
    \begin{align}   \label{6.19a}
    \begin{split}
R_{\mu\nu}^{a}
&:= X_\mu(\Gamma_\nu^a) - X_\nu(\Gamma_\mu^a)
- C_{\mu\nu}^{a}
    - \Gamma_\mu^b C_{\nu b}^{a} + \Gamma_\nu^b C_{\mu b}^{a}
\\
&\hphantom{:=\ }
    + \Gamma_\lambda^a
      ( -C_{\mu\nu}^{\lambda} + \Gamma_\mu^bC_{\nu b}^{\lambda}
			      - \Gamma_\nu^bC_{\mu b}^{\lambda} )
    + \Gamma_\mu^b\Gamma_\nu^d C_{bd}^{a}
\\
S_{\mu\nu}^{\lambda}
&:= C_{\mu\nu}^{\lambda} + \Gamma_\mu^b C_{\nu b}^{\lambda}
             - \Gamma_\nu^b C_{\mu b}^{\lambda}
    \end{split}
\left.
\begin{matrix} \\ \\ \\ \\ \end{matrix}
\right\}
\\          \label{6.19b}
\lindex[\Gamma]{}{\circ}{}_{b\mu}^{a}
&:= - X_b(\Gamma_\mu^a) - C_{\mu b}^{a} + \Gamma_\mu^d  C_{db}^{a}
    - \Gamma_\lambda^a C_{\mu b}^{\lambda}
\\          \label{6.19c}
[e_{I},e_{J}]_{\_}
&=: C_{{I}{J}}^{{K}} e_{K}
 =  C_{{I}{J}}^{{a}} e_{a} + C_{{I}{J}}^{{\lambda}} e_{\lambda} .
    \end{align}
    \end{subequations}
    \end{Prop}

    \begin{Proof}
Insert~\eref{6.13a} into the corresponding commutators, use the
definition~\eref{6.19c} of the components of the anholonomy object of
$\{e_{I}\}$, and apply~\eref{6.13a}. Notice, as $\{e_a\}$ is a basis for the
integrable distribution $\Delta^v$, we have $[e_a,e_b]_{\_}\in\Delta^v$ and
consequently $C_{ab}^{\lambda}\equiv 0$.
    \end{Proof}

    The functions $R_{\mu\nu}^{a}$ are the \emph{fibre components of the
curvature} of $\Delta^h$ and $\lindex[\Gamma]{}{\circ}{}_{b\mu}^{a}$ are the
\emph{fibre coefficients} of $\Delta^h$ in the adapted frame $\{X_{I}\}$;
if $e_{I}=\frac{\pd}{\pd u^{I}}$ for some bundle coordinates
$\{u^{I}\}$ on $E$, they reduce to~\eref{3.24a} and~\eref{3.24b},
respectively. From~\eref{6.18}, we immediately derive
    \begin{Cor} \label{Cor6.1}
A connection $\Delta^h$ is integrable iff in some (and hence any) adapted
frame:
    \begin{equation}    \label{6.20}
R_{\mu\nu}^{a} = 0.
    \end{equation}
    \end{Cor}

    \begin{Cor} \label{Cor6.2}
An adapted frame $\{X_{I}\}$ is (locally) holonomic iff in it
    \begin{equation}    \label{6.21}
R_{\mu\nu}^{a}
= \lindex[\Gamma]{}{\circ}{}_{b\mu}^{a}
=S_{\mu\nu}^{\lambda}
=C_{ab}^{d}
=C_{\mu b}^{\lambda}
=0 .
    \end{equation}
    \end{Cor}

    If the initial frame $\{e_{I}\}$ is changed into~\eref{6.15}, then
the transformation laws of the quantities~\eref{6.19} follow from~\eref{6.18}
and~\eref{6.16}; in particular, the curvature components transform according
to the tensor equation~\eref{3.15b}.

    Let us now pay attention to the case when $(E,\pi,M)$ is a \emph{vector}
bundle endowed with a connection $\Delta^h$.

    According to the above-said in this section, any \emph{adapted} frame
$\{X_{I}\}=\{X_\mu,X_a\}$ in $T(E)$ is equivalent to a pair of frames in
$T(M)$ and $E$ according to
    \begin{equation}    \label{6.22}
\{X_\mu,X_a\} \leftrightarrow
( \{E_\mu=\pi_*|_{\Delta^h}(X_\mu)\} , \{E_a=v^{-1}(X_a)\} ) .
    \end{equation}
Therefore the vertical and horizontal lifts are given by (\cf
lemma~\ref{Lem4.1}, \eref{4.6-2a} and~\eref{4.6-5})
    \begin{subequations}    \label{6.23}
    \begin{align}   \label{6.23a}
\Sec(E,\pi,M)\ni Y=Y^aE_a
& \xrightarrow{v} v(Y) := Y^v = (Y^a\circ\pi)X_a \in \Delta^v
\\          \label{6.23b}
\mathcal{X}(M)\ni F=F^\mu E_\mu
& \xrightarrow{h} h(F) := F^h =(F^\mu\circ\pi)X_\mu \in \Delta^h .
    \end{align}
    \end{subequations}
Thus, we have the linear isomorphism
    \begin{equation}    \label{6.24}
    \begin{split}
(h,v)\colon\mathcal{X}(M)\times\Sec(E,\pi,M) &\to \mathcal{X}(E)
\\
(h,v)\colon (F,Y) & \mapsto (F^h,Y^v)
    \end{split}
    \end{equation}
which explains why the covariant derivatives (see definition~\ref{Defn4.2})
represent an equivalent description of the linear connections in vector
bundles. Since any vector field
$\xi=(\xi^{I}\circ\pi)X_{I}\in\mathcal{X}(E)$ has a unique
decomposition $\xi=\xi^h\oplus\xi^v$, with
 $\xi^h=(\xi^\mu\circ\pi)X_\mu$ and $\xi^v=(\xi^a\circ\pi)X_a$,
we have
    \begin{equation}    \label{6.25}
(h,v)^{-1}(\xi)
= ( \pi_*|_{\Delta^h}(\xi^h) , v^{-1}(\xi^v) )
= (\xi^\mu E_\mu,\xi^a E_a) .
    \end{equation}

    Suppose $\{X_{I}\}$ and $\{\tilde{X}_{I}\}$ are two adapted frames.
Then they are connected by (cf.~\eref{6.3a} and~\eref{6.16})
    \begin{equation}    \label{6.26}
\tilde{X}_\mu = (B_\mu^\nu\circ\pi) X_\nu
\quad
\tilde{X}_a = (B_a^b\circ\pi) X_b ,
    \end{equation}
where $[B_\mu^\nu]$ and $[B_a^b]$ are some non-degenerate matrix-valued
functions on $M$. The pairs of frames corresponding to them, in
accordance with~\eref{6.22}, are related via
    \begin{equation}    \label{6.27}
\tilde{E}_\mu = B_\mu^\nu E_\nu
\quad
\tilde{E}_a = B_a^b E_b
    \end{equation}
and \emph{vice versa}.

    \begin{Prop}    \label{Prop6.1}
Let $\Delta^h$ be a \emph{linear} connection on a \emph{vector} bundle
$(E,\pi,M)$ and $\{X_\mu\}$ be the frame adapted for $\Delta^h$ to a
frame $\{e_{I}\}$ such that $\{e_a\}$ is a basis for $\Delta^v$ and
    \begin{equation}    \label{6.28}
(e_\mu,e_a)|_U
=
(\pd_\nu,\pd_b)
\cdot
    \begin{bmatrix}
B_\mu^\nu\circ\pi   & 0 \\
(B_{c\mu}^{b}\circ\pi)\cdot E^c & B_a^b\circ \pi
    \end{bmatrix}
=
\bigl(
(B_\mu^\nu\circ\pi)\pd_\nu + ((B_{c\mu}^{b}\circ\pi)\cdot E^c)\pd_b ,
 (B_a^b\circ\pi)\pd_b
\bigl) ,
    \end{equation}
where $\pd_{I}:=\frac{\pd}{\pd u^{I}}$ for some local bundle
coordinates $\{u^{I}\}=\{u^\mu=x^\mu\circ\pi,u^b=E^b\}$ on $U\subseteq E$,
$[B_\mu^\nu]$ and $[B_a^b]$ are non\ndash degenerate matrix\ndash valued
functions on $U$, $B_{c\mu}^{b}\colon U\to\field$, and $\{E^a\}$ is the
coframe dual to $\{E_a=v^{-1}(X_a)\}$. Then the 2\ndash index coefficients
$\Gamma_\mu^a$ of $\Delta^h$ in $\{X_{I}\}$ have the representation
(cf.~\eref{4.16})
    \begin{equation}    \label{6.29}
\Gamma_\mu^a = - (\Gamma_{b\mu}^{a}\circ\pi)\cdot E^b
    \end{equation}
on $U$ for some functions $\Gamma_{b\mu}^{a}\colon U\to\field$, called
\emph{3\ndash index coefficients of $\Delta^h$ in $\{X_{I}\}$}.
    \end{Prop}

    \begin{Rem} \label{Rem6.1}
The representation~\eref{6.29} is not valid for frames more general than the
ones given by~\eref{6.28}. Precisely, equation~\eref{6.29} is valid if and
only if ~\eref{6.28} holds for some local coordinates $\{u^{I}\}$ on $U$
--- see~\eref{6.17}.
    \end{Rem}

    \begin{Proof}
Writing~\eref{6.17} for the transformation
$\{\pd_{I}\}\mapsto\{e_{I}\}$, with $\{e_{I}\}$ given
by~\eref{6.28}, we get~\eref{6.29} with
\[
\Gamma_{b\mu}^{a}
= ( [B_d^e]^{-1})_c^a
  ( \lindex[\Gamma]{}{\pd}{}_{b\nu}^{c} B_\mu^\nu + B_{b\mu}^{c} ) ,
\]
where $\lindex[\Gamma]{}{\pd}{}_{b\nu}^{a}$ are the 3-index coefficients of
$\Delta^h$ in the frame adapted to the coordinates $\{u^{I}\}$
(see~\eref{4.16}).
    \end{Proof}

    Let $\{X_{I}\}$ and $\{\tilde{X}_{I}\}$ be frames adapted to
$\{e_{I}\}$ and $\{\tilde{e}_{I}\}$, respectively, with
(cf.~\eref{6.28})
    \begin{equation}    \label{6.30}
(\tilde{e}_\mu,\tilde{e}_a)
=
(e_\nu,e_b) \cdot
    \begin{bmatrix}
B_\mu^\nu\circ\pi   & 0 \\
(B_{c\mu}^{b}\circ\pi)\cdot E^c & B_a^b\circ \pi
    \end{bmatrix} \ ,
    \end{equation}
in which $\Delta^h$ admits 3-index coefficients. Then, due to~\eref{6.17}
and~\eref{6.29}, the 3\ndash index coefficients $\Gamma_{b\mu}^{a}$ and
$\tilde{\Gamma}_{b\mu}^{a}$ of $\Delta^h$ in  respectively $\{X_{I}\}$ and
$\{\tilde{X}_{I}\}$ are connected by (cf.~\eref{4.25})
    \begin{equation}    \label{6.31}
\tilde{\Gamma}_{b\mu}^{a}
= \bigl([B_f^e]^{-1}\bigr)_c^a
  (\Gamma_{d\nu}^{c} B_\mu^\nu + B_{d\mu}^{c}) B_b^d .
    \end{equation}

    \begin{Exrc}    \label{Exrc6.3}
Prove that the transformation
$\{e_{I}\}\mapsto\{\tilde{e}_{I}\}$, with $\{\tilde{e}_{I}\}$ given
by~\eref{6.30}, is the most general one that preserves the existence of
3\ndash index coefficients of $\Delta^h$ provided they exist in
$\{e_{I}\}$.
    \end{Exrc}

	Introducing the matrices
$\Gamma_\mu:=[\Gamma_{b\mu}^{a}]_{a,b=n+1}^{n+r}$,
$\tilde{\Gamma}_\mu:=[\tilde{\Gamma}_{b\mu}^{a}]_{a,b=n+1}^{n+r}$,
$B:=[B_b^a]$, and $B_\mu:=[B_{b\mu}^{a}]$, we rewrite~\eref{6.31} as
(cf.~\eref{4.25'})
    \begin{equation}
    \tag{\protect\ref{6.31}$^\prime$}   \label{6.31'}
\tilde{\Gamma}_\mu
=
B^{-1}\cdot (\Gamma_\nu B_\mu^\nu +B_\mu)\cdot B.
    \end{equation}
A little below (see the text after equation~\eref{6.33}), we shall prove that
the compatibility of the developed formalism with the theory of covariant
derivatives requires further restrictions on the general transformed
frames~\eref{6.15} to the ones given by~\eref{6.30} with
    \begin{equation}    \label{6.32}
B_\mu= \tilde{E}_\mu(B)\cdot B^{-1} = B_\mu^\nu E_\nu(B) \cdot B^{-1} ,
    \end{equation}
where
\(
\tilde{E}_\mu
:=\pi_*|_{\Delta^h}(\tilde{X}_\mu)
=\pi_*|_{\Delta^h}((B_\mu^\nu\circ\pi)X_\nu)
= B_\mu^\nu E_\nu .
\)
In this case,~\eref{6.31'} reduces to (cf.~\eref{4.25'})
    \begin{equation}    \label{6.33}
\tilde{\Gamma}_\mu
= B_\mu^\nu B^{-1}\cdot (\Gamma_\nu\cdot B + E_\nu(B))
= B_\mu^\nu(B^{-1}\cdot\Gamma_\nu - E_\nu(B^{-1})) \cdot B .
    \end{equation}

    At last, a few words on the covariant derivative operators $\nabla$
are in order. Without lost of generality, we define such an
operator~\eref{4.35} via the equations~\eref{4.40}. Suppose $\{E_\mu\}$ is a
basis for $\mathcal{X}(M)$ and $\{E_a\}$ is a one for $\Sec^1(E,\pi,M)$.
Define the \emph{components} $\Gamma_{b\mu}^{a}\colon M\to\field$ of $\nabla$
in the pair of frames $(\{E_\mu\},\{E_a\})$ by (cf.~\eref{4.41})
    \begin{equation}    \label{6.35}
\nabla_{E_\mu}(E_b) = \Gamma_{b\mu}^{a} E_a .
    \end{equation}
Then~\eref{4.40} imply (cf.~\eref{4.37})
\[
\nabla_F Y = F^\mu( E_\mu(Y^a) +\Gamma_{b\mu}^{a} Y^b) E_a
\]
for $F=F^\mu E_\mu\in\mathcal{X}(M)$ and $Y=Y^aE_a\in\Sec^1(E,\pi,M)$. A
change $(\{E_\mu\},\{E_a\})\mapsto(\{\tilde{E}_\mu\},\{\tilde{E}_a\})$, given
via~\eref{6.27}, entails
    \begin{equation}    \label{6.36}
\Gamma_{b\mu}^{a}\mapsto \tilde{\Gamma}_{b\mu}^{a}
=
B_\mu^\nu \bigl([B_f^e]^{-1}\bigr)_c^a
                (\Gamma_{d\nu}^{c} B_b^d + E_\nu(B_b^c)) ,
    \end{equation}
as a result of~\eref{6.35}. In a more compact matrix form, the last result
reads
    \begin{equation}
    \tag{\protect\ref{6.36}$^\prime$}   \label{6.36'}
\tilde{\Gamma}_\mu = B_\mu^\nu B^{-1}\cdot (\Gamma_\nu\cdot B + E_\nu(B))
    \end{equation}
with $\Gamma_\mu:=[\Gamma_{_b\mu}^{a}]$,
$\tilde{\Gamma}_\mu:=[\tilde{\Gamma}_{b\mu}^{a}]$, and $B:=[B_b^a]$.

    Thus, if we identify the 3-index coefficients of $\Delta^h$, defined
by~\eref{6.29}, with the components of $\nabla$, defined by~\eref{6.35},~%
\footnote{~%
Such an identification is justified by the definition of $\nabla$ via the
parallel transport assigned to $\Delta^h$ (see proposition~\ref{Prop4.2}) or
via a projection, generated by $\Delta^h$, of a suitable Lie derivative on
$\mathfrak{X}(E)$ (see definition~\ref{Defn4.2}).%
}
then the quantities~\eref{6.31'} and~\eref{6.36'} must coincide, which
immediately leads to the equality~\eref{6.32}. Therefore
    \begin{equation}    \label{6.37}
(e_\mu,e_a) \mapsto (\tilde{e}_\mu,\tilde{e}_a)
=
(e_\nu,e_b) \cdot
    \begin{bmatrix}
B_\mu^\nu\circ\pi & 0 \\
\bigl( (B_\mu^\nu E_\nu(B_d^b) (B^{-1})_c^d)\circ\pi \bigr) E^c    &
B_a^b\circ\pi
    \end{bmatrix}
\bigg|_{B=[B_a^b]}
    \end{equation}
is the most general transformation between frames in $T(E)$ such that the
frames adapted to them are compatible with the linear connection and the
covariant derivative corresponding to it. In
particular, such are all frames $\bigl\{\frac{\pd}{\pd
u^{I}}\bigr\}$ in $T(E)$ induced by vector bundle coordinates
$\{u^{I}\}$ on $E$ --- see~\eref{4.23} and~\eref{3.1}--\eref{3.3}; the
rest members of the class of frames mentioned are obtained from them
via~\eref{6.37} with $e_{I}=\frac{\pd}{\pd u^{I}}$ and non\ndash degenerate
matrix\ndash valued functions $[B_\mu^\nu]$ and $B$.

    If $\{X_{I}\}$ (resp.\ $\{\tilde{X}_{I}\}$) is the frame
adapted to a frame $\{e_{I}\}$ (resp.\ $\{\tilde{e}_{I}\}$), then
the change $\{e_{I}\} \mapsto\{\tilde{e}_{I}\}$, given
by~\eref{6.37}, entails $\{X_{I}\} \mapsto\{\tilde{X}_{I}\}$
with $\{\tilde{X}_{I}\}$ given by~\eref{6.26} (see~\eref{6.15}
and~\eref{6.16}). Since the last transformation is tantamount to
the change
    \begin{equation}    \label{6.38}
( \{E_\mu\}, \{E_a\} ) \mapsto ( \{\tilde{E}_\mu\},\{\tilde{E}_a\} )
    \end{equation}
of the basis of $\mathcal{X}(M)\times\Sec(E,\pi,M)$ corresponding
to $\{X_{I}\}$ via the isomorphism~\eref{6.24}
(see~\eref{6.22}, \eref{6.26}, and~\eref{6.27}), we can say that
the transition~\eref{6.38} induces the change~\eref{6.36} of the
3\ndash index coefficients of the connection $\Delta^h$. Exactly
the same is the situation one meets in the
literature~\cite{K&N-1,Warner,Poor} when covariant derivatives are
considered (and identified with connections).

    Regardless that the change~\eref{6.37} of the frames in $T(E)$ looks
quite special, it is the most general one that, through~\eref{6.16}
and~\eref{6.22}, is equivalent to an arbitrary change~\eref{6.38} of a basis
in $\mathcal{X}(M)\times\Sec(E,\pi,M)$, \ie of a pair of frames in $T(M)$ and
$E$.

    We would like to remark that, in the general case, equation~\eref{4.45}
also holds with $F=F^\mu E_\mu$, $G=G^\mu E_\mu$, and
    \begin{equation}    \label{6.39}
(R(E_\mu,E_\nu))(E_b) = R_{b\mu\nu}^{a} E_a ,
    \end{equation}
so that
    \begin{equation}    \label{6.40}
R_{b\mu\nu}^{a}
=
E_\mu(\Gamma_{b\nu}^{a}) - E_\nu(\Gamma_{b\mu}^{a})
- \Gamma_{b\mu}^{c} \Gamma_{c\nu}^{a}
+ \Gamma_{b\nu}^{c} \Gamma_{c\mu}^{a}
- \Gamma_{b\lambda}^{a} C_{\mu\nu}^{\lambda} ,
    \end{equation}
where the functions $C_{\mu\nu}^{\lambda}$ define the anholonomy object of
$\{E_\mu\}$ via $[E_\mu,E_\nu]_{\_}=:C_{\mu\nu}^{\lambda}E_\lambda$.

	The above results, concerning linear connections on vector bundles,
can be generalized for affine connections on vector bundles. For instance, the
analogue of propositions~\ref{Prop6.1} reads.

    \begin{Prop}    \label{Prop6.2}
Let $\Delta^h$ be an \emph{affine} connection on a \emph{vector} bundle
$(E,\pi,M)$ and $\{X_\mu\}$ be the frame adapted for $\Delta^h$ to a
frame $\{e_{I}\}$ such that $\{e_a\}$ is a basis for $\Delta^v$ and
    \begin{equation}    \label{6.41}
(e_\mu,e_a)|_U
=
(\pd_\nu,\pd_b)
\cdot
    \begin{bmatrix}
B_\mu^\nu\circ\pi   & 0 \\
(B_{c\mu}^{b}\circ\pi)\cdot E^c & B_a^b\circ \pi
    \end{bmatrix}
=
\bigl(
(B_\mu^\nu\circ\pi)\pd_\nu + ((B_{c\mu}^{b}\circ\pi)\cdot E^c)\pd_b ,
 (B_a^b\circ\pi)\pd_b
\bigl) ,
    \end{equation}
where $\pd_{I}:=\frac{\pd}{\pd u^{I}}$ for some local bundle
coordinates $\{u^{I}\}=\{u^\mu=x^\mu\circ\pi,u^b=E^b\}$ on $U\subseteq E$,
$[B_\mu^\nu]$ and $[B_a^b]$ are non\ndash degenerate matrix\ndash valued
functions on $U$, $B_{c\mu}^{b}\colon U\to\field$, and $\{E^a\}$ is the
coframe dual to $\{E_a=v^{-1}(X_a)\}$. Then the 2\ndash index coefficients
$\Gamma_\mu^a$ of $\Delta^h$ in $\{X_{I}\}$ have the representation
(cf.~\eref{4.61})
    \begin{equation}    \label{6.42}
\Gamma_\mu^a = - (\Gamma_{b\mu}^{a}\circ\pi)\cdot E^b + G_\mu^a\circ\pi
    \end{equation}
on $U$ for some functions $\Gamma_{b\mu}^{a},G_\mu^a\colon U\to\field$.
    \end{Prop}

    \begin{Rem} \label{Rem6.2}
The representation~\eref{6.42} is not valid for frames more general than the
ones given by~\eref{6.41}. Precisely, equation~\eref{6.42} is valid if and
only if ~\eref{6.41} holds for some local coordinates $\{u^{I}\}$ on $U$
--- see~\eref{6.17}.
    \end{Rem}

    \begin{Proof}
Writing~\eref{6.17} for the transformation
$\{\pd_{I}\}\mapsto\{e_{I}\}$, with $\{e_{I}\}$ given
by~\eref{6.41}, we get~\eref{6.42} with
\[
\Gamma_{b\mu}^{a}
= ( [B_d^e]^{-1})_c^a
  ( \lindex[\Gamma]{}{\pd}{}_{b\nu}^{c} B_\mu^\nu + B_{b\mu}^{c} )
\quad
G_\mu^a = ( [B_d^e]^{-1})_b^a \lindex[G]{}{\pd\mspace{-3mu}}_\nu^b B_\mu^\nu
\]
where $\lindex[\Gamma]{}{\pd}{}_{b\nu}^{a}$ and
$\lindex[G]{}{\pd\mspace{-3mu}}_\nu^b$ are defined via the 2-index coefficients
$\lindex[\Gamma]{}{\pd}_\mu^a$ of $\Delta^h$ in the frame adapted to the
coordinates $\{u^{I}\}$ via
\(
\lindex[\Gamma]{}{\pd}_\mu^a
= - (\lindex[\Gamma]{}{\pd}_{b\mu}^{a}\circ\pi)\cdot E^b
  + \lindex[G]{}{\pd\mspace{-3mu}}_\mu^a\circ\pi
\)
(see theorem~\ref{Thm4.3}).
    \end{Proof}

    Let $\{X_{I}\}$ and $\{\tilde{X}_{I}\}$ be frames adapted to
$\{e_{I}\}$ and $\{\tilde{e}_{I}\}$, respectively, with
(cf.~\eref{6.41})
    \begin{equation}    \label{6.43}
(\tilde{e}_\mu,\tilde{e}_a)
=
(e_\nu,e_b) \cdot
    \begin{bmatrix}
B_\mu^\nu\circ\pi   & 0 \\
(B_{c\mu}^{b}\circ\pi)\cdot E^c & B_a^b\circ \pi
    \end{bmatrix} \ ,
    \end{equation}
in which~\eref{6.42} holds for $\Delta^h$. Then, due to~\eref{6.17}
and~\eref{6.42}, the pairs $(\Gamma_{b\mu}^{a},G_\mu^a)$ and
$(\tilde{\Gamma}_{b\mu}^{a},\tilde{G}_\mu^a)$ for $\Delta^h$ in respectively
$\{X_{I}\}$ and $\{\tilde{X}_{I}\}$ are connected by (cf.~\eref{4.61}
and~\eref{4.62})
    \begin{subequations}    \label{6.44}
    \begin{align}    \label{6.44a}
\tilde{\Gamma}_{b\mu}^{a}
& = \bigl([B_f^e]^{-1}\bigr)_c^a
  (\Gamma_{d\nu}^{c} B_\mu^\nu + B_{d\mu}^{c}) B_b^d
\\		    \label{6.44b}
\tilde{G}_{\mu}^{a}
& = \bigl([B_f^e]^{-1}\bigr)_b^a  G_{\nu}^{b} B_\mu^\nu
    \end{align}
    \end{subequations}

    \begin{Exrc}    \label{Exrc6.4}
Prove that the transformation
$\{e_{I}\}\mapsto\{\tilde{e}_{I}\}$, with $\{\tilde{e}_{I}\}$ given
by~\eref{6.43}, is the most general one that preserves the existence of
the relation~\eref{6.42} for $\Delta^h$ provided it holds in $\{e_{I}\}$.
    \end{Exrc}

	Further one can repeat \emph{mutatis mutandis}  the text after
exercise~\ref{Exrc6.3} to the paragraph containing equation~\eref{6.38}
including.


\section {Conclusion}
\label{Conclusion}

	In this paper we have presented a short (and partial) review of (one
of the approaches to) the connection theory on bundles whose base and bundle
spaces are ($C^2$) differentiable manifolds. Special attention was paid to
connections, in particular linear ones, on vector bundles, which find wide
applications in physics~\cite{Mangiarotti&Sardanashvily,Eguchi&et_al.}.
However, many other approaches, generalizations, alternative descriptions,
particular methods, etc.\ were not mentioned at all. In particular, these
include: connections on more general (\eg topological) bundles, connections
on principal bundles (which are important in the gauge field theories),
holonomy groups, flat connections, Riemannian connections, etc., etc. The
surveys~\cite{Lumiste-1971,Alekseevskii&et_al.} contain essential information
on these and many other items. Consistent and self\ndash contained exposition
of such problems can be found in~\cite{Lichnerowicz,K&N,Greub&et_al.,Poor}.

	If additional geometric structures are added to the theory considered
in Sect.~\ref{Sect3}, there will become important connections compatible
with these structures. In this way arise many theories of particular
connections; we have demonstrated that on the example of linear connections
on vector bundles (Sect.~\ref{Sect4}). Here are two more such cases.

	If a free right action $R\colon g\mapsto R_g\colon E\to E$, $g\in G$,
of a Lie group $G$ on the bundle space $E$ of a bundle $(E,\pi,M)$ is given
and $\pi\colon E\to M=E/G$ is the canonical projection, we have a principal
bundle $(E,\pi,M,G)$. The connections that respect the right action $R$ are
the most important ones on principal bundles. Such a connection $\Delta^h$ is
defined by definition~\ref{Defn3.1} to which the condition
	\begin{equation}	\label{Con.1}
(R_g)_* (\Delta_p^h) = \Delta_{R_g(p)}^{h}
\qquad g\in G\quad p\in E
	\end{equation}
is added and is called a principal connection. Alternatively, one can require
the parallel transport $\Psf$ generated by $\Delta^h$ to commute with $R$, \viz
	\begin{equation}	\label{Con.2}
R_g\circ\Psf^\gamma = \Psf^\gamma\circ R_g
\qquad g\in G\quad \gamma\colon [\sigma,\tau]\to M .
	\end{equation}
The theory of connections satisfying~\eref{Con.1} is very well developed;
see, e.g.,~\cite{K&N-1,Greub&et_al.}.

	Suppose a real bundle $(E,\pi,M)$ is endowed with a bundle metric $g$,
\ie $g\colon x\mapsto g_x$, $x\in M$, with
$g_x\colon\pi^{-1}(x)\times\pi^{-1}(x)\to\field[R]$
being  bilinear and non\ndash degenerate mapping for all $x\in M$.
The equality
	\begin{equation}	\label{Con.3}
g_{\gamma(\sigma)}
= g_{\gamma(\tau)} \circ \bigl( \Psf^{\gamma}\times\Psf^\gamma\bigr)
\qquad \gamma\colon[\sigma,\tau,]\to M,
	\end{equation}
which expresses the preservation of the $g$-scalar products by the parallel
transport $\Psf$ assigned to a connection $\Delta^h$, specifies the class of
$g$\ndash compatible (metric\ndash compatible) connections on $(E,\pi,M)$.
Such are the Riemannian connections on a Riemannian manifold $M$, which are
 $g$\ndash compatible connections on the tangent bundle $(T(M),\pi_T,M)$;
see, for instance,~\cite{K&N-1,Poor}.

	The consideration of arbitrary (co)frames in Sect.~\ref{Sect6} may
seem slightly artificial as the general theory can be developed without them.
However, this is not the generic case when one starts to apply the connection
theory for investigation of particular problems. It may happen that some
problem has solutions in general (co)frames but it does not possess solutions
when (co)frames generated directly by local coordinates are involved. For
example~\cite{bp-Frames-n+point}, local coordinates (holonomic frames) normal
at a given point for a covariant derivative operator (linear connection)
$\nabla$ on a manifold exist if $\nabla$ is torsionless at that point, but
anholonomic frames normal at a given point for $\nabla$ exist always.


\section*{Acknowledgments}

	This work was partially supported by the National Science Fund of
Bulgaria under Grant No.~F~1515/2005.


\addcontentsline{toc}{section}{References}
\bibliography{bozhopub,bozhoref}
\bibliographystyle{unsrt}
\addcontentsline{toc}{subsubsection}{This article ends at page}

\end{document}